\pdfoutput=1
\documentclass[twocolumn]{aastex631}
\usepackage{graphicx}
\usepackage{amsmath} 
\usepackage{xcolor}
\usepackage{subfigure}
\usepackage{tikz}
\tikzset{
  load/.style   = {ultra thick,-latex},
  stress/.style = {-latex},
  dim/.style    = {latex-latex},
  axis/.style   = {-latex},
}
\usepackage{hyperref}

\newcommand{\vb}[1]{\boldsymbol{#1}}
\newcommand{\nablab}{\boldsymbol{\nabla}}

\newcommand{\dpartial}[2]{\frac{\partial #1}{\partial #2}}

\newcommand{\dtotal}[2]{\frac{\mathrm{d} #1}{\mathrm{d} #2}}

\newcommand{\tw}{\the\textwidth}
\newcommand{\cw}{\the\columnwidth}
\graphicspath{{./figures/}{}}

\defcitealias{Pontin2023}{paper~1}

\submitjournal{ApJ}
\received{Aug 15, 2023}
\revised{Oct 25, 2023}
\accepted{Nov 6, 2023}
\shorttitle{Tides in rotating and stratified planets}
\shortauthors{Pontin, Barker \& Hollerbach}

\begin{document}
\title{Tidal dissipation in stably stratified and semi-convective regions of rotating giant planets: incorporating Coriolis forces}

\author{Christina M. Pontin}
\affiliation{Department of Applied Mathematics, School of Mathematics, University of Leeds, Leeds, LS2 9JT, UK}

\author[0000-0003-4397-7332]{Adrian J. Barker}
\affiliation{Department of Applied Mathematics, School of Mathematics, University of Leeds, Leeds, LS2 9JT, UK}
\correspondingauthor{Adrian Barker}
\email{A.J.Barker@leeds.ac.uk}

\author[0000-0001-8639-0967]{Rainer Hollerbach}
\affiliation{Department of Applied Mathematics, School of Mathematics, University of Leeds, Leeds, LS2 9JT, UK}
\email{R.Hollerbach@leeds.ac.uk}

\begin{abstract}
We study how stably stratified or semi-convective layers alter tidal dissipation rates associated with the generation of inertial, gravito-inertial, interfacial and surface gravity waves in rotating giant planets. We explore scenarios in which stable (non-convective) layers contribute to the high rates of tidal dissipation observed for Jupiter and Saturn in our solar system. Our model is an idealised spherical Boussinesq system incorporating Coriolis forces to study effects of stable stratification and semi-convective layers on tidal dissipation. Our detailed numerical calculations consider realistic tidal forcing and compute the resulting viscous and thermal dissipation rates. The presence of an extended stably stratified fluid core significantly enhances tidal wave excitation of both inertial waves (due to rotation) in the convective envelope and gravito-inertial waves in the dilute core. We show that a sufficiently strongly stratified fluid core enhances inertial wave dissipation in a convective envelope much like a solid core does. We demonstrate that efficient tidal dissipation rates (and associated tidal quality factors $Q'$) -- sufficient to explain the observed migration rates of Saturn’s moons -- are predicted at the frequencies of the orbiting moons due to the excitation of inertial or gravito-inertial waves in our models with stable layers (without requiring resonance-locking). Stable layers could also be important for tidal evolution of hot and warm Jupiters, and hot Neptunes, providing efficient tidal circularisation rates. Future work should study more sophisticated planetary models that also account for magnetism and differential rotation, as well as the interaction of inertial waves with turbulent convection.
\end{abstract}


\section{Introduction}\label{sec:intro}

Jupiter and Saturn in our solar system are the best studied giant planets in the Universe. They have been explored by recent and ongoing space missions such as Cassini and Juno, thereby allowing us to probe their gravity and magnetic fields, and to observe fascinating features in their ring systems \citep[e.g.][]{Bolton2017,Wahl2017,Guillot2018,Durante2020,Ingersoll2020,Connerney2022}. This wealth of data has constrained models of the interior structures of these bodies, and we have found that classical models consisting of a solid rocky core with two distinct, chemically homogeneous layers above, are inconsistent with observations. For Jupiter, gravity field measurements from Juno have inferred a dilute fluid core (which may or may not be stably stratified/non-convective) containing heavy elements extending out to 40-50\% of Jupiter's radius \citep{Wahl2017}. Saturn has had an extended stably-stratified layer inferred in its interior, probably produced by compositional gradients, and possibly extending out to 60\% of the planetary radius, based on features observed in its rings thought to be produced by resonances with global oscillation modes within the planet \citep{Marley1993,Hedman2013,Fuller2014,Mankovich2021}. These recent findings strongly motivate new studies of giant planets containing extended stably-stratified fluid layers in their interiors \citep[e.g.][]{Vazan2018}. Such studies could also shed light on the structures and properties of extrasolar planets including hot and warm Jupiters, as well as hot Neptunes \citep[e.g.][]{Guillot2022}.

The dissipative tidal responses of Jupiter and Saturn have been probed using astrometric observations over the past century or so that measure the orbital migration of the Galilean and Saturnian moons \citep{Lainey2009,Lainey2012,Lainey2017,Lainey2020}. To explain the observed rates of outward migration of these moons, we require much more efficient tidal dissipation in these planets than previously estimated \citep{GS1966}. Such efficient rates of tidal dissipation are not currently understood theoretically, but have motivated an increasing number of works to explore tides in giant planets. Possible dissipative mechanisms that have been proposed include: inertial waves (restored by Coriolis forces) in convective regions \citep{Ogilvie2004,Wu2005b,Goodman2009}, gravity or gravito-inertial waves in stable layers (restored by buoyancy forces, and also by rotation) -- which might be locked in resonance \citep{Fuller2016} -- and interactions of equilibrium tides with turbulent convection, though the latter mechanism is not widely believed to be important and is more uncertain \citep{Goldreich1977,Duguid2020,T2021,BA2021,deVries2023}. Finally, dissipation in the visco-elastic rocky/icy cores of giant planets have also been proposed to be important \citep[e.g.][]{Remus2012,Storch2014,Remus2015,Storch2015,Lainey2017}, and are worthy of future study, even if fluid mechanisms are typically favoured currently due to the instability and mixing of such cores in the high temperature and pressure environments near the centres of giant planets \citep[e.g.][]{Mazevet2015}.

The efficiency of tidal dissipation is often quantified by the modified tidal quality factor
\begin{equation}
Q' = \frac{3}{2k_2} \frac{2\pi E_0}{\int D \mathrm{d} t},
\end{equation}
where $k_2$ is the quadrupolar Love number (essentially a measure of the degree of central concentration of mass), $E_0$ is the stored tidal energy, $D$ is the energy dissipation rate, and the integral represents the energy dissipated over one tidal period. Observations require $Q'\approx (0.94\pm 0.44)\times 10^4$ for Saturn and $Q'\approx (1.59 \pm 0.25)  \times 10^5$ for Jupiter, for most tidal frequencies probed \citep{Lainey2009,Lainey2012,Lainey2017}.

Stabilising compositional gradients can lead to the inhibition of ordinary convection if they can compete with the destabilising thermal gradients. Since planetary interiors have more efficient thermal diffusion of heat than viscous diffusion of momentum \citep[i.e.~they have thermal Prandtl numbers smaller than one,][]{Guillot2004}, convection instead takes the form of double-diffusive convection \citep[an oscillatory linear instability or ``overstability" involving excitation of internal gravity waves, e.g.][]{Garaud2018}, whose nonlinear evolution can produce layering or density staircases, involving well-mixed steps separated by sharp, diffusive interfaces \citep{Wood2013}. Whether or not this outcome is expected for conditions relevant in giant planets is an open question \citep{Fuentes2022}. Nevertheless, the possible presence and effects of stably-stratified layers in giant planets, which may or may not involve density staircases, are important to study and are likely to strongly influence the planet's tidal response. Indeed, the influence of interior stably-stratified layers on planetary tidal flows is currently poorly understood, and has motivated our prior work on this problem \citep{Andre2017,Andre2019,Pontin2020,Pontin2023}.

We present new theoretical models of giant planets containing stable layers similar to those constrained observationally for Saturn, and hypothesised for Jupiter, to explore dissipation of tidal flows inside these giant planets. Our idealised (Boussinesq) model of a rotating and tidally-forced planet analyses the dissipative fluid response in a spherical shell using both linear theoretical analysis and numerical calculations obtained with high-resolution spectral methods. In this paper, we build upon \cite{Pontin2023} (hereafter paper 1) to study rotating and stratified planets, by fully incorporating Coriolis forces to study the effects of rotation on the tidal response. Most notably, the incorporation of Coriolis forces allows for inertial (restored by rotation) and gravito-inertial (restored by both buoyancy and rotation) waves to be excited by tidal forcing and subsequently dissipated. Since Coriolis forces are important for planetary applications, including the tidal responses of Jupiter and Saturn at the forcing frequencies of their moons, our model is more realistic than the one considered in \citetalias{Pontin2023}. However, we continue to neglect centrifugal deformations and to consider a spherical body for simplicity. Incorporating centrifugal deformations considerably complicates the analysis \citep[e.g.][]{Braviner2014,Braviner2015,Barker2016,Dewberry2022} but is unlikely to change our results substantially, and we leave its study to future work.

Our work is complementary to the recent studies of \cite{Lin2023} and \cite{Dewberry2023} who study more sophisticated compressible and self-gravitating planetary models numerically, even incorporating centrifugal deformations in the case of the latter, which are important advances. We have chosen to focus on a simpler Boussinesq model here to allow a much wider exploration of parameter space, and also to permit a deeper understanding of the effects of stable layers, and semi-convective ones, on the tidal response. Our Boussinesq (incompressible) model incorporates buoyancy forces, and hence allows us to study gravito-inertial and inertial wave excitation and dissipation, but it neglects the density variation expected in giant planets. The overall features of inertial wave excitation in spherical shells are however broadly similar in incompressible and compressible polytropic models \citep[e.g.][]{Ogilvie2013}, so while the particular linear predictions for resonant frequencies are likely to depend strongly on the particular model, much of the behaviour we observe here is likely to be similar in more complex models. Here we study the tidal response of semi-convective layers for the first time in global models with rotation, and perform a detailed exploration of parameter space for interior stable layers.

The structure of this paper is as follows. \S~\ref{sec:model} presents our model, including the governing equations, energetics, planetary density/entropy profiles and numerical methods adopted (see \citetalias{Pontin2023} for further details). \S~\ref{results} presents an overview of our numerical exploration of parameter space, analysing the frequency-dependent tidal response and its dissipative properties. In \S~\ref{varyparam} we vary the parameters (rotation rate, core sizes, smooth vs layered density profiles, diffusivities) and also analyse an integrated measure of the response, given by the frequency-averaged dissipation rate, to determine how this correspondingly varies. In \S~\ref{dilute}, we compare various models of dilute stably-stratified fluid cores with a rigid core model. Finally, we apply our results to a Saturn-like model in \S~\ref{Saturn} and conclude in \S~\ref{conclusions}.

\section{Model}\label{sec:model}

\subsection{Governing equations}

We briefly recap the model adopted in \citetalias{Pontin2023} and highlight our extensions to incorporate rotation. We consider the linearised momentum equation in the Boussinesq approximation, in the frame rotating at the rate $\vb{\Omega}=\Omega\vb{e}_z$,
\begin{equation}\label{eq:mtm}
\frac{\partial \vb{u}}{\partial t}  +2 \vb{\Omega} \times \vb{u}=-\frac{1}{\rho_0}\nablab p - b \vb{g} - \nablab \psi + \nu \nablab^2 \vb{u},
\end{equation}
where $\vb{u}$, $p$ are velocity and Eulerian pressure perturbations, $\rho_0$ is the reference density and $\psi$ is the tidal potential. We use spherical polar coordinates $(r,\theta,\phi)$ centred on the planet, and the rotation axis (along $z$) corresponds to $\theta=0$. We define a (non-standard) dimensionless buoyancy variable $b =-\rho/\rho_0$, where $\rho$ is the Eulerian density perturbation, and adopt a gravitational acceleration \hbox{$\vb{g}=-g(r) \boldsymbol{\hat{r}}$}. The quantity $b$ is proportional to the entropy perturbation in our Boussinesq model. We focus predominantly on a homogeneous body with constant density $\rho_0$, for which $g=g_0 r$, where $g_0$ is the surface gravity with $r$ measured in units of the planetary radius (see  \citetalias{Pontin2023} for exploration of different profiles). Self-gravity is neglected (i.e.~we adopt the Cowling approximation) to allow us to develop more detailed understanding and because it is only likely to lead to a moderate linear effect on any quantitative results. The perturbation of the gravitational potential due to surface deformations of our homogeneous model would also lead to unrealistically large effects compared with more realistic centrally-condensed models, further motivating our neglect of its effects here. The flow is assumed to be incompressible, so
\begin{equation}\label{eq:incompres}
\nablab \cdot \vb{u} = 0.
\end{equation}
The heat equation written in terms of $b$ is
\begin{equation}\label{eq:heat}
\dpartial{b}{t} +\frac{u_r}{g}N^2 = \kappa \nablab^2 b,
\end{equation}
where $N^2$ is the (radially varying) squared Brunt-V\"ais\"al\"a or buoyancy frequency, defined to be, 
\begin{equation}
N^2=g\left(\frac{1}{\Gamma_{1}}\frac{\mathrm{d} \ln p_0}{\mathrm{d}r}-\frac{\mathrm{d}\ln\rho_0}{\mathrm{d}r}\right)\propto \frac{\mathrm{d} s}{\mathrm{d}r},
\end{equation}
where $\Gamma_1=\left(\frac{\partial \ln p_0}{\partial \ln \rho_0}\right)_{\mathrm{ad}}$ is the first adiabatic exponent and $s(r)$ is the specific entropy.

We have adopted a single scalar field for the density ($b$) with a single ``thermal diffusivity" ($\kappa$) to model our density/entropy stratification. If compositional gradients are important then the density should strictly consist of two distinct components (thermal and compositional) with two associated (unequal) diffusivities. This is necessary for double-diffusive convection to operate in giant planet interiors and potentially produce layered density structures \citep[e.g.][]{Garaud2018}, but we leave exploration of double-diffusive effects here to future work.

We introduce tidal forcing by considering the dominant component of the tidal potential
\begin{equation}\label{eq:forcing}
\psi = \psi_0~r^2 Y^2_2(\theta,\phi) e^{-i \omega t},
\end{equation}
where $\psi_0 =\sqrt{\frac{6 \pi}{5}}\frac{M_2}{M}\omega_d^2 \big(\frac{r_0}{a} \big)^3$. We take the $l=m=2$ dimensional tidal amplitude for a circular orbit, as defined in \cite{Ogilvie2014}, where $M$ is the mass of the planet, $M_2$ is the companion mass, $a$ is the orbital semi-major axis, and the dynamical frequency $\omega_d=\sqrt{GM/r_0^3}$, where $G$ is the gravitational constant. The forcing frequency is $\omega=2(\Omega_o-\Omega)$, where $\Omega_o$ is the orbital frequency of the satellite and $\Omega$ is the spin frequency of the planet -- the most relevant one for a circular and aligned orbit of a non-synchronously orbiting moon. Note that we solve for the entire linear tidal response to $\psi$ directly; we do not split up the tide into an equilibrium and a dynamical tide, though only the dynamical/wavelike response is typically important for dissipation in our models.

We expand perturbations using spherical harmonics with a harmonic time-dependence such that, 
\begin{equation}\label{eq:u_r}
u_r(r,\theta,\phi,t)= \sum_{l=m}^{\infty} \tilde{u}_r^l(r) Y^{m}_{l}(\theta,\phi)e^{-i\omega t}, 
\end{equation}
\begin{equation}\label{u_theta}
u_{\theta}(r,\theta,\phi,t)= r \sum_{l=m}^{\infty} \bigg[ \tilde{u}_b^l(r) \dpartial{}{\theta} + \frac{\tilde{u}_c^l(r)}{\sin \theta} \dpartial{}{\phi} \bigg] Y^{m}_{l}(\theta,\phi)e^{-i\omega t},
\end{equation}
\begin{equation}\label{u_phi}
u_{\phi}(r,\theta,\phi,t)= r \sum_{l=m}^{\infty} \bigg[ \frac{\tilde{u}_b^l(r)}{\sin \theta} \dpartial{}{\phi} -\tilde{u}_c^l(r) \dpartial{}{\theta} \bigg] Y^{m}_{l}(\theta,\phi)e^{-i\omega t},
\end{equation}
\begin{equation}
p(r,\theta,\phi,t)= \sum_{l=m}^{\infty} \tilde{p}^l(r) Y^{m}_{l}(\theta,\phi)e^{-i\omega t},
\end{equation}
\begin{equation}
b(r,\theta,\phi,t)= \sum_{l=m}^{\infty} \tilde{b}^l(r) Y^{m}_{l}(\theta,\phi)e^{-i\omega t},
\end{equation}
where we adopt the standard normalisation 
\begin{equation*}
\int_0^{2\pi} \int_0^\pi  [Y_{l'}^{m'} (\theta,\phi)]^* Y_l^m(\theta,\phi)  \sin^2\theta\, \mathrm{d} \theta\, \mathrm{d} \phi=\delta_{l,l'}\, \delta_{m,m'}.
\end{equation*}

The resulting differential equations in radius for each $l$ and $m$ are (dropping tildes) 
\begin{align}
\nonumber
(-i \omega) u_r^l + 2 \Omega r \Big( -im u_b^l + (l-1)q_l u_c^{l-1}-(l+2)q_{l+1}u_c^{l+1} \Big) \\ = -\frac{1}{\rho_0} \dtotal{p^l}{r} + g b^l - \dtotal{\psi^l}{r}\delta_{l,2} -\nu \frac{l(l+1)}{r^2} \Big[u_r^l - \dtotal{}{r}(r^2 u_b^l)\Big],
\label{eq:ur_nondim}
\end{align} \vspace{-0.7cm}
\begin{align}
\nonumber
&(-i \omega) r^2 u_b^l + 2 \Omega r^2 \left( \frac{-i m}{l(l+1)}\left( \frac{u_r^l}{r}+u_b^l \right) \right. \\ \nonumber&\left.\hspace{2.5cm}+ \frac{l-1}{l}q_l u_c^{l-1}+\frac{l+2}{l+1}q_{l+1}u_c^{l+1} \right) \\ \nonumber &\hspace{2cm}= -\frac{p^l}{\rho_0}-\psi^l \delta_{l,2} + \nu\, \left[ \frac{2 u_r^l}{r} +\frac{1}{r^2}\dtotal{}{r} \bigg( r^4 \dtotal{u_b^l}{r} \bigg) \right.\\ &\left.\hspace{3cm}-(l-1)(l+2) u_b^l \right],
\end{align} \vspace{-0.7cm}
\begin{align}
\nonumber
&(-i \omega) r^2 u_c^l + 2 \Omega r^2 \bigg(\frac{-i m}{l(l+1)}u_c + \frac{q_l}{l} \frac{u_r^{l-1}}{r}-\frac{q_{l+1}}{l+1}\frac{u_r^{l+1}}{r} \\ \nonumber&\hspace{3cm}+\frac{l-1}{l}q_l u_b^{l-1}-\frac{l+2}{l+1}q_{l+1}u_b^{l+1}\bigg) \\ &\hspace{1.2cm}=\nu \Bigg[\frac{1}{r^2}\dtotal{}{r}\bigg(r^4 \dtotal{u_c^l}{r}\bigg)-(l-1)(l+2)u_c^l \Bigg],
\end{align} \vspace{-0.3cm}
\begin{equation}
\frac{1}{r^2}\dtotal{}{r}(r^2 u_r^l)-l(l+1)u_b^l=0,
\end{equation} \vspace{-0.6cm}
\begin{equation} 
(-i \omega)b^l+N^2\frac{u_r^l}{g} = \kappa\, \Bigg[ \frac{1}{r^2}\dtotal{}{r}\bigg(r^2 \dtotal{b^l}{r} \bigg) - \frac{l(l+1)}{r^2}b^l \Bigg],
\label{eq:b_nondim}
\end{equation}
where $q_l=\big(\frac{l^2-m^2}{4 l^2-1}\big)^{\frac{1}{2}}$. These are consistent with the equations in \cite{Ogilvie2009} when $b=0$. The equations for each $m$ are uncoupled due to the axisymmetric basic state, but the Coriolis term couples components with different angular wavenumbers $l$. Hence, we solve our system for $m=2$ and $l\in[2,l_{max}]$, where $l_{max}$ is taken to be sufficiently large such that the solution converges. Typically, $l_{max}=100$, though larger values are required for the smallest diffusivities explored.

We non-dimensionalise our system with the planetary radius $r_0$ as the unit of length, the reference density $\rho_0$ for density, and $\omega_d^{-1}$ as the time unit. Therefore, we introduce the following dimensionless variables: $r=r_0\hat{r}$, $u_r=r_0 \omega_d \hat{u}_r$, $u_b=\omega_d \hat{u}_b$, $u_c=\omega_d \hat{u}_c$, $\psi=\psi_0\hat{\psi}$, $b=\hat{b}$, $g=g_0\hat{g}$, $p=\rho_0 r_0 g\hat{p}$, $\omega=\sqrt{\frac{g_0}{r_0}}\hat{\omega}$, where $g_0=\omega_d^2 r_0$ is the surface gravity. We henceforth drop hats on all variables. This leads to four dimensionless parameters for each frequency $\omega$ (in units of $\omega_d$),
\begin{equation*}
\small
\frac{\Omega}{\omega_d},~~~
\frac{N^2}{\omega_d^2},~~~
\frac{\nu}{r_0^2 \omega_d}~~~\textrm{and}~~~\frac{\kappa}{r_0^2 \omega_d}~~~\bigg(\textrm{alternatively } \mathrm{Pr}=\frac{\nu}{\kappa}\bigg).
\end{equation*}\noindent
These are varied in later analysis, and are referred to simply as $\Omega$, $N^2$ (which can depend on $r$ with mean value $\bar{N}^2$), $\nu$ and Pr, since we use units such that $r_0=\omega_d=1$.

The molecular values of $\nu$ and $\kappa$ expected in giant planets, as well as the values of Pr, are somewhat uncertain and are likely to depend on radius. It is believed that Pr ranges from 0.01 to 0.1 in deep giant planet interiors \citep[e.g.][albeit for Jupiter]{Guillot2004,French2012}, with smaller values in the thin outermost atmosphere. Values of the Ekman number $\nu/(r_0^2 \Omega)$ are likely to be very small, on the order of $10^{-15}$ or smaller using molecular values. It is not clear whether molecular values are appropriate though or whether tidal waves should be damped by turbulent diffusivities instead, which may be expected to be much larger \citep[though still small, as discussed in e.g.][]{deVries2023}. We assume $\nu$ and $\kappa$, and hence Pr, are constant in radius, and explore the widest range of these parameters that is computationally accessible, though we are unable to study values as small as the microscopic ones in planetary interiors -- common with many other problems in astrophysics, thereby requiring us to understand how $\nu$ and $\kappa$ affect our results before we can potentially extrapolate to real planets. Another uncertainty is if there are semi-convective layers, turbulent motions within convective layers may lead to $\mathrm{Pr}\sim1$, whereas the diffusive interfaces between them may have molecular values of $\mathrm{Pr}\sim10^{-2}$. This matter is very speculative however, so we prefer to study models with constant diffusivities in this work.

\subsection{Boundary Conditions}\label{sec:boundary}

We assume an idealised, perfectly rigid, solid core that is impermeable at $r=\alpha r_0$, thus 
\begin{equation}
u_r^l= 0 \quad \mathrm{at}  \quad  r=\alpha r_0.
\end{equation} 
At the planetary radius $r=r_0$ there is a perturbed free surface with vanishing normal stress, requiring \citepalias{Pontin2023}
\begin{equation}
W^l - \frac{g_0}{(-i \omega)} u_r^l - 2 \nu \frac{\mathrm{d} u_r^l}{\mathrm{d} r} = \psi_{0} \delta_{l2} \quad \mathrm{at}   \quad  r=r_0,
\end{equation}
where $W^l=\frac{p^l}{\rho_0}+\psi$.  At both boundaries, we apply stress-free conditions (no tangential stress). This is an approximation at $r=\alpha r_0$, but is less computationally costly than applying a no-slip condition, and it is unlikely to produce significant differences in our results. Hence, we require
\begin{equation}
\frac{\mathrm{d} u_b^l}{\mathrm{d}r}+\frac{u_r^l}{r^2}=\frac{\mathrm{d} u_c^l}{\mathrm{d}r}=0 \quad \mathrm{at}  \quad  r=\alpha r_0\quad \mathrm{and} \quad  r=r_0.
\end{equation}
The inner core has fixed entropy, such that
\begin{equation}
b^l=0 \quad \mathrm{at} \quad r=\alpha r_0, 
\end{equation}
and we adopt vanishing perturbations to the buoyancy flux through the surface, i.e.
\begin{equation}
\frac{\partial b^l}{\partial r}=0  \quad \mathrm{at} \quad r=r_0.
\end{equation}
Our results are not particularly sensitive to choices of thermal boundary conditions though.

\subsection{Energetics} \label{sec:energy}

We define the mean rate of energy injection by the tidal forcing
\begin{equation}
I= \int_V \rho_0 \vb{u}\cdot (- \nabla \psi) \mathrm{d}V = -\oint_S \rho_0 \psi \vb{u} \cdot \mathrm{d}\vb{S}.
\end{equation}
By taking the scalar product of equation \ref{eq:mtm} with $\rho_0 \vb{u}$, using equation \ref{eq:heat} and integrating over the volume, we find
\begin{equation}\label{eq:energy_balance}
I = \frac{\mathrm{d} E_K}{\mathrm{d} t}+\frac{\mathrm{d} E_{PE}}{\mathrm{d} t} +\frac{1}{V} \oint_S p \vb{u} \cdot \mathrm{d}\vb{S}+D_{ther}+D_{visc},
\end{equation}
where
\begin{equation}\label{eq:KE} 
E_K= \int_V \frac{\rho_0}{2}  |\vb{u}|^2 \mathrm{d}V,
\end{equation}
\begin{equation}\label{eq:PE}
E_{PE} = \int_V \frac{g^2}{2 N^2} b^2 \mathrm{d} V,
\end{equation}
are the kinetic and potential energies of the system, respectively, except when $N=0$, in which case $E_{PE}=0$ and $\dtotal{E_{PE}}{t}=0$. The volume integrated viscous dissipation rate is given by
\begin{equation}\label{eq:Dvisc}
D_{visc}=-\int_V \rho_0 \nu \vb{u} \cdot \nabla^2 \vb{u}\, \mathrm{d} V,
\end{equation}
and volume integrated thermal dissipation rate is written
\begin{equation}\label{eq:Dther}
D_{ther}=-\int_V \rho_0 \kappa \frac{g^2}{N^2} b \nabla^2 b\, \mathrm{d} V,
\end{equation}
and when $N=0$, $D_{ther}=0$.

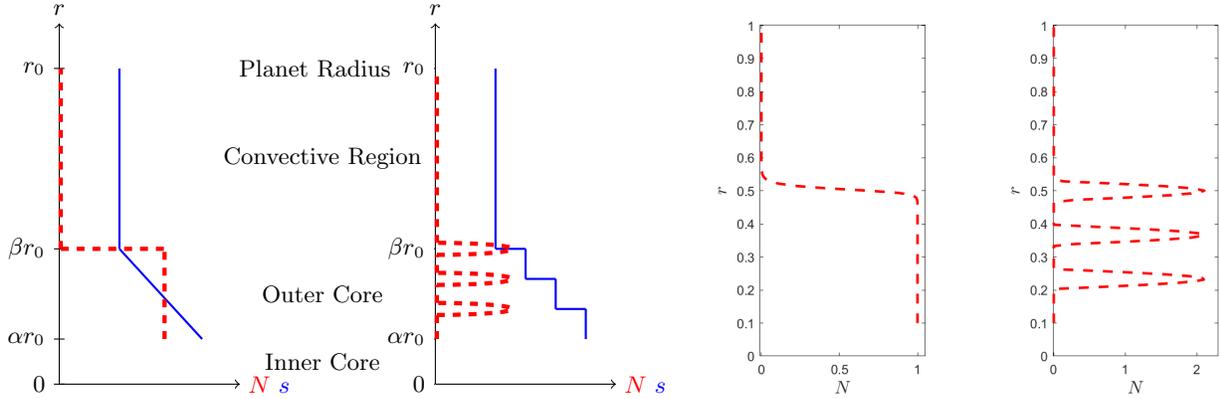
\begin{figure*}[t]\centering
       \subfigure{
		\begin{tikzpicture}[scale=2]
\tikzstyle{every node}=[font=\small]
		\draw[->] (0,0) -- (1.2,0) coordinate (x axis) node[right]{\textcolor{red}{$N$} \textcolor{blue}{$s$}};
		\draw[->] (0,0) -- (0,2.4) coordinate (y axis) node[above]{$r$};
		\foreach \y/\ytext in {0/0, 0.3/ \alpha r_0, 0.9/ \beta r_0, 2.1/r_0} 
		\draw (1pt,\y cm) -- (-1pt,\y cm) node[anchor=east,fill=white] {$\ytext$};
		\draw[thick,blue] (0.4,0.9) -- (0.95,0.3);
		\draw[thick,blue] (0.4,0.9) -- (0.4,2.1);
		\draw[ultra thick,red,dashed] (0.7,0.3) -- (0.7,0.9);
		\draw[ultra thick,red,dashed] (0.01,0.9) -- (0.7,0.9);
		\draw[ultra thick,red,dashed] (0.01,0.9) -- (0.01,2.1);
		\node at (1.75,0.15) {Inner Core};		
		\node at (1.75,0.6) {Outer Core};
		\node at (1.75,1.5) {Convective Region};
		\node at (1.7,2.1) {Planet Radius};
		\draw[->] (2.5,0) -- (3.7,0) coordinate (x axis) node[right]{\textcolor{red}{$N$} \textcolor{blue}{$s$}};
		\draw[->] (2.5,0) -- (2.5,2.4) coordinate (y axis) node[above]{$r$};
		\foreach \y/\ytext in {0/0, 0.3/ \alpha r_0, 0.9/ \beta r_0, 2.1/r_0} 
		\draw (2.49,\y cm) -- (2.49,\y cm) node[anchor=east,fill=white] {$\ytext$};
		\draw[thick,blue] (3.5,0.5) -- (3.5,0.3);
		\draw[thick,blue] (3.3,0.5) -- (3.5,0.5);
		\draw[thick,blue] (3.3,0.7) -- (3.3,0.5);		
		\draw[thick,blue] (3.1,0.7) -- (3.3,0.7);
		\draw[thick,blue] (3.1,0.9) -- (3.1,0.7);
		\draw[thick,blue] (2.9,0.9) -- (3.1,0.9);
		\draw[thick,blue] (2.9,0.9) -- (2.9,2.1);
		\draw[ultra thick,red,dashed] (2.51,0.46) arc (-90:90:0.5cm and 0.04cm);
		\draw[ultra thick,red,dashed] (2.51,0.66) arc (-90:90:0.5cm and 0.04cm);
		\draw[ultra thick,red,dashed] (2.51,0.86) arc (-90:90:0.5cm and 0.04cm);
		\draw[ultra thick,red,dashed] (2.51,0.3) -- (2.51,0.46);
		\draw[ultra thick,red,dashed] (2.51,0.54) -- (2.51,0.66);
		\draw[ultra thick,red,dashed] (2.51,0.74) -- (2.51,0.86);
		\draw[ultra thick,red,dashed] (2.51,0.94) -- (2.51,2.1);
		\end{tikzpicture}}
	\subfigure{\includegraphics[width=0.21\textwidth, trim={3.5cm 0 3.5cm 0},clip]{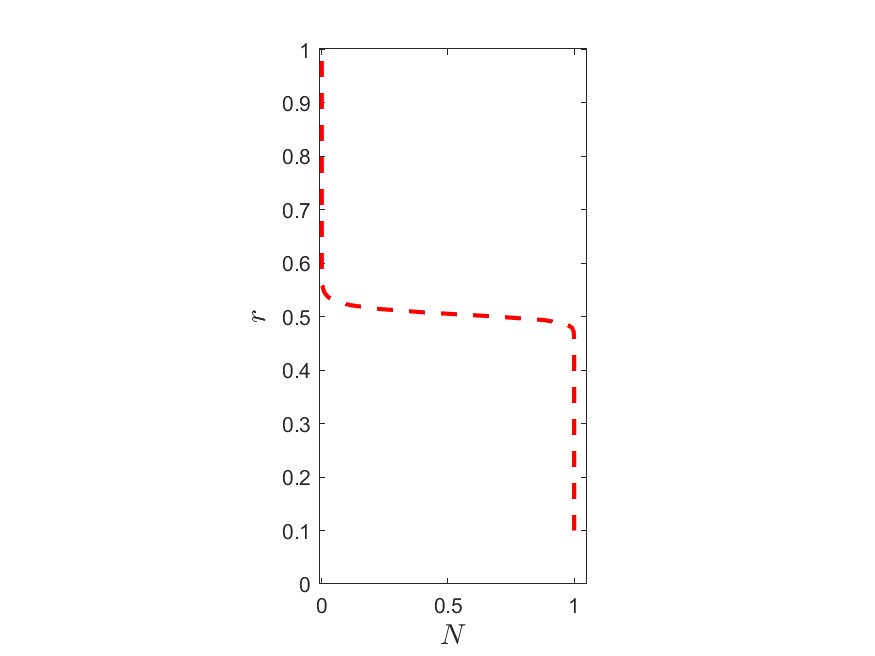}}
	\subfigure{\includegraphics[width=0.21\textwidth, trim={3.5cm 0 3.5cm 0},clip]{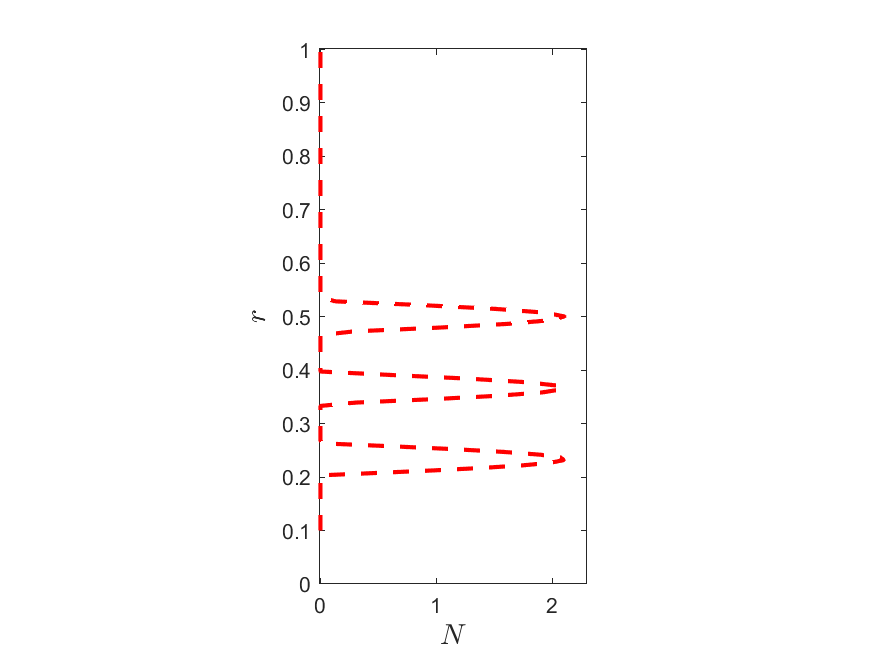}}
	\caption{Left panels: illustrative entropy profile and Brunt-V\"ais\"al\"a frequency ($N^2$) profiles for a continuously stratified layer and for semi-convective layers. A solid inner core extends to $\alpha r_0$ and the stable layer (``dilute core") extends to $\beta r_0$, outside of which is a convective envelope to planetary radius $r_0$. Right panels: examples of the numerical profiles adopted.}\label{fig:profile_ex}
\end{figure*}

In \citetalias{Pontin2023}, we showed how the viscous dissipation can be separated into bulk and boundary contributions. The former can be written \citep[e.g.][]{Ogilvie2009},
\begin{multline}\label{eq:bulk_visc}
\hspace{-0.5cm}D_{interior}= \frac{\rho_0 \nu}{2} \sum_l \Bigg(l(l+1)\Bigg( \bigg|\frac{u_r^{l}}{r}+r\dtotal{u_b^l}{r}\bigg|^2 + \bigg|r\dtotal{u_c^l}{r}\bigg|^2\Bigg) \\ 
\hspace{0.5cm}+ 3 \bigg|\dtotal{u_r^l}{r}\bigg|^2 +(l-1)l(l+1)(l+2)(|u_b^l |^2 + |u_c^l |^2) \Bigg),
\end{multline}
and is typically by far the dominant contribution for the parameters we study. We subsequently refer to the viscous dissipation using equation \ref{eq:Dvisc}, but the differences are negligible for our purposes.

In a steady state, averaged over the tidal period $2 \pi/\omega$, the injection rate $I$ balances the total viscous and thermal dissipation $D_{total}=D_{visc}+D_{ther}$ (ignoring a small boundary term discussed in \citetalias{Pontin2023}). In this study we will analyse how the dissipation rates $D_{visc}$ and $D_{ther}$ depend on the properties of the system.

\subsection{Modelling stratified layers}

We consider stably-stratified regions in a giant planet, or ones with semi-convective layers (``density staircases"), as might be produced by double-diffusive convection. We model stable layers both near a small solid core to model a dilute stably-stratified fluid core, and ones further out to represent a layer near a potential H/He molecular to metallic transition where helium rain may occur. Firstly, we model a continuous stably stratified region, with a constant $N$ which sits above the solid core extending a defined distance into the planetary envelope. This is shown on the left-hand side of Figure~\ref{fig:profile_ex} and is described by a step function in $N^2(r)$ that is non-zero from $\alpha r_0$ to $\beta r_0$. When $\beta \neq 1$, we consider a smoothly varying profile, 
\begin{equation}\label{eq:N2_layer}
N^2(r)=\frac{\bar{N}^2}{2} \Big( \tanh \big( \Delta(\beta r_0-r) \big)+1 \Big),
\end{equation}
with $\Delta=100$ unless otherwise specified, and set $N^2=0$ when $N^2(r) < \frac{\bar{N}^2}{10^7}$. For cases where $\beta=1$ we simply set $N^2(r)=\bar{N}^2$.

To model a semi-convective ``dilute fluid core" with $n_{max}$ steps within the layer we consider a series of $\delta$-like-functions for $N^2$ to give a staircase-like density profile, as shown on the right-hand side of Figure \ref{fig:profile_ex}. We use
\begin{equation}\label{eq:N2_steps}
N^2(r)=
\begin{cases}
\frac{N_0^2}{2} \bigg(1+\cos\Big(2 \pi \frac{r-r_n}{\delta r}\Big)\bigg)  \quad & |r-r_n| < \frac{\delta r}{2},\\
0 \quad & \text{otherwise},
\end{cases}
\end{equation}
for $1<n<n_{max}$, where $r_n=\alpha r_0+nd$, $d=\frac{(\beta-\alpha)r_0}{n_{max}}$, and $\epsilon=\delta r/d$. $N_0^2$ is set to a value which gives a mean stratification equivalent to a constant stratification $\bar{N}^2$,
$
N_0^2 = \bar{N}^2 \frac{(\beta - \alpha) r_0}{n_{max}~\delta r}.
$
This allows for comparisons between stratified layers and semi-convective regions. If $\beta=1$, we omit the final step at the planetary radius (i.e.~$0<n<n_{max}-1$), and $N_0$ is altered accordingly to maintain the mean stratification. 

To model a stratified layer at the metallic/molecular transition zone, we consider one wide ``interface":
\begin{equation}
N^2(r)=
\begin{cases}
\frac{N_0^2}{2} \bigg(1+\cos\Big(2 \pi \frac{r-\beta}{\delta r} \Big)\bigg)   \quad &  |r-\beta| < \frac{\delta r}{2},\\
0 \quad & \text{otherwise},
\end{cases}
\end{equation}
and $N_0^2 = \bar{N}^2 \frac{(r_0 - \alpha) r_0}{\delta r}$.
This models an isolated stable layer embedded within a convective medium.

\subsection{Frequency averaged dissipation}\label{freqavg}

The tidal response is expected to be strongly frequency-dependent when wavelike tides are excited, but we have found it helpful to define a single quantitative measure of the tidal dissipation that can be compared as we vary our parameters. To do so, we define a frequency-averaged dissipation measure for a given dissipation rate $D$
\begin{equation}\label{eq:freq_avg}
\bar{D} = \int^{\omega_{max}}_{\omega_{min}} \frac{D(\omega)}{\omega}~\mathrm{d} \omega,
\end{equation} 
which gives more emphasis to lower frequencies. Numerically, we adopt a small non-zero lower bound $\omega_{min}$, and unless otherwise stated, $\omega_{max}=\bar{N}$. This allows analysis of the low frequency regime while slightly reducing the contribution of the surface gravity modes (f-modes) that are not our primary focus. We also use the different weighting
\begin{equation}\label{eq:freq_avg_2}
\bar{D}_2 = \int^{\omega_{max}}_{\omega_{min}} \frac{D(\omega)}{\omega^2}~\mathrm{d} \omega.
\end{equation} 
This quantity is directly relevant for comparison with \cite{Ogilvie2013}, where in the low frequency limit, for the unstratified case with a solid core, they used impulsive forcing to calculate the associated energy transfer into inertial modes $\hat{E}$ analytically. This quantity is related to tidal dissipation rates $D$ by
\begin{equation}
\hat{E} = \frac{1}{2 \pi} \int^{\infty}_{-\infty} \frac{D}{\omega^2}~\mathrm{d} \omega,
\end{equation}
which provides motivation for using an $\omega^{-2}$ weighting. Other weighting factors could be used instead, and it is not at all clear what is the most useful one for astrophysical purposes, so we will explore various possibilities.

\subsection{Numerical method for the forced problem}

We solve the system of ordinary differential equations in radius, equations~\ref{eq:ur_nondim} to \ref{eq:b_nondim} for each $l$, using a Chebyshev collocation method, where the ordinary differential equations in $r$ are converted into a linear system of equations on a Chebyshev extrema grid. We consider points in radius as a set of $(n_{cheb}+1)$ Gauss-Lobatto-Chebyshev points, which are defined as,
\begin{equation}
x_j=\cos\bigg( \frac{j \pi }{ N_{c}} \bigg), \quad j=0,\cdots, N_{c},
\end{equation}
where $N_c$ is taken as the (smallest) appropriate value for which numerical convergence is found. This value varies depending on the parameters, but we typically take $N_c=100$ to $N_c=400$. This method is well-suited for many non-periodic problems as it is a spectral method free of the Runge phenomenon, which converges exponentially fast with resolution $N_c$ for smooth solutions \citep{Boyd2001}. We then have a linear algebra problem which we solve using the inbuilt MATLAB routine “mldivide”, where matrices are stored in sparse form to reduce memory requirements. The solutions for $\vb{u}$, $p$ and $b$ can then be used in equations \ref{eq:Dvisc} and \ref{eq:Dther}.

\subsection{Eigenvalue problem} \label{sec:eigs}

In \citetalias{Pontin2023} \citep[see also][]{Pontin2020} we studied analytically the non-diffusive free gravity modes in our model without rotation, and we also explored numerically the eigenvalues of the dissipative system. In the presence of rotation, analytical progress is severely hampered by the $l$-coupling caused by Coriolis forces, so in almost all cases we must resort to numerical calculations to obtain eigenfrequencies. This is particularly the case when inertial modes are excited, i.e.~for $|\omega|<2\Omega$ in convective regions (or for gravito-inertial modes in stable layers), where the inviscid (non-diffusive) problem becomes mathematically ill-posed. To compute the eigenvalues (free modes) of the dissipative problem numerically, we set $\psi_0=0$, and manipulate our system of equations into a linear generalised eigenvalue problem with eigenvalue $(-i\omega)$, i.e.,
 \begin{equation}
(-i \omega) \mathrm{A}
\begin{bmatrix}
\vb{u} \\ p \\ b 
\end{bmatrix}
= \mathrm{B} 
\begin{bmatrix}
\vb{u} \\ p \\ b 
\end{bmatrix},
\end{equation}
where $\mathrm{A}$ and $\mathrm{B}$ are matrices that describe equations \ref{eq:ur_nondim} to \ref{eq:b_nondim} with the same boundary conditions as above. This is solved numerically to obtain eigenvalues ($-i\omega$), and corresponding eigenvectors for $\vb{u}$, $p$, $b$. We use the iterative \textit{eigs} solver in MATLAB to scan the relevant frequency range, as a non-iterative method would be prohibitive in its memory requirements for even modest spatial resolutions, since this is a two-dimensional eigenvalue problem in $r$ and $\theta$. The frequency of a mode is $\operatorname{Re}[\omega]$ and its corresponding damping rate is $\operatorname{Im}[\omega]$ (there are no unstable modes).

\section{Overview of the tidal response of rotating planets}\label{results}

In \citetalias{Pontin2023} we studied in detail the response of non-rotating planets containing stable layers (or semi-convective layers), which consists of internal, surface and interfacial gravity modes. When incorporating rotation, the additional excitation of low frequency inertial waves in convective (neutrally stratified) regions, and the modification of internal gravity modes to become gravito-inertial modes, means we see new resonances at low frequencies that align with these modes compared with calculations without rotation. 

\begin{figure*}
	\centering
	\subfigure[$\alpha=0.5$, $\bar{N}^2=0$]{\includegraphics[width=0.49\textwidth, trim={0cm 0 0cm 0},clip]{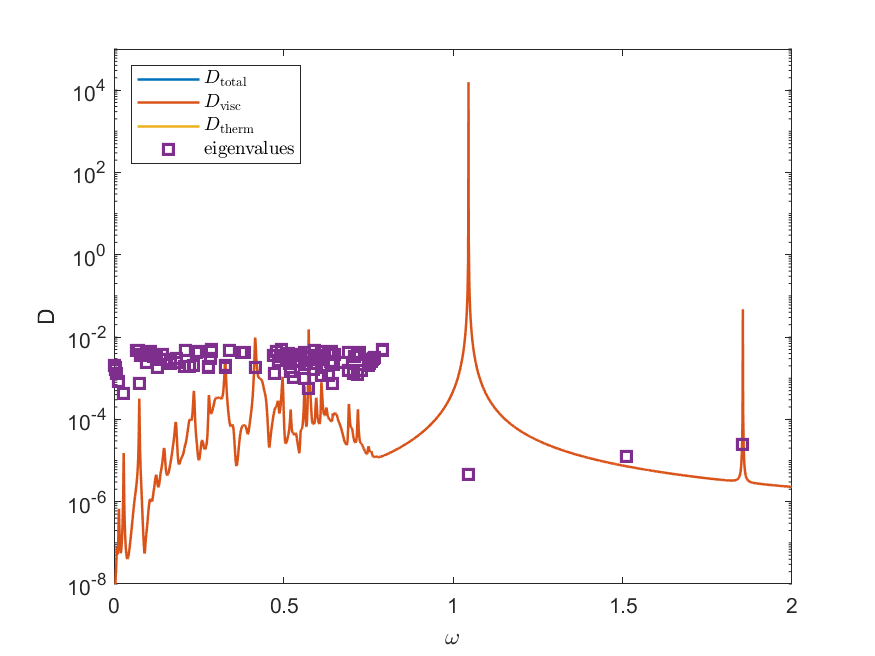}\label{fig:diss_over_omega}}
	\subfigure[$\alpha=0.5$, $\bar{N}^2=0$]{\includegraphics[width=0.47\textwidth,height=0.36\textwidth, trim=0mm 0mm 23mm 140mm, clip]{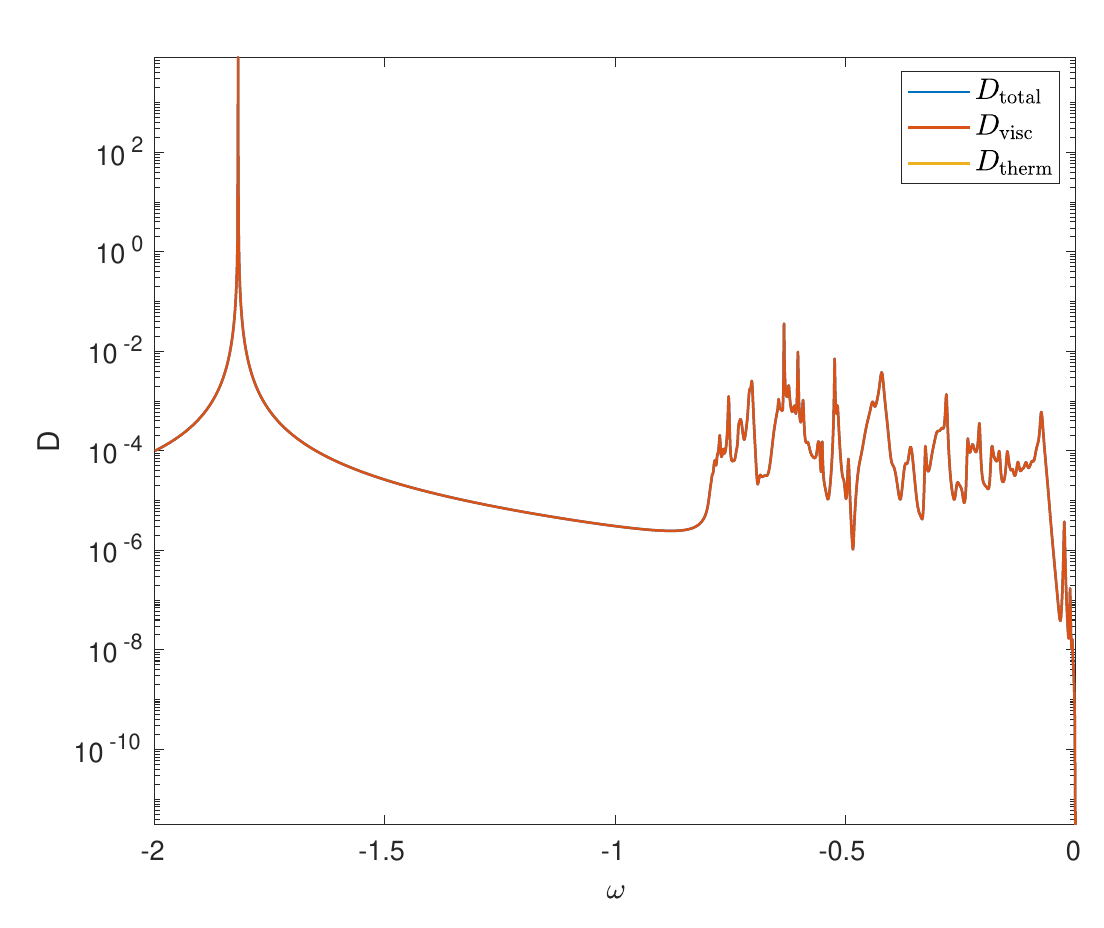} \label{fig:diss_pos_neg}}
	\subfigure[$\alpha=0.1$, $\beta=0.5$, $\bar{N}^2=1$]{\includegraphics[width=0.49\textwidth]{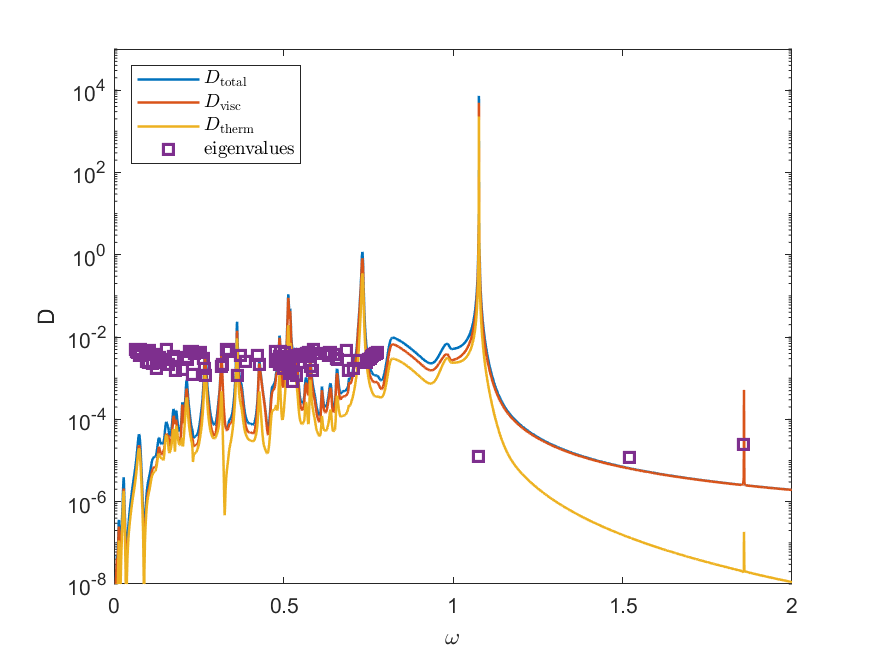}\label{fig:diss_over_omegaN}}
	\caption{Illustrative examples of tidal dissipation as a function of frequency in two rotating models, restricted to positive frequencies in panels (a) and (c), with numerically-obtained eigenvalues over-plotted for both models. The $y$-axis for the eigenvalues indicates their decay rates (imaginary parts). One example of an entirely convective fluid envelope outside a solid inner core (panels \protect\subref{fig:diss_over_omega} and \protect\subref{fig:diss_pos_neg}, where the latter shows negative frequencies) and one with a stably stratified layer extending to half the planetary radius and a tiny solid core (panel \protect\subref{fig:diss_over_omegaN}). In both cases we set $\Omega=0.4$ and $\nu=\kappa=10^{-6}$.} \label{fig:diss_over_rot}
\end{figure*}

We illustrate the dissipative response of a rotating planet as a function of frequency in Figure~\ref{fig:diss_over_omega}. These plots show the viscous ($D_{visc}$), thermal ($D_{therm}$), and total ($D_{total}$) dissipation rates. In this example, we show results for a rotating body with a large solid core of size $\alpha=0.5$ surrounded by a well-mixed convective fluid envelope with $N=0$, thereby allowing pure inertial waves to be excited (hence $D_{therm}=0$). We adopt $\Omega=0.4$ here to approximately model Saturn's rotation rate, and take $\nu=\kappa=10^{-6}$ for this illustration (for molecular diffusion we expect much smaller values in reality but this value is readily accessible computationally). For this example we only plot positive frequencies, but note that with rotation the response at a frequency $\omega$ will in general differ from that at $-\omega$. 

Pure inertial waves are excited at low frequencies for which $|\omega| < 2 \Omega =0.8$, and surface gravity modes associated with the free surface are excited at higher frequencies $|\omega|\gtrsim 1$. We observe peaks of enhanced dissipation close to resonances with inertial modes for low frequencies, which produce the irregular frequency dependence characteristic of these modes. Here we use a large core size to enhance the appearance of inertial waves in the tidal response, to more clearly illustrate their properties \citep{Ogilvie2009,Rieutord2009,Rieutord2010}, though this can be thought to model strongly stably stratified dilute cores, as we will illustrate later. The numerically-obtained eigenvalues (purple squares) agree well with the locations and heights of the peaks, where the least damped modes typically correspond to the tallest resonances, with the strongest dissipation. We have plotted only the least damped modes, with a cut-off value of $\mathrm{Im}\left[\omega\right]$ chosen for aesthetic reasons.

As in the non-rotating cases in \citetalias{Pontin2023}, we observe a strong surface gravity mode resonance around $\omega\sim 1$, but here it is shifted to a significantly lower frequency. We show in Appendix \ref{analyticalfmode} \citep[see also][]{Lebovitz1961,Braviner2014,Barker2016}, that rotation causes the splitting of this mode for a given $l$ for different $m$ values. In the limit of slow rotation, the $l=2$ surface gravity mode is split into five modes, with frequencies $\omega=\sqrt{2}\omega_d- \frac{m}{2} \Omega$, where $m=-2,-1,0,1,2$ instead of the non-rotating frequency $\sqrt{2}\omega_d$ \cite[in agreement with][for a Maclaurin spheroid in this limit]{Lebovitz1961}. As we consider $m=2$ forcing, we only observe the $m=2$ surface gravity mode, with a frequency $\omega\approx 1.07$ that is nicely predicted by (\ref{l2modes}) (the small $\Omega$ expression predicts $1.01$). There is an additional resonance close to $\omega=1.8$, caused by the coupling of different harmonic degrees appearing close to a resonance with the $l=4$, $m=2$, surface gravity mode (again shifted due to splitting). Around $\omega=1.5$ there is an eigenvalue solution that does not align with any peak, corresponding to the $l=3$ surface gravity mode. However, this mode is not actually excited, even though it is a solution to the unforced eigenvalue problem. The reason for this is that while all $l$'s are coupled by the Coriolis force, the equatorial symmetry imposed by the $l=2$ tidal forcing means that only modes with this symmetry (equatorially symmetric for $u_r$, therefore only with even $l$'s for $m=2$) are excited.

Figure~ \ref{fig:diss_over_omegaN} shows a similar but potentially more realistic case where instead of a large solid core there is a stably stratified layer extending to half the planetary radius, also with $\Omega=0.4$ and $\nu=\kappa=10^{-6}$. We consider a buoyancy profile as in equation (\ref{eq:N2_layer}) with $\alpha=0.1$, $\beta=0.5$, and $\bar{N}=1$. This case is the same as Fig.~2(b) in \citetalias{Pontin2023} but with the addition of rotation. In this case, as well as exciting inertial waves in the outer convective envelope, as observed in the solid core case, we also excite gravito-inertial waves in the extended stably stratified layer over a larger range of frequencies\footnote{The range of frequencies for gravito-inertial wave propagation is explained in e.g.~\cite{Rieutord2009}. Alternatively, please see \cite{Andre2017} equations 2.14 (or 2.33) for this upper bound in the case of plane waves when the rotation axis is perpendicular to the local gravity vector.} $|\omega| <  \sqrt{\bar{N}^2 +4\Omega^2} =1.28$.

\begin{figure}
	\centering
	\subfigure[$\omega=0.181$]{\includegraphics[width=0.22\textwidth, trim=80mm 0mm 25mm 62mm, clip]{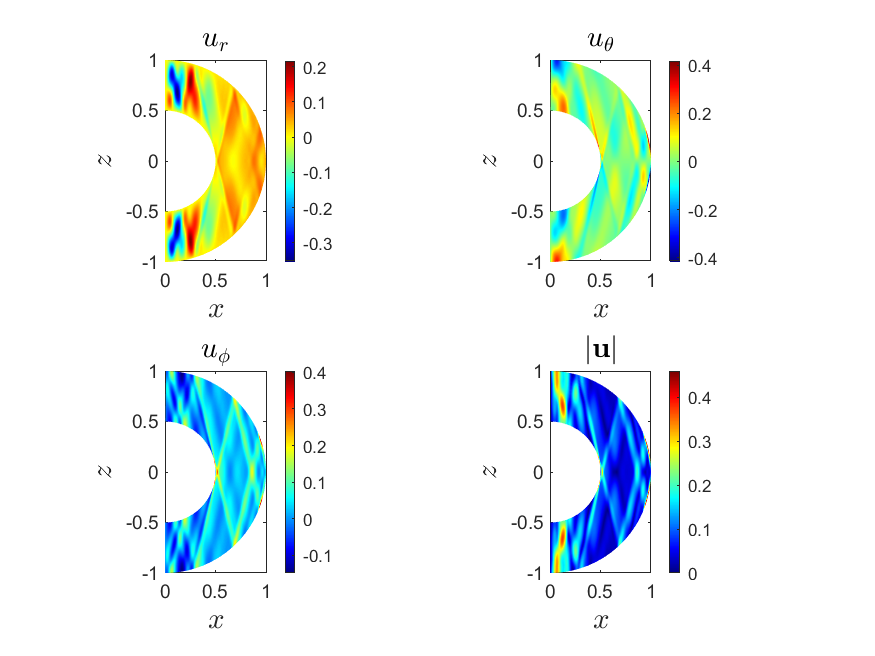}
	\label{fig:i-mode_1} }
	\subfigure[$\omega=0.561$]{
	\includegraphics[width=0.22\textwidth, trim=80mm 0mm 25mm 62mm, clip]{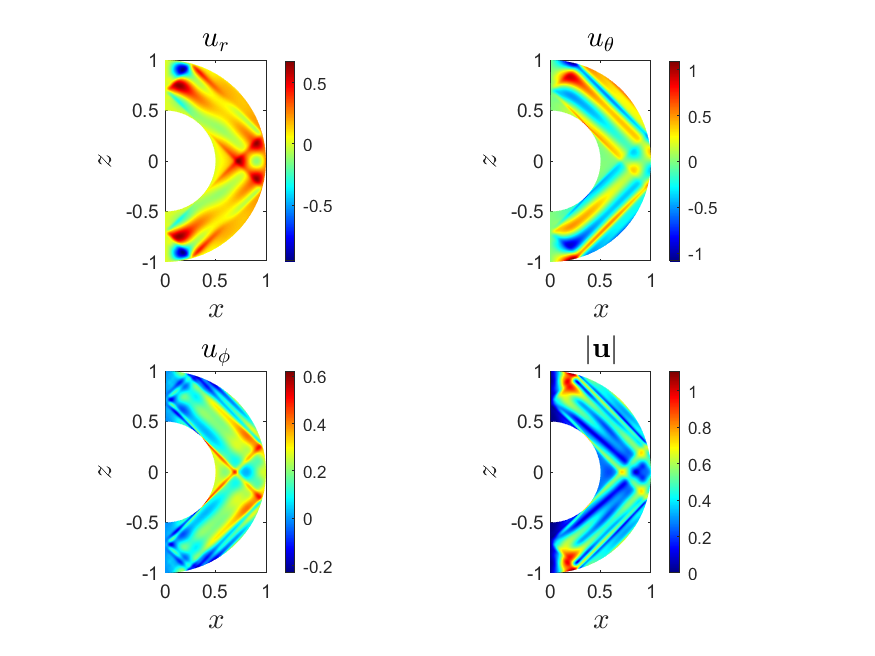}\label{fig:i-mode_2}}
		\caption{Illustrative examples of $|\boldsymbol{u}|$ for two different forcing frequencies of the forced response in a convective envelope with $\alpha=0.5$, $\bar{N}^2=0$, $\Omega=0.4$, and $\nu=\kappa=10^{-6}$. Both panels show examples of tidally-forced inertial waves.}\label{fig:mode_i}
\end{figure}

\begin{figure}
	\centering
	\subfigure[$\omega=0.514$]{
	\includegraphics[width=0.22\textwidth, trim=80mm 0mm 25mm 62mm, clip]{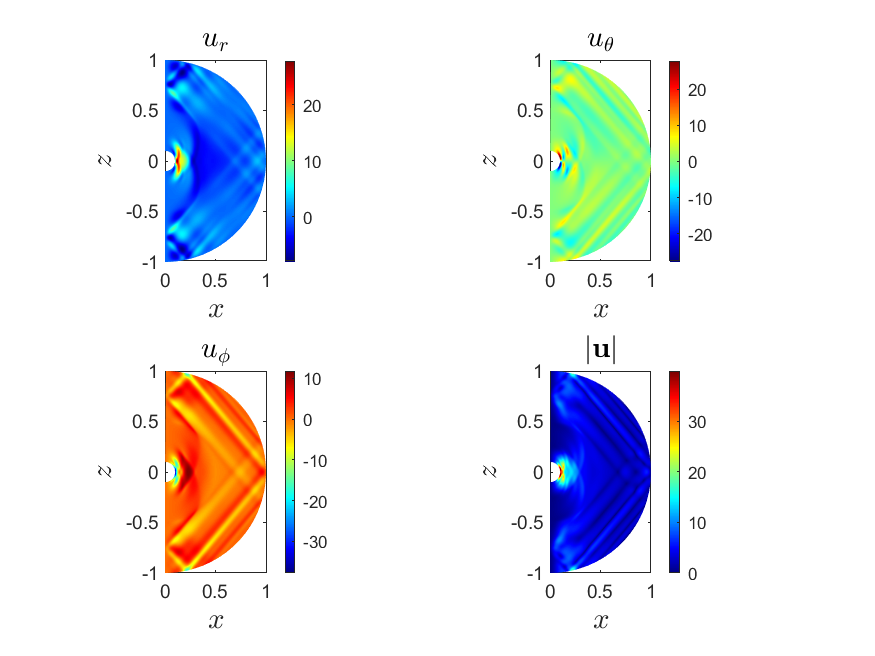}\label{fig:gi-mode_1}}
\subfigure[$\omega=0.95$]{
	\includegraphics[width=0.22\textwidth, trim=80mm 0mm 25mm 62mm, clip]{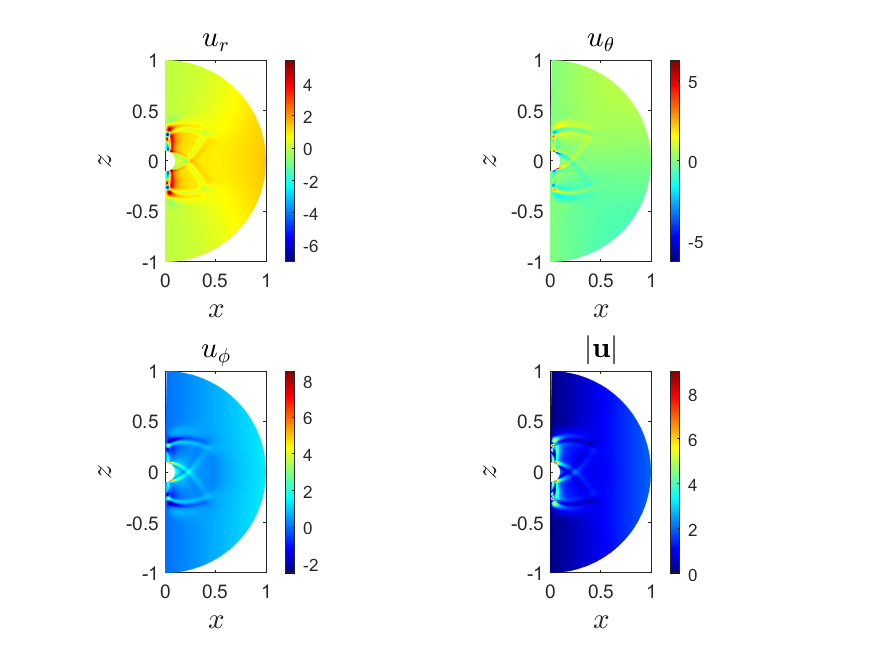}
\label{fig:gi-mode_2}}
\caption{Illustrative examples of $|\boldsymbol{u}|$ for the forced response within a planet with a stably stratified core and convective envelope, with $\alpha=0.1$, $\beta=0.5$, $\bar{N}^2=1$, $\Omega=0.4$, $\nu=\kappa=10^{-6}$. Panel \protect\subref{fig:gi-mode_1} shows a gravito-inertial wave response in the stratified layer and inertial wave response in the convective envelope. Panel \protect\subref{fig:gi-mode_2} shows gravito-inertial wave in the stratified layer only.} \label{fig:mode_gi_shell}
\end{figure}

In Figures~\ref{fig:mode_i} and \ref{fig:mode_gi_shell} we show the spatial structure of the forced response, which displays features similar to those reported in previous studies for the free and forced modes in rotating and stratified planets \citep[e.g.][]{Rieutord1997,Dintrans1999,Ogilvie2004,Rieutord2009}. In Figure~\ref{fig:mode_i}, we show solutions at two different forcing frequencies for the case shown in Figure~\ref{fig:diss_over_omega}. We observe inertial wave beams in the convective envelope that propagate with the angle $\theta_i$ between the rotation axis along $z$ and the wavevector $\vb{\hat{k}}$ as predicted by $\cos{\theta_i} = \omega/(2 \Omega)$, thus satisfying the dispersion relation $\omega^2=4\Omega^2\cos\theta_i$, with the group velocity (along which energy travels) propagating perpendicular to this direction and lying along the visible wave beams.

In Figure~\ref{fig:gi-mode_1} we consider the spatial structure of the forced solution at a frequency of $\omega=0.514$ for the case in Figure~\ref{fig:diss_over_omegaN}, a frequency which is within the range of both gravito-inertial waves in the stratified region and inertial waves in the convective region. We observe both these waves to be excited, the gravito-inertial waves within the stratified core, and the inertial waves within the convective envelope. At this point we note the similarity between the spatial structure in the outer envelope with the example shown in Figure~\ref{fig:i-mode_2}, where the forcing frequency is similar. When we consider a higher frequency $\omega=0.95$, outside the inertial wave range but within the gravito-inertial wave range, shown in Figure~\ref{fig:gi-mode_2}, we can see that only gravito-inertial waves within the stably stratified layer are excited and the response is evanescent above.
This demonstrates how in a region with stable stratification, instead of the straight lines characteristic of an inertial wave we observe the curved lines characteristic of gravito-inertial waves \citep[e.g.][though their profile has $N^2\propto r$]{Dintrans1999}. 

One significant difference between the tidal response in rotating and non-rotating models is that rotation breaks the symmetry between positive and negative forcing frequencies. Figure~\ref{fig:diss_pos_neg} shows the dissipation rate for negative forcing frequencies in a convective envelope with a solid core for the same model as Figure~\ref{fig:diss_over_omega}, demonstrating that $D(\omega) \neq D(-\omega)$ in general. At small negative frequencies we have additional Rossby mode resonances (also known as planetary waves). Rossby modes are a subset of inertial modes obtained by considering conservation of vorticity, and in the absence of a background flow and stratification, have a dispersion relation \citep{Papaloizou1978, Zaq2021}, 
\begin{equation}
\omega=-\frac{2 m \Omega}{l(l+1)},
\end{equation} 
which has strictly the opposite sign to rotation, indicating that they propagate in the retrograde direction. These are more evident in thin shells (large $\alpha$) and tend to excite modes with $l=3$ and $m=2$, for which $\omega=-1/3$. At higher frequency magnitudes, the direction of the shift in frequency of the surface gravity mode discussed above depends on the sign of $\omega$, as predicted in Appendix \ref{analyticalfmode}, and which we confirm numerically here.

\section{Variation of the parameters}
\label{varyparam}

We now begin to explore how the dissipation depends on the parameters of our model. To do so, we follow \citetalias{Pontin2023} and employ a frequency-averaged measure of the dissipation, but now using the two different weightings introduced in \S~\ref{freqavg}. In many of our parameter ranges, the highest frequency gravito-inertial and surface gravity modes have comparable frequencies, therefore it is not always possible to separate their behaviour using these integrated measures. Without rotation, we can always take an upper limit on our integrals to be $\omega_{max}=\bar{N}$ to include internal gravity modes and exclude surface gravity modes, but we are unable to do the same here because gravito-inertial waves propagate up to $\omega_{max}=\sqrt{\bar{N}^2+4 \Omega^2}$, which overlaps with surface gravity modes (which are themselves shifted to lower frequencies for positive $\omega$). The frequency-averaged measures are helpful to explore how the dissipative properties depend on the various parameters, but we emphasise that results involving these quantities by themselves should be interpreted with caution. This is partly because they use a single number to represent an entire complicated spectrum, and partly because this measure is less robust to the frequency interval considered than in the non-rotating cases explored in \citetalias{Pontin2023}.

\subsection{Variation of rotation rate $\Omega$}

First, we vary $\Omega$ to determine how the planetary rotation rate affects its dissipative properties. We focus on two models: one with an entirely stably stratified planet \hbox{$\beta=1.0$} and one with an extended stably stratified ``dilute core" extending to half the planetary radius $\beta=0.5$, with a well-mixed convective region above. 

\begin{figure}
	\centering
	\subfigure[$\bar{D}=\int D / \omega~\mathrm{d} \omega$]{\includegraphics[width=0.49\textwidth]{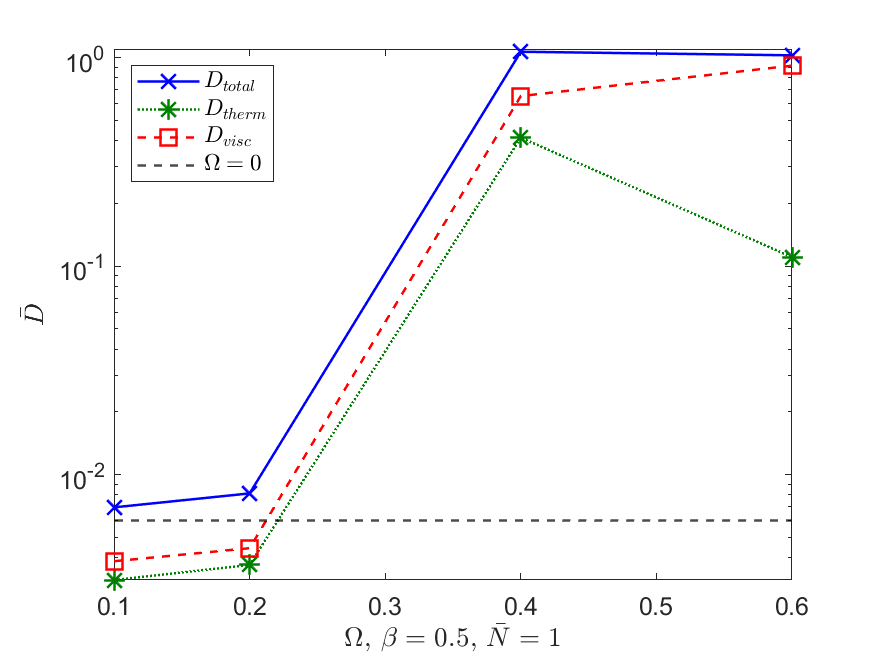}}
        \subfigure[$\bar{D}=\int D / \omega^2~\mathrm{d} \omega$]{\includegraphics[width=0.49\textwidth]{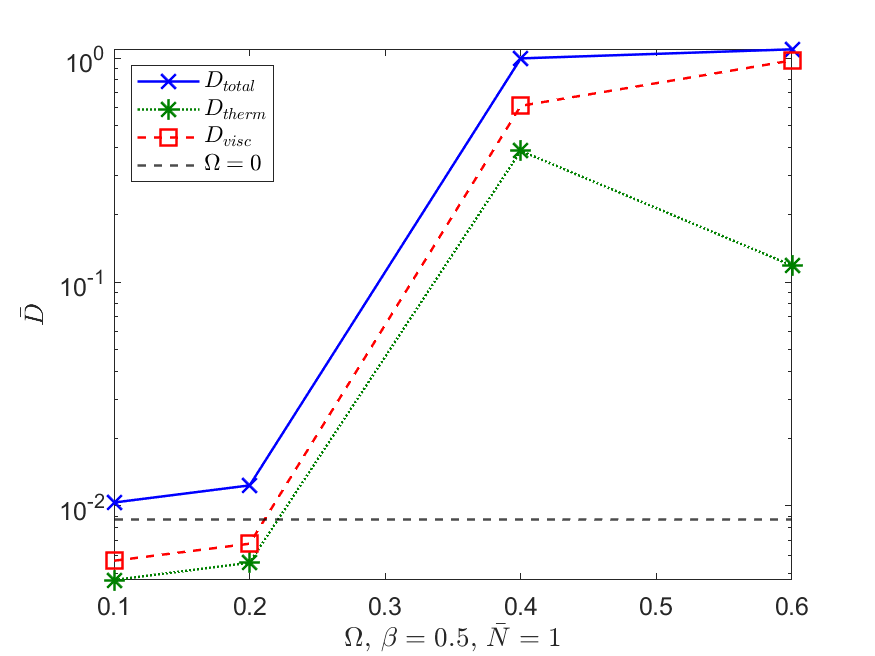}}
	\subfigure[$\bar{D}= (1/ \Delta \omega) \int D ~\mathrm{d} \omega$]{\includegraphics[width=0.49\textwidth]{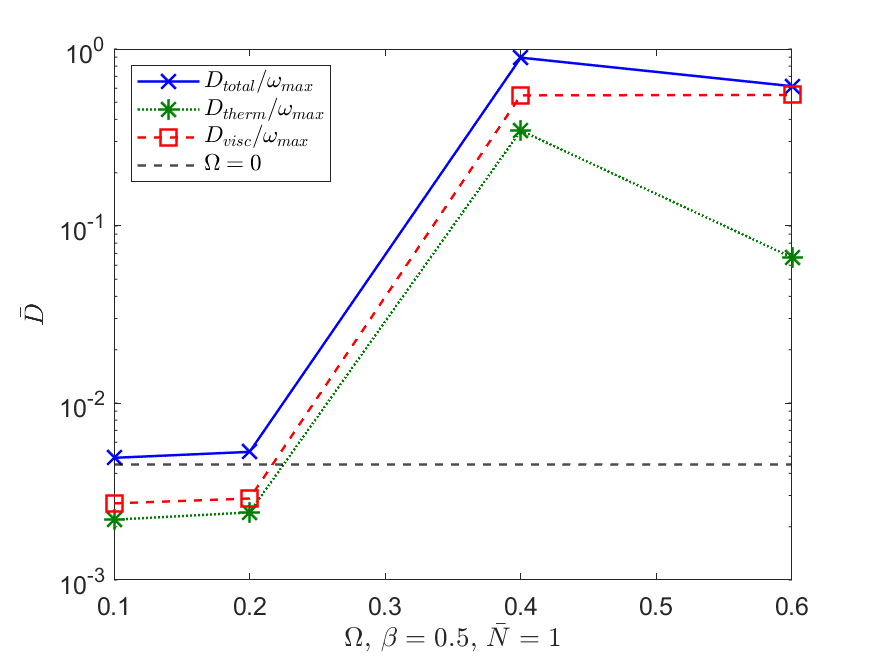}}
	\caption{Comparison of different frequency weightings for the frequency-averaged dissipation as a function of rotation rate. Other parameters are fixed at $\alpha=0.1$, $\beta=0.5$, $\bar{N}=1$, $\nu=\kappa=10^{-6}$. In all cases the integration limits have been taken to be $\omega_{min}=0$ and $\omega_{max}=\sqrt{\bar{N}^2+4 \Omega^2}$.}\label{fig:diss_Omega_b05_weightings}
\end{figure}	

\begin{figure}
	\centering
	\subfigure[$\beta=0.5$]{
	\includegraphics[width=0.49\textwidth]{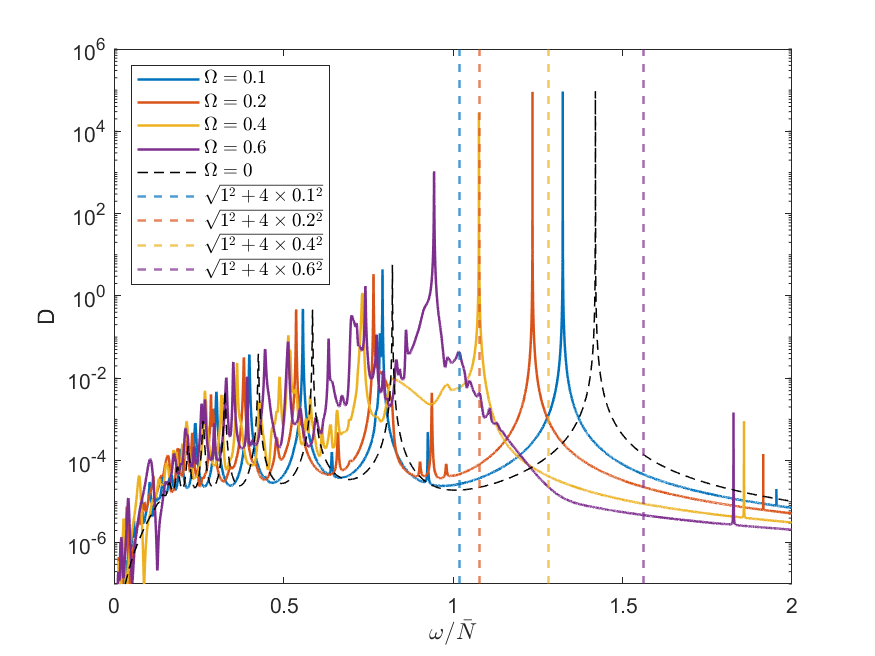}\label{fig:diss_omega_b05}}
	\subfigure[$\beta=1$]{
	\includegraphics[width=0.49\textwidth]{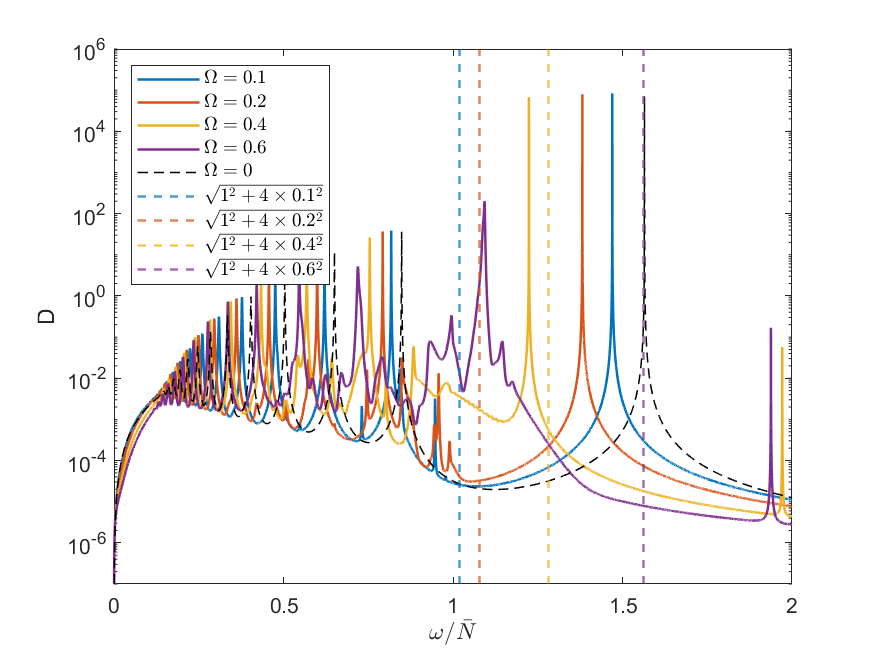}\label{fig:diss_omega_b1}}
	\subfigure[$\beta=1$]{
	\includegraphics[width=0.49\textwidth]{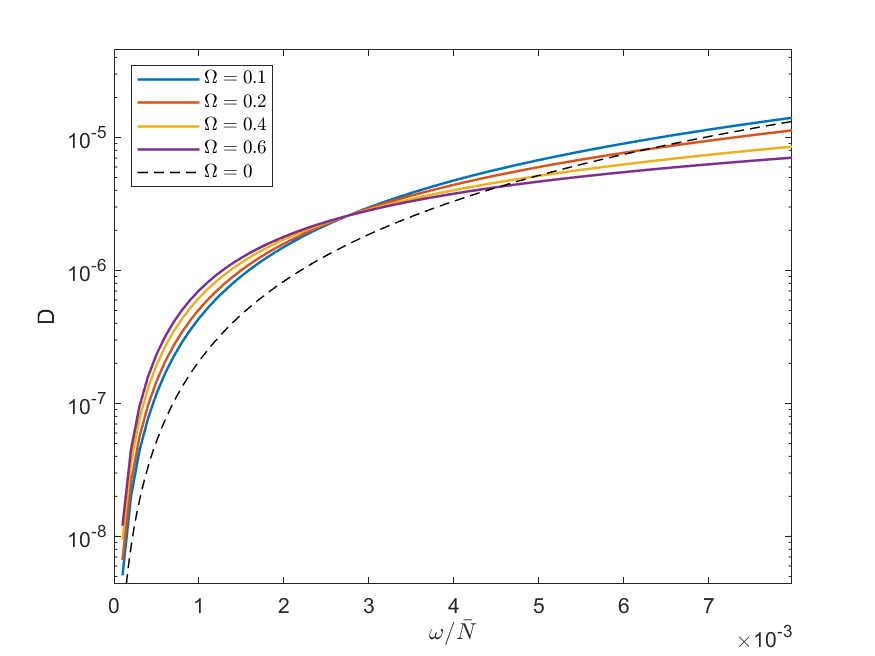}\label{fig:diss_omega_lowfreq}}
	\caption{Dissipation rate as a function of $\omega$ for varying rotation rates $\Omega$, with $\alpha=0.1$, $\bar{N}=1$, $\nu=\kappa=10^{-6}$ in all cases. Panel \protect\subref{fig:diss_omega_b05} shows a stratified layer extending to half the planetary radius, panels \protect\subref{fig:diss_omega_b1} and \protect\subref{fig:diss_omega_lowfreq} show a uniform layer extending to the planetary radius, with \protect\subref{fig:diss_omega_lowfreq} showing just the low frequency regime (travelling wave).}\label{fig:diss_over_Omega}
\end{figure}

In Figure~\ref{fig:diss_Omega_b05_weightings} we consider $\beta=0.5$, and first compare the outcomes of three different definitions of the frequency-averaged dissipation. We consider weightings with $\frac{1}{\omega}$, $\frac{1}{\omega^2}$ and a ``mean average", which we define to be $\bar{D}=\frac{1}{\omega_{max}-\omega_{min}} \int D~\mathrm{d} \omega$. In all three cases we have kept the integration limits the same as the range for gravito-inertial waves, i.e.~\hbox{$\omega_{min}=0$} and $\omega_{max}=\sqrt{\bar{N}^2+4 \Omega^2}$; the same quantity for the corresponding non-rotating case has been plotted for reference in each case. There is a slight quantitative difference between these three definitions but no qualitative differences. The similarity in all three measures implies that we can focus on only one of these for an overview of the dissipative response. 

In all three cases the total dissipation is larger than obtained without rotation and there is a large jump between $\Omega=0.2$ and $\Omega=0.4$. Since $\omega_{max}=\sqrt{N^2+4 \Omega^2}$ is not a robust limit to separate the surface and internal modes, it is also instructive to consider the frequency-dependent solution, which we plot in Figure~\ref{fig:diss_over_Omega} for all of these cases. Considering the vertical lines in Figure~\ref{fig:diss_over_Omega}, which mark the gravito-inertial wave limit, we can see that for the rotation rates $\Omega=0,0.1,0.2$, the surface gravity mode is well above $\omega_{max}$, whereas for $\Omega=0.4,0.6$, it is below this limit. In fact, we see that in the case of $\Omega=0.6$, the surface gravity mode no longer appears as an isolated peak. It is therefore likely that the surface gravity mode is to a large part dictating the trends observed in Figure~\ref{fig:diss_Omega_b05_weightings}.

\begin{figure}
	\subfigure[$\bar{D}=\int D / \omega^2~\mathrm{d} \omega$, $\omega_{min}=0$, $\omega_{max}=1$]{
	\includegraphics[width=0.49\textwidth]{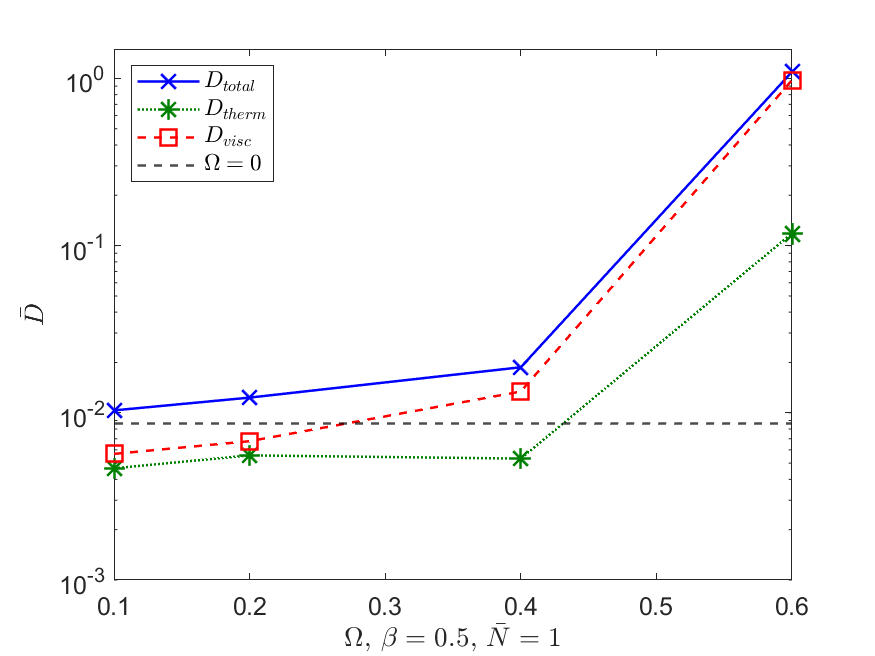}}
	\subfigure[$\bar{D}=\int D / \omega^2~\mathrm{d} \omega$, $\omega_{min}=0$, $\omega_{max}=1.8$]{
	\includegraphics[width=0.49\textwidth]{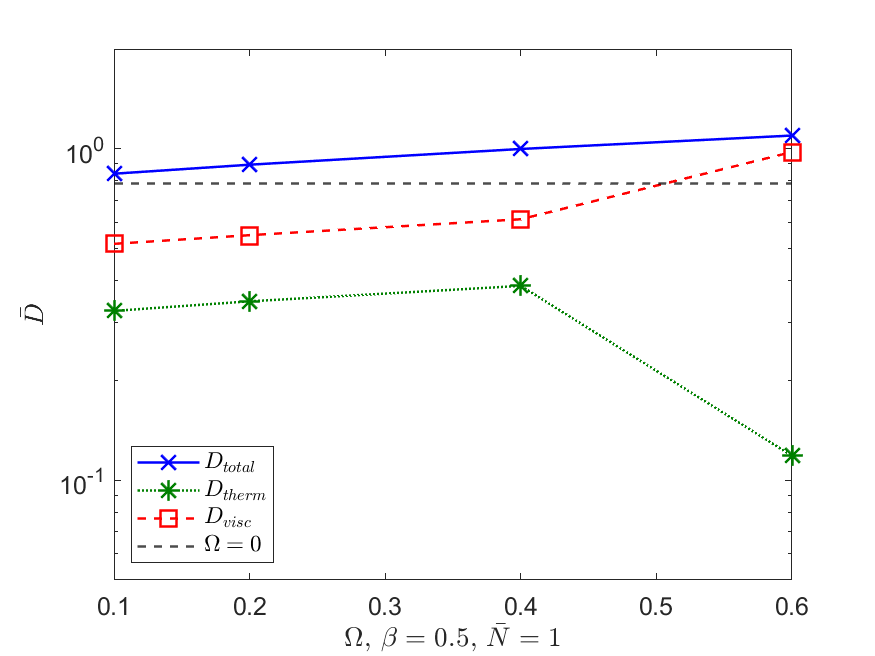}}
	\caption{Comparison of different integration limits for frequency-averaged dissipation with $1/\omega^2$ weighting as a function of rotation rate. Other parameters are fixed at $\alpha=0.1$, $\beta=0.5$, $\bar{N}=1$, $\nu=\kappa=10^{-6}$.}\label{fig:diss_Omega_b05_limits}
\end{figure}

In Figure~\ref{fig:diss_Omega_b05_limits} we consider different integration limits, but show results only for the $1/\omega^2$ frequency weighting. In the top panel, we consider $\omega_{max}=1$, which removes the surface gravity mode from most cases (for $\Omega=0.6$ it is still included). The frequency-averaged quantities increase away from the non-rotating baseline as rotation rate increases and the inertial wave response is enhanced. In the bottom panel, we consider a higher integration limit $\omega_{max}=1.8$, which incorporates the surface gravity mode behaviour in all cases. We observe that more rapid rotation leads to more efficient dissipation according to this measure. Using the higher limit, we find that there are some differences between the different weightings (not shown) and those that give the most emphasis to inertial waves exhibit an increase as rotation rate is increased.

\begin{figure}
	\subfigure[$\bar{D}=\int D / \omega~\mathrm{d} \omega$, $\omega_{min}=0$, $\omega_{max}=1$]{
	\includegraphics[width=0.49\textwidth]{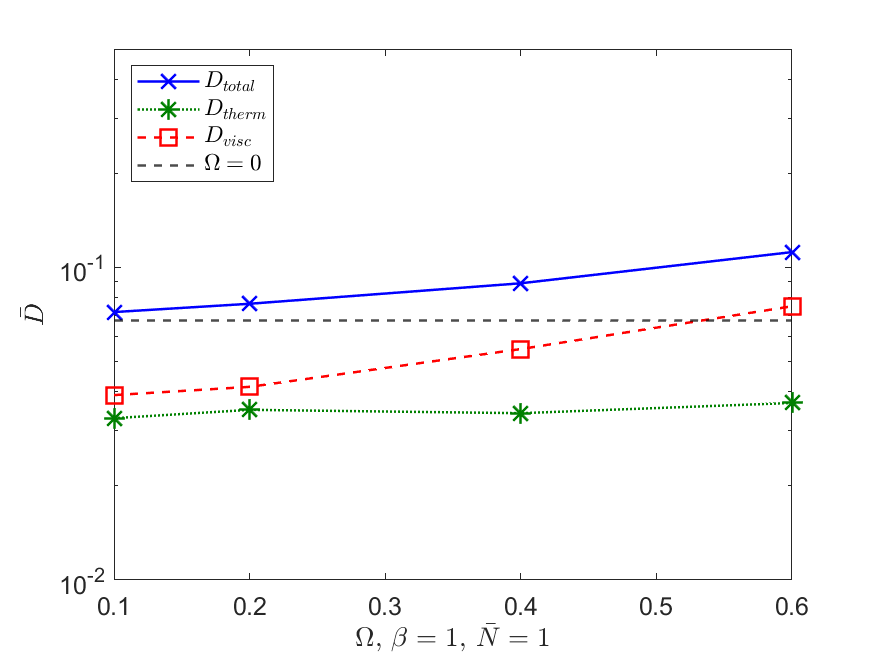}}
	\subfigure[$\bar{D}=\int D / \omega~\mathrm{d} \omega$, $\omega_{min}=0$, $\omega_{max}=\sqrt{N^2+4\Omega^2}$]{
	\includegraphics[width=0.49\textwidth]{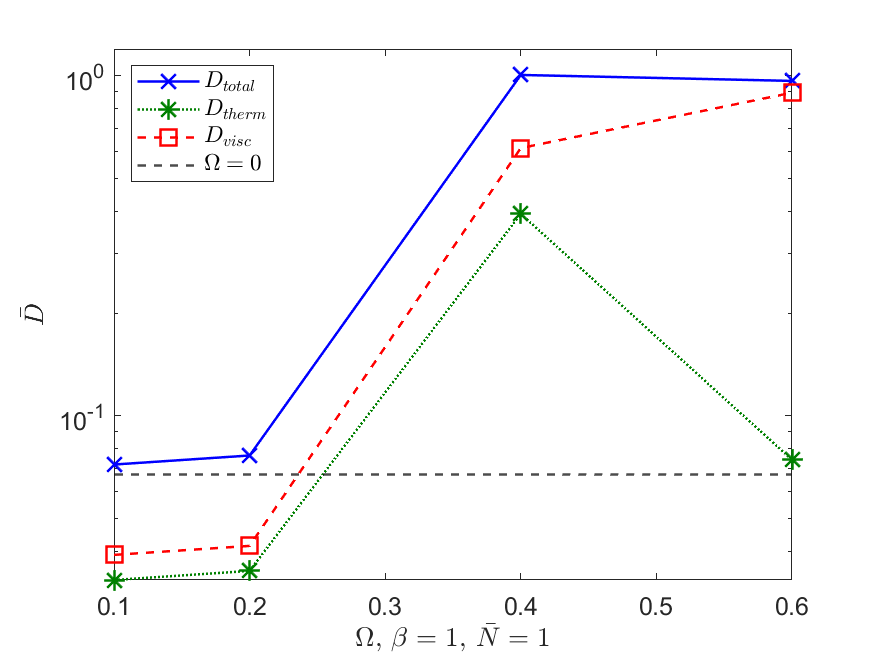}}
	\subfigure[$\bar{D}=\int D / \omega~\mathrm{d} \omega$, $\omega_{min}=0$, $\omega_{max}=1.8$]{
	\includegraphics[width=0.49\textwidth]{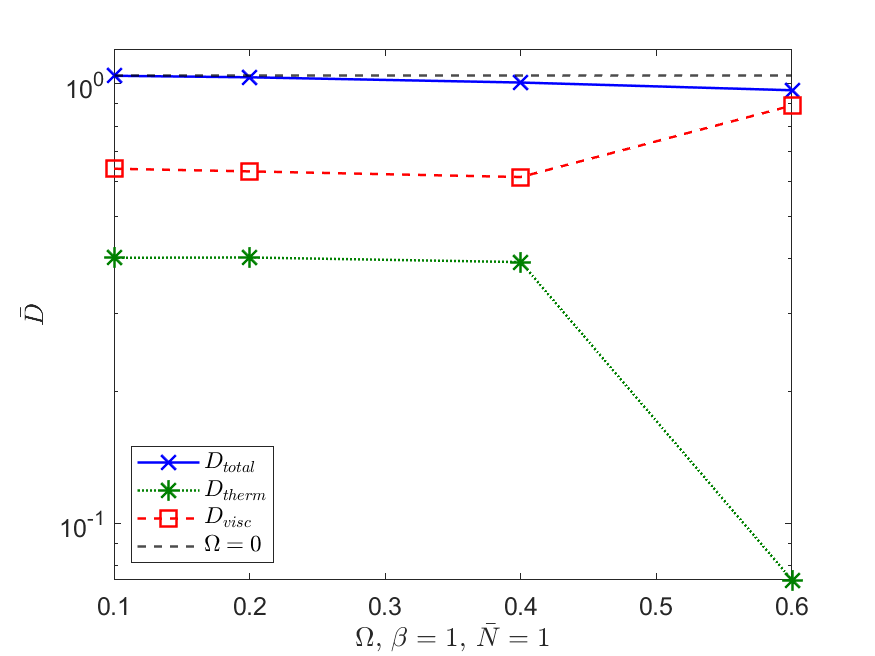}}
	\caption{Comparison of integration limits for the frequency-averaged dissipation as a function of rotation rate for a fully stratified planet. Other parameters at fixed at $\alpha=0.1$, $\beta=1$, $\bar{N}=1$, $\nu=\kappa=10^{-6}$.}\label{fig:diss_Omega_b1}
\end{figure}

We now turn to explore a case with a fully stratified planet with $\beta=1$ in Figure~\ref{fig:diss_Omega_b1}, with all other parameters fixed at $\alpha=0.1$, $\bar{N}=1$, $\nu=\kappa=10^{-6}$. Due to the similarities between the weightings, we show here just the frequency-averaged measure used in the non-rotating case in \citetalias{Pontin2023} ($1/\omega$), but we have included all three limits of integration discussed so far in this section. We see that trends observed for $\beta=0.5$ also hold here. The only key difference is that, for data points in which the surface gravity mode is not included in the integration, the overall dissipation is lower than for $\beta=0.5$, due to the absence of a convective envelope permitting inertial mode excitation.

\begin{figure*}\centering
\subfigure[$\Omega=0.2$, $\omega=0.25$]{
        \includegraphics[width=0.24\textwidth, trim=86mm 9.5mm 25mm 61.8mm, clip]{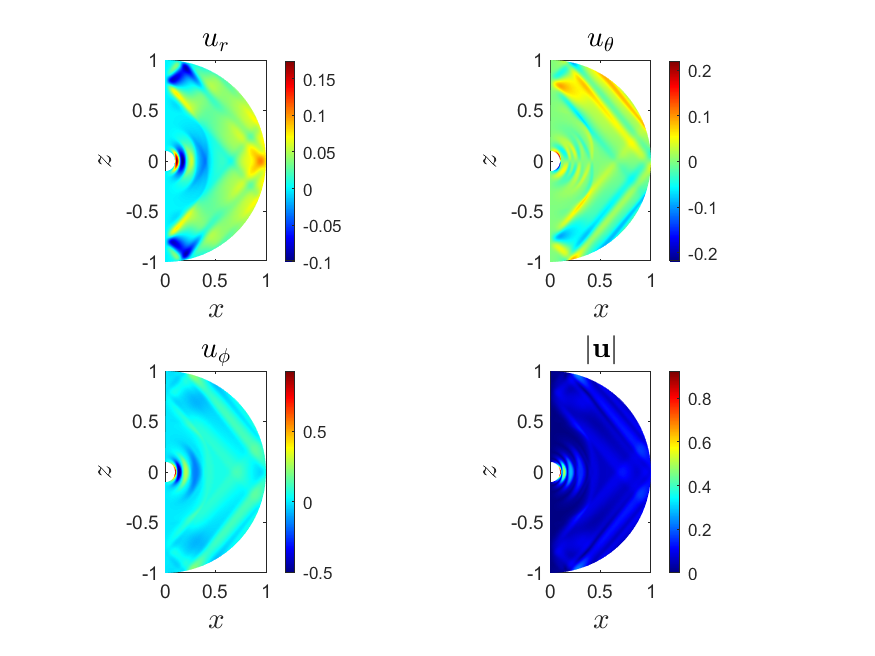}}
\subfigure[$\Omega=0.2$, $\omega=0.95$]{
        \includegraphics[width=0.24\textwidth, trim=86mm 9.5mm 25mm 61.8mm, clip]{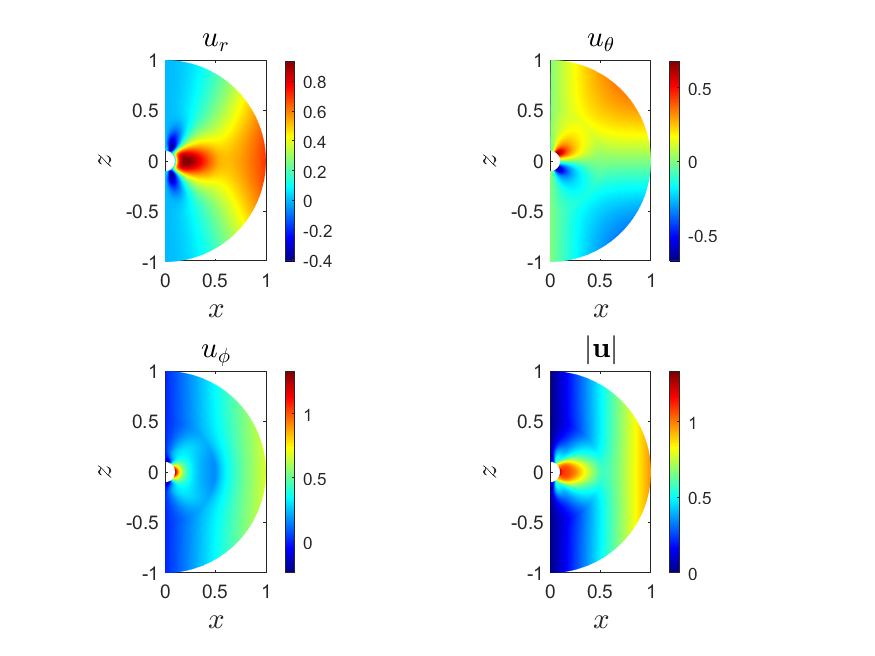}}
\subfigure[$\Omega=0.2$, $\omega=1.1$]{
        \includegraphics[width=0.24\textwidth, trim=86mm 9.5mm 25mm 61.8mm, clip]{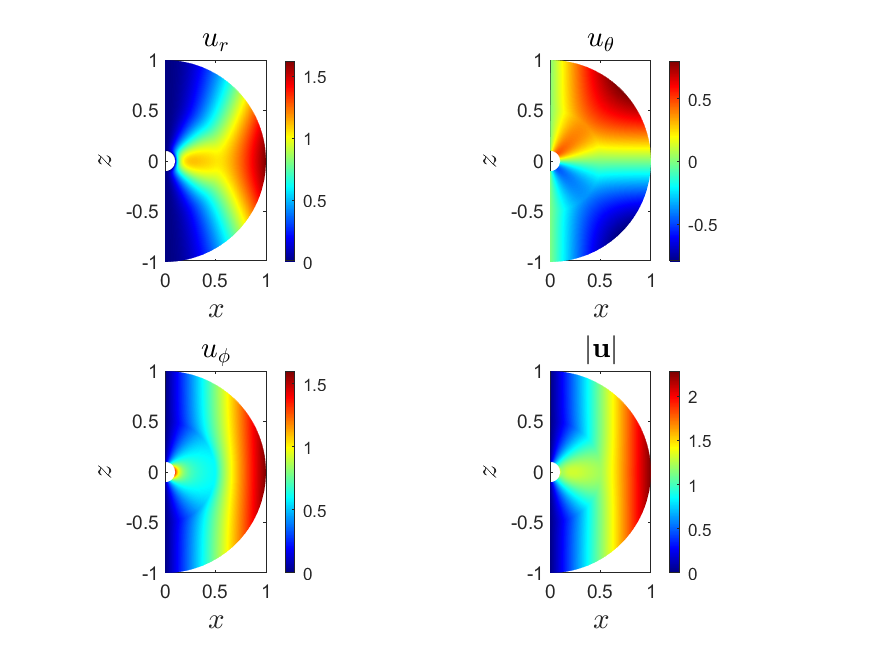}}\\
\subfigure[$\Omega=0.4$, $\omega=0.25$]{
        \includegraphics[width=0.24\textwidth, trim=86mm 9.5mm 25mm 61.8mm, clip]{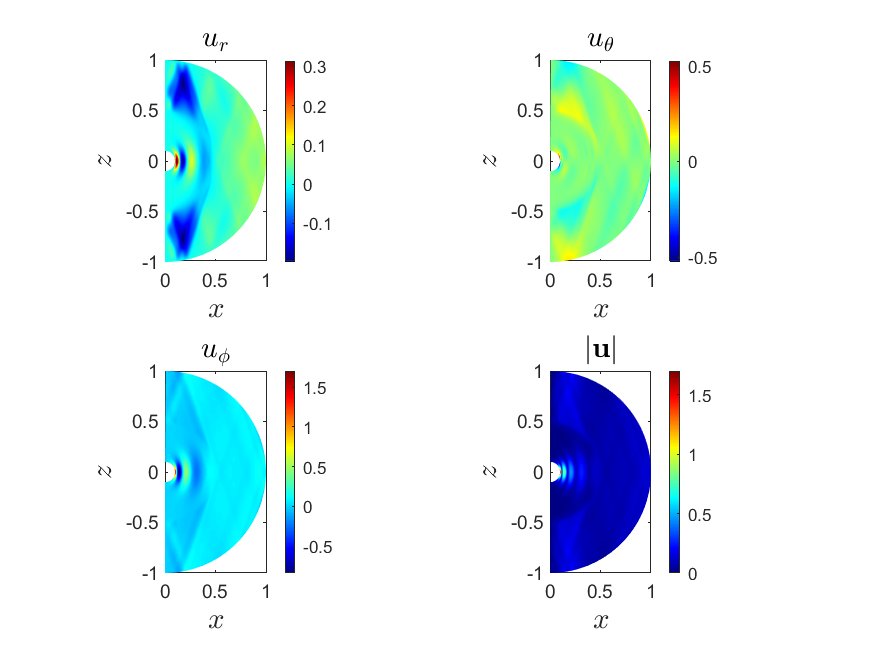}}
        \subfigure[$\Omega=0.4$, $\omega=0.95$]{
        \includegraphics[width=0.24\textwidth, trim=86mm 9.5mm 25mm 61.8mm, clip]{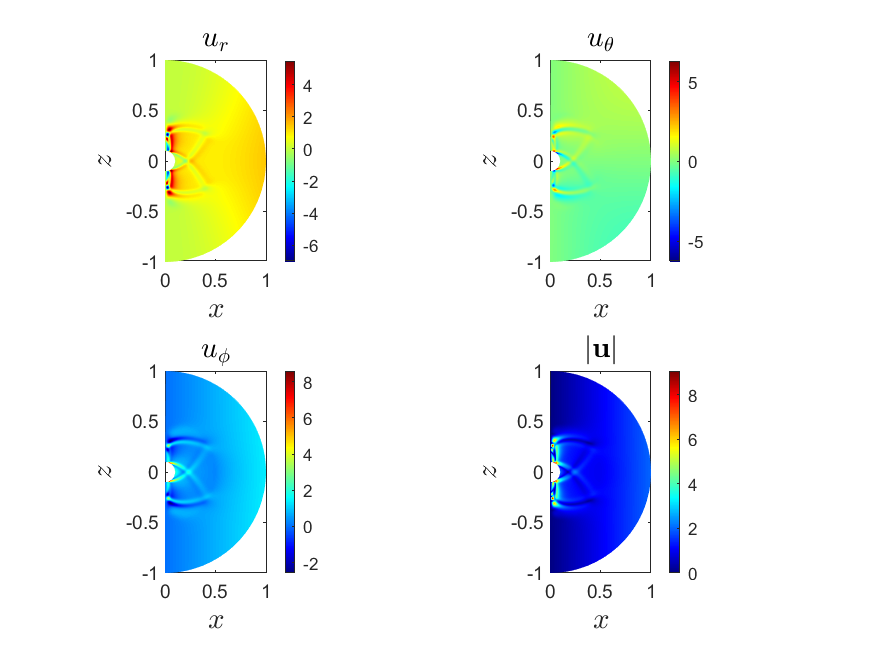}}
\subfigure[$\Omega=0.4$, $\omega=1.1$]{
        \includegraphics[width=0.24\textwidth, trim=86mm 9.5mm 25mm 61.8mm, clip]{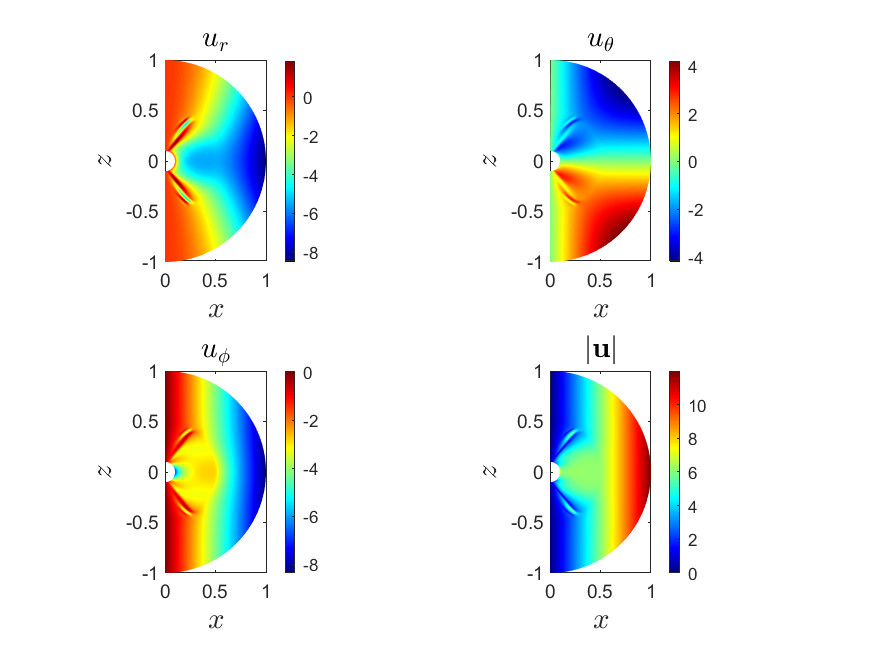}}\\
	\subfigure[$\Omega=0.6$, $\omega=0.25$]{
        \includegraphics[width=0.24\textwidth, trim=86mm 9.5mm 25mm 61.8mm, clip]{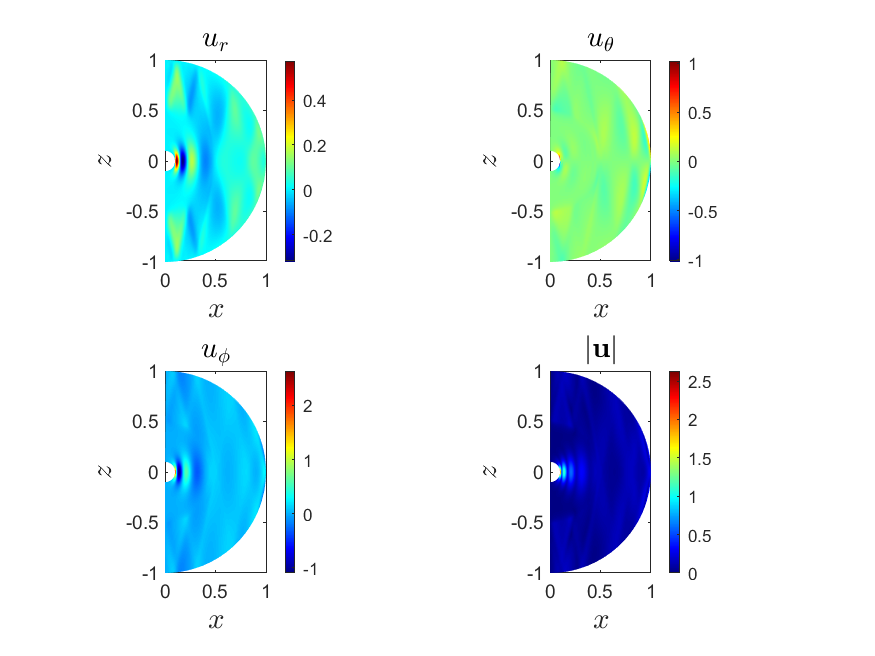}}
        \subfigure[$\Omega=0.6$, $\omega=0.95$]{
         \includegraphics[width=0.24\textwidth, trim=86mm 9.5mm 25mm 61.8mm, clip]{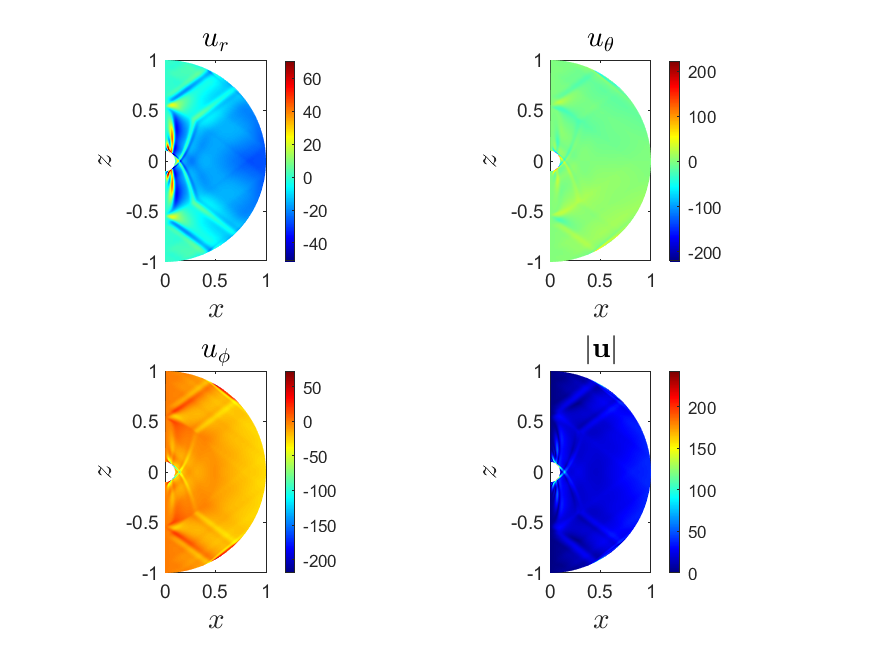}}
         \subfigure[$\Omega=0.6$, $\omega=1.1$]{
         \includegraphics[width=0.24\textwidth, trim=86mm 9.5mm 25mm 61.8mm, clip]{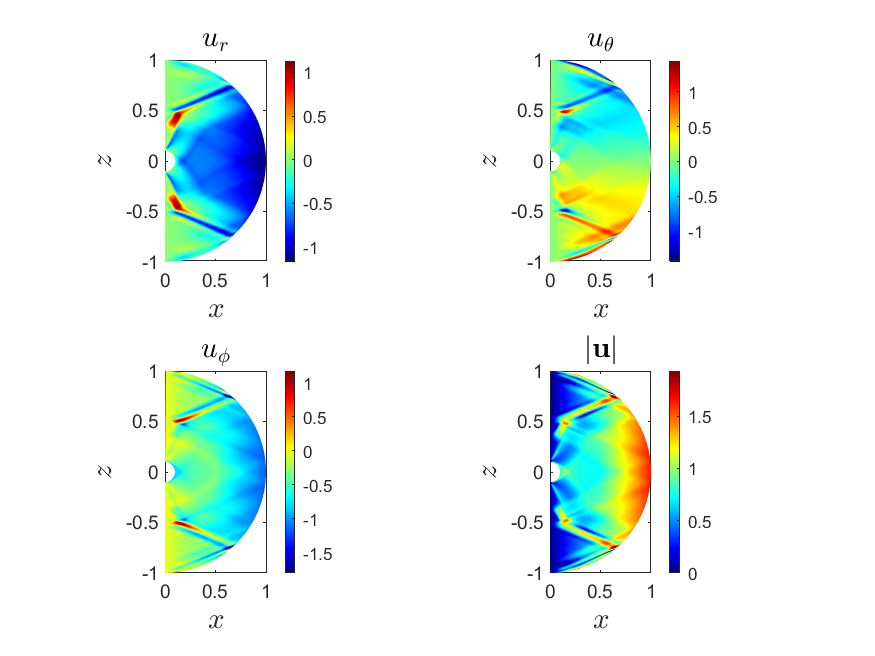}}
	\caption{Examples of the spatial structure for different rotation rates and forcing frequencies; in all cases we consider a stably stratified layer extending to half the planetary radius $\alpha=0.1$, $\beta=0.5$, $\bar{N}=1$, $\nu=\kappa=10^{-6}$.}\label{fig:Omega_structure}
\end{figure*}

Returning to the frequency-dependent dissipation in Figures~\ref{fig:diss_omega_b05} and~\ref{fig:diss_omega_b1}, which show the total dissipation ($D_{total}$) for four different rotation rates, as well as the corresponding non-rotating cases from \citetalias{Pontin2023} (black-dashed line); the cases shown are for $\beta=1$ and $\beta=0.5$, respectively. We can see that there is a non-trivial balance between the roles of the buoyancy and Coriolis forces. In all rotating cases we see gravito-inertial waves in the expected range $\omega < \sqrt{\bar{N}^2 + 4 \Omega^2}$. As well as observing the increasing range of modes we also see that the irregular pattern characteristic of inertial modes for $\beta=0.5$ is more pronounced as rotation rate increases. Indeed, at a low rotation rate, $\Omega=0.1$, the profile is similar to that of $\Omega=0$, as buoyancy forces appear to dominate. When comparing the two figures, we note that for a partially stratified planet, varying rotation has a more pronounced effect. This is because the stratified layer acts similarly to a large solid core in enhancing the excitation of inertial waves, which are not excited in a homogeneous full sphere \citep[e.g.][]{Ogilvie2009}. 

Figure~\ref{fig:diss_omega_lowfreq} shows the low frequency range for $\beta=1$. Based on our analysis in \citetalias{Pontin2023} we interpret this regime as being the travelling wave regime, for which tidally-excited inwardly-propagating gravito-inertial waves are excited in the stably stratified core by perturbations (by non-wavelike tides plus inertial waves) at $r\sim \beta r_0$, which are then subsequently fully damped by viscous and thermal diffusion before they reflect from the solid core and return to their launching sites. We can see that at low frequencies there is a clear $O(1)$ dependence on $\Omega$, though it does not typically lead to order of magnitude variations. It is possible that a similar analysis to that carried out in the travelling wave regime in \citetalias{Pontin2023} \citep[and considered by][in different models]{Papaloizou1997,Ogilvie2004,Ivanov2013} could explain this dependence (e.g. adopting the ``traditional approximation") but we leave this to future work as it only leads to $O(1)$ differences in the dissipation.

Turning our attention to the spatial structure of the velocity magnitude in the $\beta=0.5$ case in Figure~\ref{fig:Omega_structure}, which shows the forced solution for four values of $\Omega$ increasing in each row, with increasing forcing frequency in the different columns. We see clear agreement in properties with the predicted ranges for inertial and gravito-inertial waves in the convective envelope and outer core, respectively, as we move between the different regimes. Additionally we observe larger magnitude inertial waves as the rotation rate $\Omega$ increases, corresponding to the larger dissipation rates. Note the changing colour bars between the different cases. Solutions in the first column are visually similar to the non-rotating solutions in Fig.3(a) of \citetalias{Pontin2023}, except that inertial waves are excited in the convective envelope. In the middle column, showing a frequency outside the inertial range in the envelope in all panels except for $\Omega=0.6$, the solution varies substantially for the different rotation rates, from g-mode like in the top panel, to being strongly modified by rotation in the core in the bottom two panels. The rightmost column shows a higher frequency for which the free surface perturbations become more important, for which $\omega=1.1$ is outside the inertial range except for $\Omega=0.6$. Panel (l) shows a mode with a complicated structure that exhibits properties of each of surface gravity, inertial and gravito-inertial modes near the surface, in the envelope and core, respectively.

Varying the rotation rate can therefore modify the modes and tidal dissipation rates substantially, depending on the relevant tidal frequency, and whether it lies relative to the free inertial, gravito-inertial and surface gravity modes. Overall, the dissipation is typically enhanced for faster rotation rates.

\subsection{Variation of core sizes $\alpha$ and $\beta$ for uniform $N$}

A large uncertainty is how deep any stable layers extend throughout planetary interiors, with current estimations and observational constraints varying significantly. We know from previous studies that the size of a solid core can significantly enhance the excitation of inertial waves in incompressible models \citep{Ogilvie2009,Goodman2009,Rieutord2010}. Here we compare two scenarios, cases with a solid core with a convective envelope above ($\bar{N}=0$, varying $\alpha$) and cases where a small solid core is surrounded by a stably stratified (``dilute core") layer again beneath a convective envelope ($\bar{N}=1$, $\alpha=0.1$, varying $\beta$). 

\begin{figure}
\subfigure[$\bar{D}=\int D / \omega~\mathrm{d} \omega$, $\omega_{min}=0$, $\omega_{max}=1$]{
	\begin{tikzpicture}
		\node at (0,0) {\includegraphics[width=0.49\textwidth]{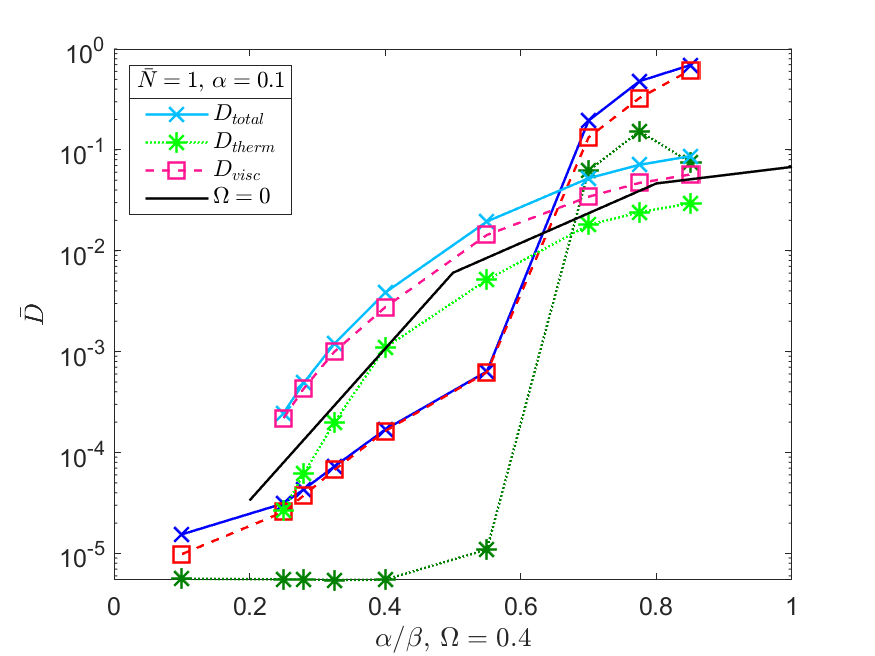}};
		\node at (2.5,-1) {\includegraphics[width=0.1\textwidth]{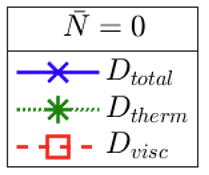}};
	\end{tikzpicture}}
	\caption{Frequency-averaged dissipation as a function of core size for both cases where a convective layer sits above a solid core ($\bar{N}=0$ and varying $\alpha$) and those where a stably stratified layer extends to the same radius ($\bar{N}=1$, $\alpha=0.1$, varying $\beta$). Other parameters kept constant at $\Omega=0.4$ and $\nu=\kappa=10^{-6}$.}\label{fig:int_diss_coresize}
\end{figure}

\begin{figure*}
	\centering
	\subfigure[$\bar{N}=0$ and varying $\alpha$]{\includegraphics[width=0.49\textwidth]{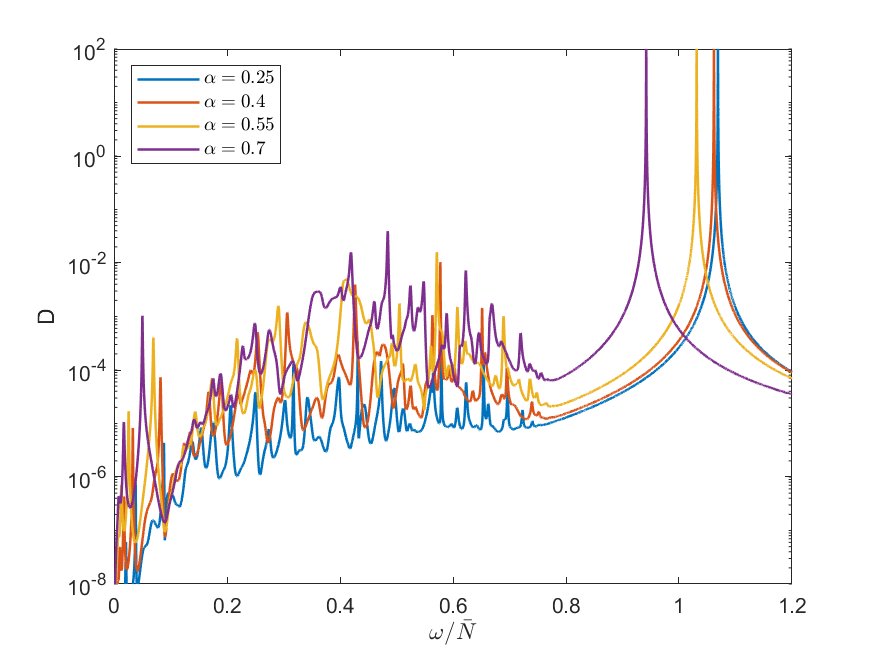}
	\label{fig:diss_alpha}}
	\subfigure[$\bar{N}=1$, $\alpha=0.1$, varying $\beta$]{
	\includegraphics[width=0.49\textwidth]{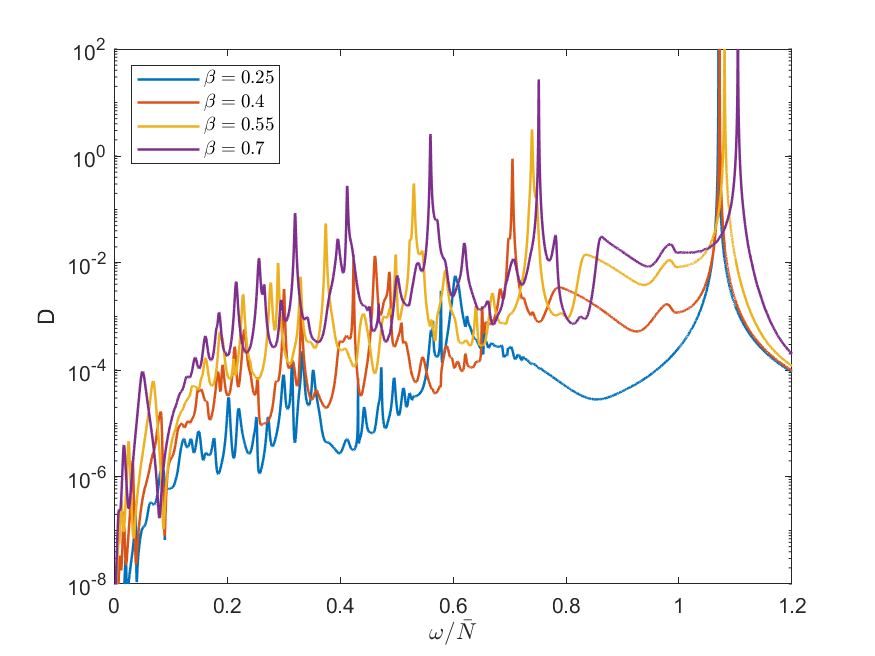}\label{fig:diss_beta}}
	\subfigure[Core extending to $0.25 r_0$]{
	\includegraphics[width=0.49\textwidth]{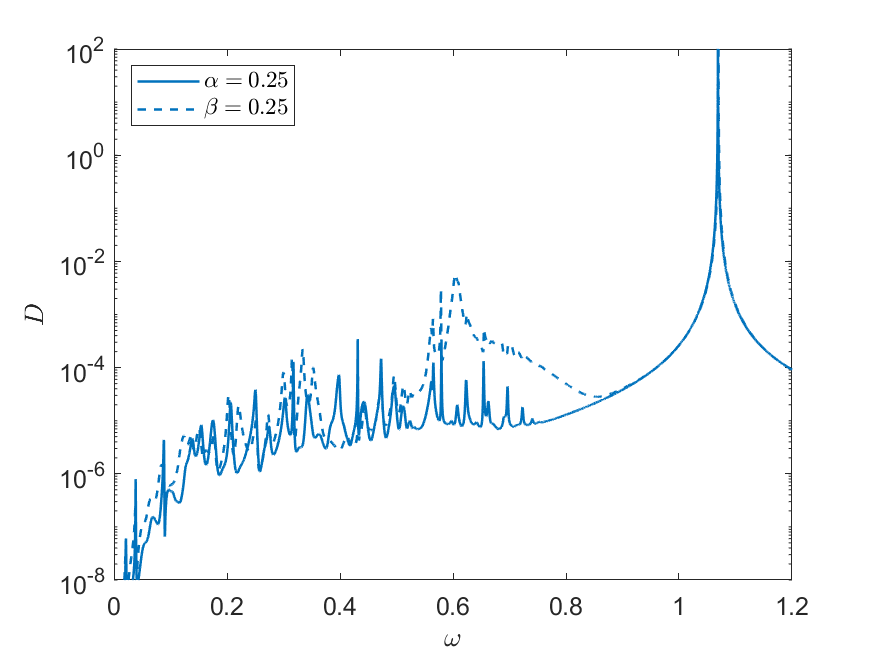}\label{fig:diss_0.25}}
	\subfigure[Core extending to $0.55 r_0$]{
	\includegraphics[width=0.49\textwidth]{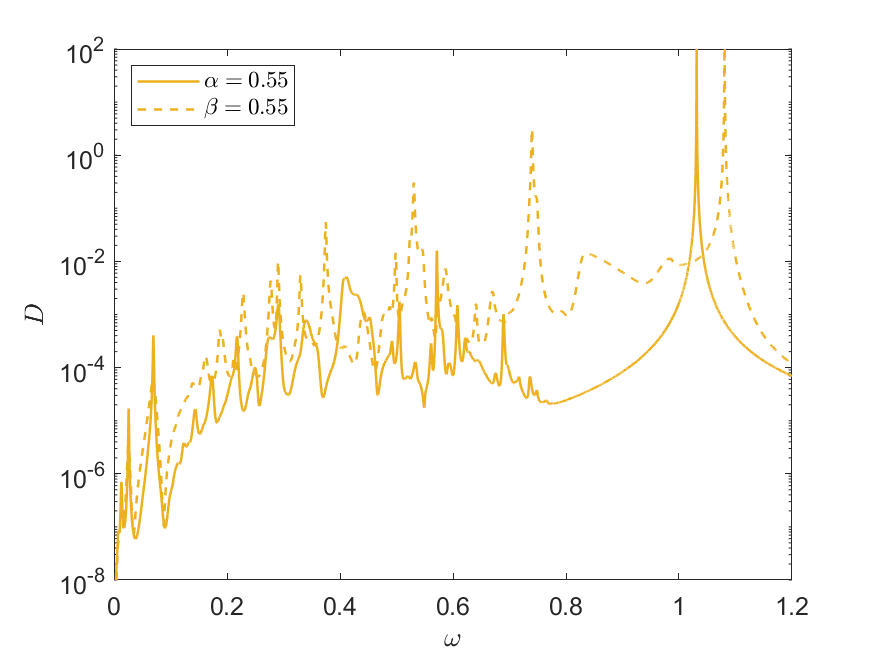}\label{fig:diss_0.55}}
	\subfigure[Core extending to $0.85 r_0$]{
	\includegraphics[width=0.49\textwidth]{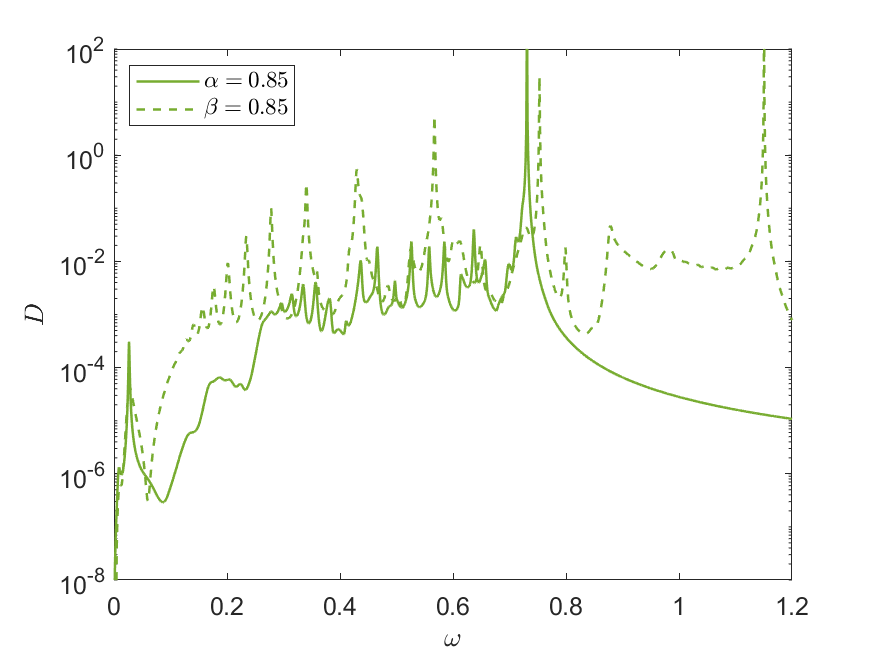}
	\label{fig:diss_0.85}}
	\caption{Frequency dependence of the dissipation rate for different core sizes, comparing both a solid core and stably stratified core, with $\Omega=0.4$, $\nu=\kappa=10^{-6}$ in all cases. Panels \protect\subref{fig:diss_alpha} and \protect\subref{fig:diss_beta} show a solid core and stratified layer, respectively, for different core sizes. Panels \protect\subref{fig:diss_0.25} to \protect\subref{fig:diss_0.85} compare a solid core to a stratified layer for three different radii.}\label{fig:diss_over_core}
\end{figure*}

Figure~\ref{fig:int_diss_coresize} shows the frequency-averaged profiles for these two types of model. The darker lines show neutrally stratified cases and the dependence on the radius of the solid core $\alpha$, whilst the lighter coloured lines show the cases varying the size of the stably stratified core $\beta$. The black line shows the equivalent trend found in the non-rotating cases, where we consider a small core with a stably stratified region above. We can see that, in both cases, the frequency-averaged dissipation shows a strong dependence on core size, but this trend varies between the two core types. For core sizes less than $0.5 r_0$ there are larger dissipation rates where there is a stably stratified core; for larger core sizes (larger than would be expected in the case of Jupiter and Saturn), we see the opposite trend. This suggests that for the smaller core sizes, the stratified layer acts as a solid core for the excitation of inertial waves, with additional contribution to the overall dissipation rate arising from the excitation of gravito-inertial waves in the stratified layer. We consider the trend found for the largest core sizes with caution however, as when considering the frequency-dependent dissipation in Figure~\ref{fig:diss_over_core}, we see that the surface gravity mode has shifted to frequencies less than the integration limit $\omega=1$ when considering a solid core, which also contributes to the dissipation rate. We again see the enhancement of the dissipation in solid core cases with rotation compared with non-rotating results.

Figure~\ref{fig:diss_over_core} shows the frequency-dependent dissipation rates, which allow us to understand more about the contribution of a stratified core. In Figures~\ref{fig:diss_alpha} and~\ref{fig:diss_beta} we compare cases with differently-sized cores for a solid core and a stratified core, respectively. In both cases, we see that the amplitudes of the resonances increase as the core sizes increase, but we note some differences. For cases with a stratified core, we observe the increased frequency range of resonances matching gravito-inertial waves rather than inertial waves. Additionally, when a stably stratified layer extends to $0.85 r_0$, there are regular discrete peaks that are characteristic of internal gravity waves that dominate around a forcing frequency of $0.2$. In Figures~\ref{fig:diss_0.25} to~\ref{fig:diss_0.85}, we compare the total dissipation rates for cases with a solid core and stably stratified core for three different radii, $0.25$, $0.55$ and $ 0.85$. We see clearly that for the smallest core size, the inertial wave response is very similar in both models, departing from each other at frequencies around $0.6-0.9$ only. However, as we increase the core size we increase the contribution of the gravito-inertial waves so this agreement between a solid core model and a dilute core is worse. Indeed, when the core extends as far as $0.85r_0$, we see at frequencies less than $0.6$ that the dissipation in a stably stratified core resembles the equivalent non-rotating system more than the equivalent solid core case. We note also that the shift in the frequency of the surface gravity mode varies between these cases as the stratified layer also affects the frequency of this mode (as predicted analytically without rotation in \citetalias{Pontin2023}). Finally, we note that when considering these thin shell convective regions (large cores) we expect negative frequency Rossby modes to be more significant than the positive frequency modes we show here, which require further investigation to explore.

\subsection{Variation of step number for a density staircase}
\label{4.3}

\begin{figure}
	\centering
	\subfigure[$\bar{D}=\int D / \omega~\mathrm{d} \omega$, $\omega_{min}=0.1$, $\omega_{max}=1$]{
	\includegraphics[width=0.48\textwidth]{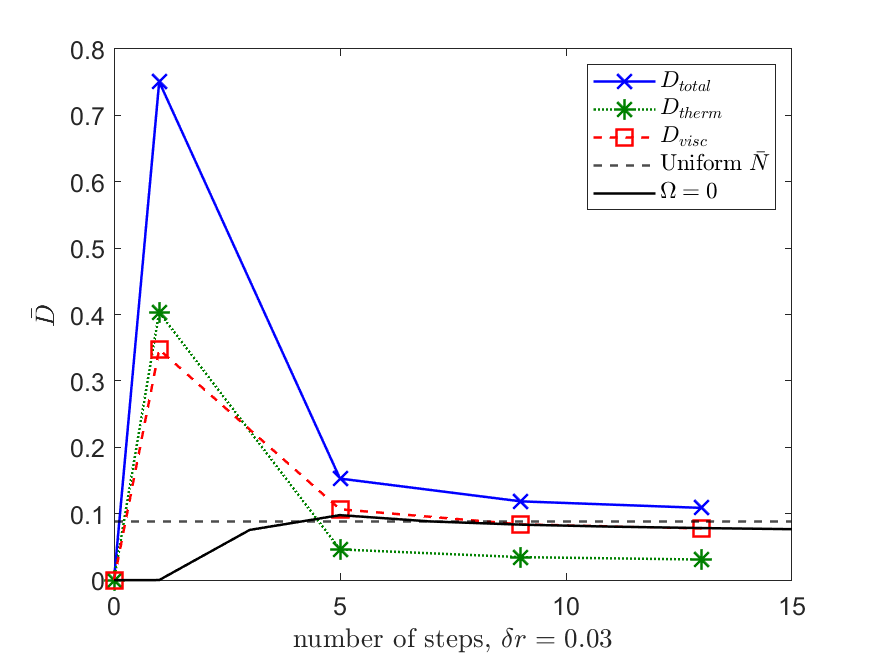}
	\label{fig:diss_int_steps_1}}
        \subfigure[$\bar{D}=\int D / \omega~\mathrm{d} \omega$, $\omega_{min}=0.1$, $\omega_{max}=1$]{
	\includegraphics[width=0.48\textwidth]{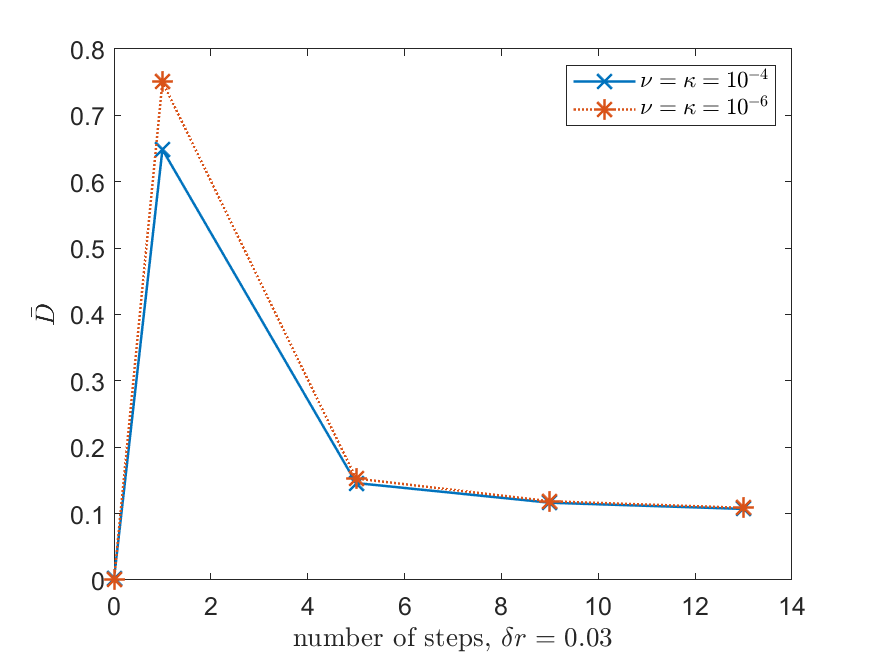}\label{fig:diss_int_steps_nu}}
	\caption{Frequency-averaged dissipation as a function of the number of steps. Other parameters kept constant at $\alpha=0.1$, $\beta=1.0$, $\bar{N}=1$, $\Omega=0.4$, $\nu=\kappa=10^{-6}$. Panel \protect\subref{fig:diss_int_steps_1} compares total, viscous and thermal dissipation with the total dissipation of the uniformly stratified case and the non-rotating case. Panel \protect\subref{fig:diss_int_steps_nu} compares two different viscosities/diffusivities for the same case.} \label{fig:diss_steps_over}
\end{figure}

\begin{figure}
	\centering
	\subfigure[$\nu=\kappa=10^{-7}$]{
	\includegraphics[width=0.49\textwidth]{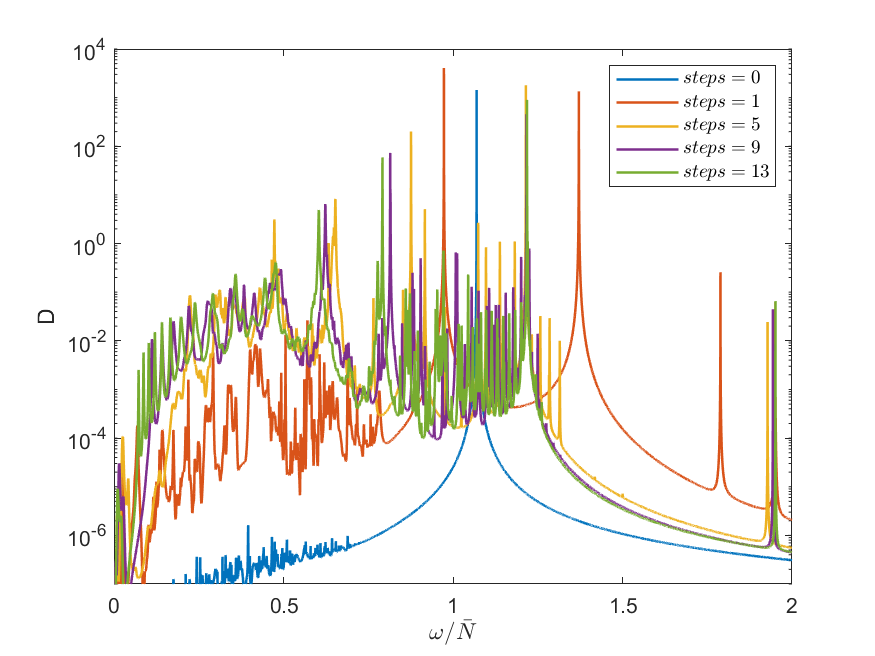}\label{fig:diss_steps_nu7}}
	\subfigure[$\nu=\kappa=10^{-6}$]{
	\includegraphics[width=0.49\textwidth]{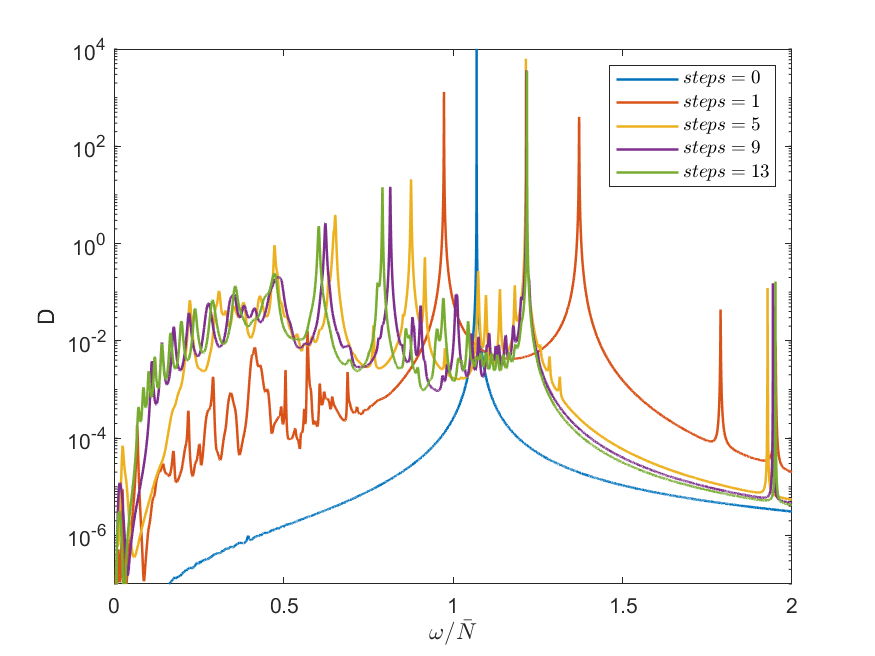}\label{fig:diss_steps_nu6}}
	\subfigure[$\nu=\kappa=10^{-4}$]{
	\includegraphics[width=0.49\textwidth]{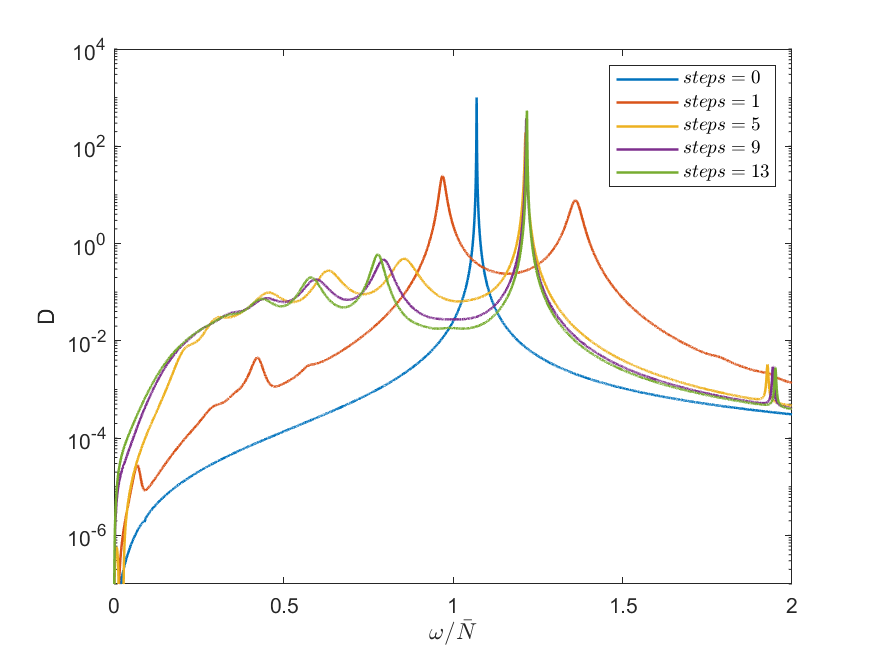}\label{fig:diss_steps_nu4}}
	\caption{Frequency dependence of dissipation for different numbers of steps in the staircase density profile. In all cases $\alpha=0.1$, $\beta=1$, $\bar{N}=1$. In panel \protect\subref{fig:diss_steps_nu6}, we fix $\nu=\kappa=10^{-6}$ and in panel \protect\subref{fig:diss_steps_nu4} we fix $\nu=\kappa=10^{-4}$.} \label{fig:diss_steps_nu}
\end{figure}

\begin{figure}
	\centering
	\subfigure[steps $=0$]{
        \includegraphics[width=0.22\textwidth, trim=80mm 0mm 25mm 55mm, clip]{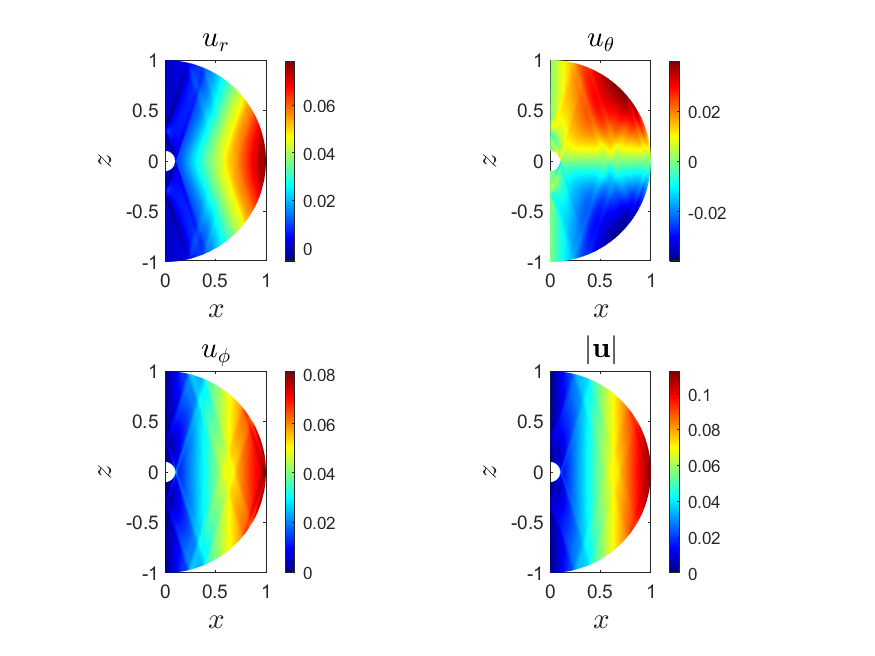}}
        \subfigure[steps $=1$]{
        \includegraphics[width=0.22\textwidth, trim=80mm 0mm 25mm 55mm, clip]{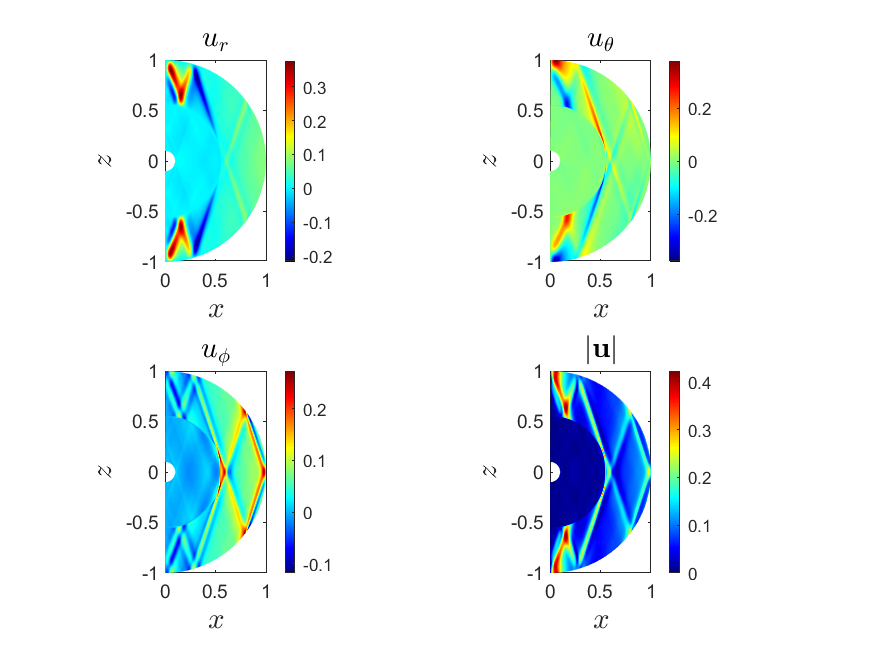}}
        \subfigure[steps $=5$]{
        \includegraphics[width=0.22\textwidth, trim=80mm 0mm 25mm 55mm, clip]{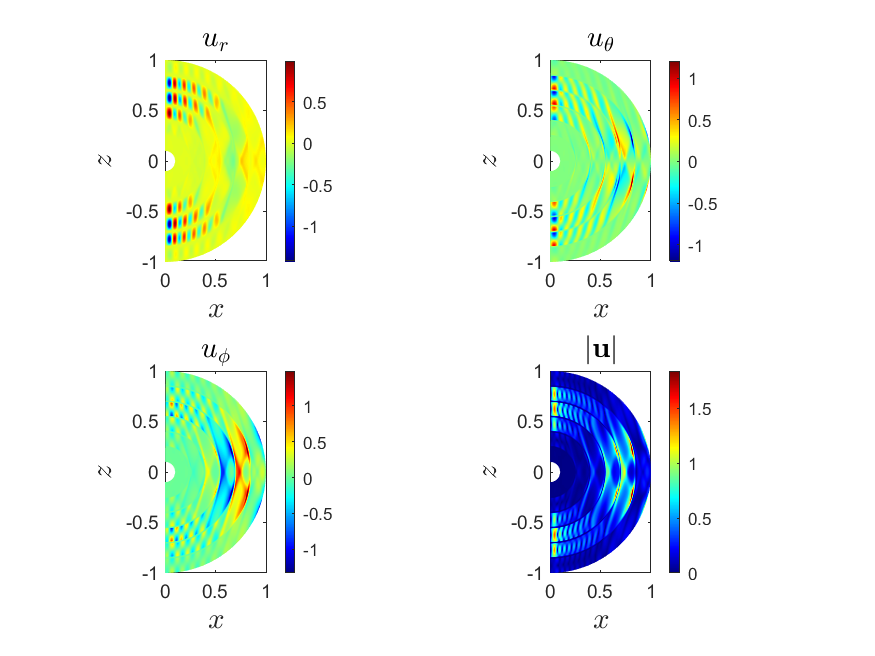}}
        \subfigure[steps $=9$]{
        \includegraphics[width=0.22\textwidth, trim=80mm 0mm 25mm 55mm, clip]{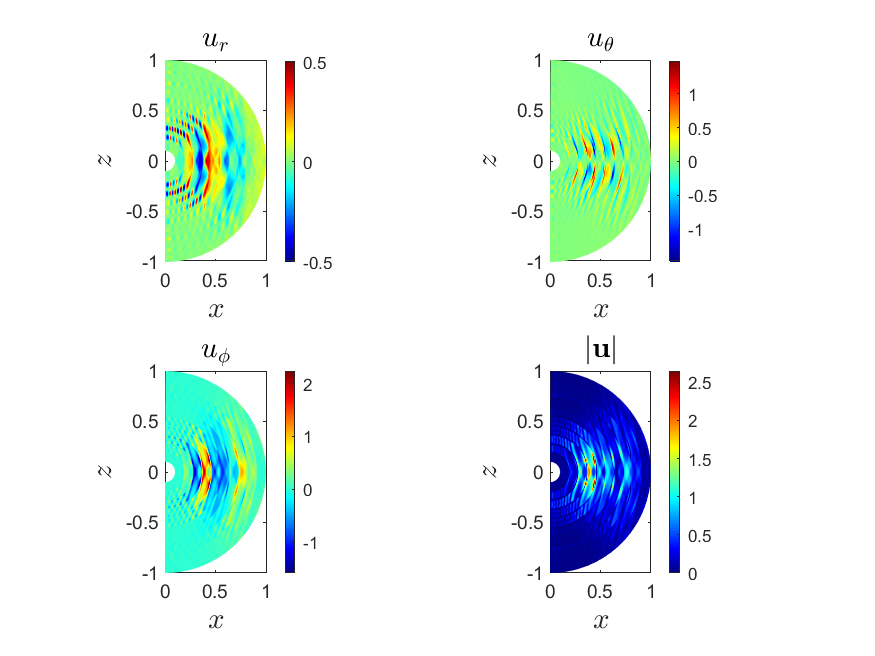}}
        \subfigure[steps $=13$]{
        \includegraphics[width=0.22\textwidth, trim=80mm 0mm 25mm 55mm, clip]{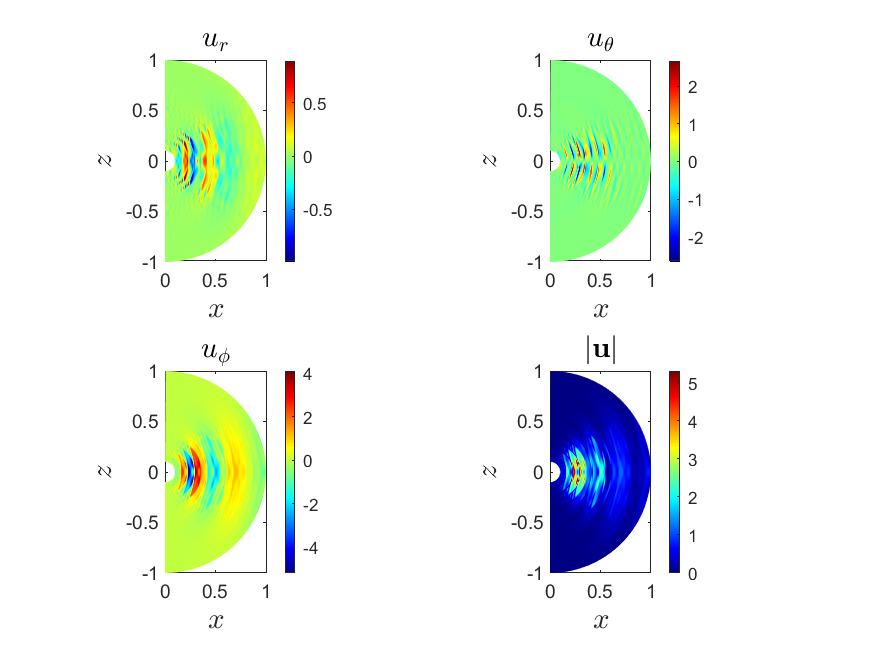}}
	\caption{Examples of the spatial structure of the forced response for a density staircase with 0 (i.e. unstratified),1,5,9 and 13 steps, in all cases with $\alpha=0.1$, $\beta=1.0$, $\bar{N}=1$, $\nu=\kappa=10^{-6}$, and a forcing frequency $\omega=0.25$.}\label{fig:steps_omega_0.25}
\end{figure}

We now turn to explore the consequences of semi-convective layers. As in non-rotating cases in \citetalias{Pontin2023}, we consider a staircase-like density profile as defined by equation~\ref{eq:N2_steps}, and we consider cases in which $\alpha=0.1$, $\beta=1.0$, $\bar{N}=1$, $\delta r=0.03$, $\Omega=0.4$ and $\nu=\kappa=10^{-6}$, with varying step numbers. We consider $\beta=1.0$ here to allow us to explore the largest number of steps possible, allowing us to resolve modes with variations on the scale of the steps while minimising computational costs. In Figure~\ref{fig:diss_int_steps_1} we show the frequency-averaged dissipation as the step number is increased; we have in this case shown just the $\frac{1}{\omega}$ weighting and used an upper integration limit of $\omega_{max}=1$ to exclude the surface gravity waves. Note, the lower limit of $\omega_{min}=0.1$ has been chosen for purely numerical reasons; this is because the lowest frequency waves require the highest resolution and contribute little to the overall dissipation using this measure. For comparison, we also show the equivalent non-rotating case as a function of step number (black-dashed line), as well as the uniformly stratified layer with equivalent mean stratification (black solid line). 

We see that the dissipation for all step numbers considered is higher than both a uniformly stratified rotating medium and cases without rotation according to this measure. Although there is a trend towards the uniformly stratified case as we increase the number of steps, our results do not converge to this result as quickly as we had found in the non-rotating cases in Figure 11 of \citetalias{Pontin2023}. The dissipation is initially nearly 8 times larger with one step as the case with a constant $N$. For large step numbers though, there is a trend towards the uniformly stratified case, implying that if the planetary interior contains a large number of steps it will behave on average similar to an equivalent continuously stratified medium. Here the staircase dissipation remains slightly larger than the uniform case even for 13 steps, though it is very similar.

In Figure~\ref{fig:diss_int_steps_nu} we explore the effects of varying viscosity and thermal diffusivity according to the same frequency-averaged measure, showing total dissipation for $\nu=\kappa=10^{-6}$ compared with $\nu=\kappa=10^{-4}$. Although it is hard to draw robust conclusions from two values of the diffusivities, these initial results suggest that our findings may be robust to varying the viscosity and thermal diffusivity to approach planetary values. We also expect that varying the Prandtl number may also be unimportant for the total dissipation, and that, as found in \citetalias{Pontin2023}, it will primarily alter the balance between viscous and thermal dissipation rates rather than the total dissipation.

Figure~\ref{fig:diss_steps_nu} shows the corresponding frequency-dependent dissipation rates for various cases with different step numbers for $\nu=\kappa=10^{-6}$ in panel (a) and $\nu=\kappa=10^{-4}$ in (b). At low frequencies the behaviour varies significantly for different numbers of steps. Note that the frequency limit for propagation of gravito-inertial waves is $\omega < \sqrt{N_{max}^2 + 4 \Omega^2}$ and when considering a staircase-like structure $N_{max} \geq \bar{N}$, thereby increasing the range slightly (but with no significant implications for the most astrophysically relevant parameter values, so we do not focus on this aspect). 

We first compare the cases with zero steps, one step and five steps, noting that zero steps consists of a small solid core with a convective envelope to the outer edge. At low frequencies, although inertial waves are excited in the convective envelopes in all three cases, these are barely visible for the case of zero steps, where there is only the small solid core to launch inertial waves from. As we increase the number of steps to one and then five, the excitation of inertial waves occurs from boundaries at additional and increasingly larger radii, leading to enhanced dissipation. Following this, we now compare the cases with five, nine and 13 steps and find there are significantly smaller differences between these cases. At this point the outer radius does not vary significantly as the number of steps is increased. The individual resonant peaks do shift as the number of steps is increased, but overall the dissipation spectrum looks similar when there are ``enough steps" (here this means more than five).

Figure~\ref{fig:steps_omega_0.25} shows the forced solutions at $\omega=0.25$ for all five step numbers explored for $\nu=\kappa=10^{-6}$. We see inertial wave beams excited in the convective layers, which have a very small amplitude response dominated by the non-wavelike tide in the case of zero steps, but increase in amplitude as we increase the number of steps and the outer interface moves outwards. The case with one step behaves visually like a solid core at the location of the interface, with inertial waves in the envelope and little activity inside the interface within the ``core". We will explore this similarity further in the next section. As we increase the number of steps, the solution appears to be dominated by inertial waves, presumably excited at the critical latitudes on each interface (where rays are tangent to it), that subsequently bounce between the interfaces to form these modes. The solution becomes increasingly dominated by wavelike tides, rather than the larger scale non-wavelike tide, as we increase the number of steps in these examples.

\begin{figure}
	\centering
	\subfigure{\includegraphics[width=0.49\textwidth]{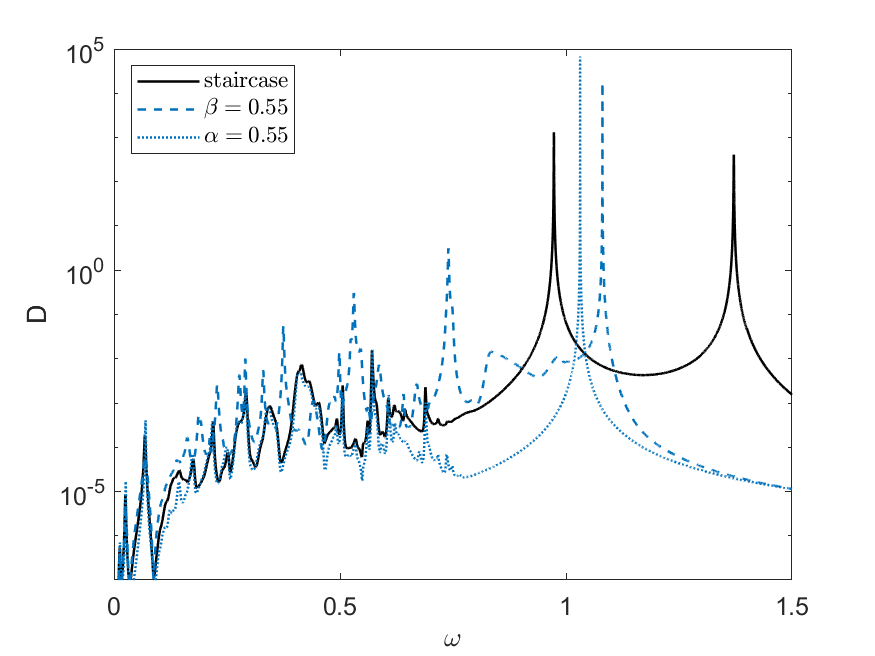}\label{fig:diss_1step}}
	\subfigure{\includegraphics[width=0.49\textwidth]{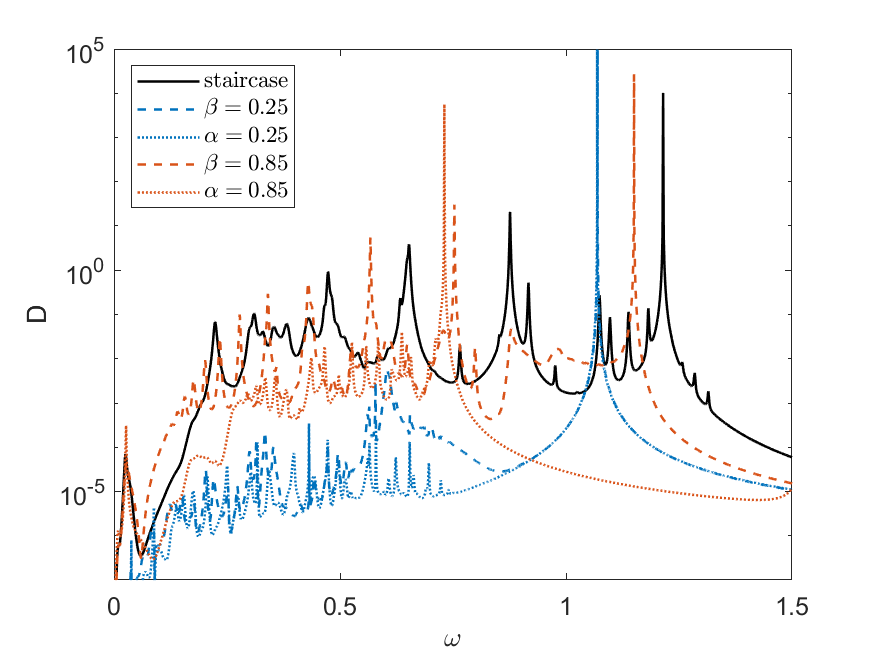}\label{fig:diss_5step}}
	\caption{Comparison of total dissipation between profiles with interfaces, uniformly stratified layers and a solid core, with $\Omega=0.4$ and $\nu=\kappa=10^{-6}$ in all cases.  The black solid lines are the single and five step cases in panels \protect\subref{fig:diss_1step} and \protect\subref{fig:diss_5step} respectively. The dashed coloured lines describe a stratified layer with $\alpha=0.1$ and $\bar{N}=1$, and solid coloured lines a solid core with $\bar{N}=0$. The blue aligns the core/stratification with the first (or only) interface and the red the last interface.} \label{fig:diss_step_int}
\end{figure}

We are beginning to see that the size of the convective envelope is key to the behaviour of the inertial waves and the corresponding dissipation, which is consistent with expectation from incompressible models of the unstratified case \citep[e.g.][]{Ogilvie2009,Goodman2009,Rieutord2010,Ogilvie2013}. Therefore, we now compare cases with a solid or stratified core that extends to the same radii as the staircase interfaces. The solid black lines on Figures~\ref{fig:diss_1step} and \ref{fig:diss_5step} show the total dissipation as a function of frequency for a single step and five steps, respectively. In Figure \ref{fig:diss_1step}, we have compared this case to both a solid core, and a stratified layer extending to that radius, i.e.~$\alpha=0.55$ and $\bar{N}=0$ and $\alpha=0.1$, $\beta=0.55$ and $\bar{N}=1$. We can see that at low frequencies there is very good agreement between these three different profiles, suggesting that the forced wave response in the outer envelope only depends weakly on what is below the envelope, provided the buoyancy frequency of the stably stratified layers is sufficiently strong. We will explore this issue further in the next section.

In Figure \ref{fig:diss_5step} we have similarly shown a solid core and a stably stratified layer, this time extending to the first interface (blue, $\alpha / \beta=0.25$) and last interface (red, $\alpha / \beta=0.85$). We notice that although the agreement is not as good as in the single step case, there is still closer agreement between the cases where the core corresponds to the last interface of the staircase. This suggests that this is the key interface in dictating the dissipation due to inertial waves. We note that dissipation is larger in the staircase model due to the additional interior interfaces where inertial waves can be excited from critical latitudes. In both figures there is significantly different behaviour between the various models in the mid to high frequencies where it is expected that the buoyancy effects dominate. This comparison suggests that the importance of buoyancy forces compared with Coriolis forces will strongly depend on the forcing frequency. 

\section{Comparison of dilute core models}
\label{dilute}

\begin{figure}
    \centering
    \includegraphics[width=0.49\textwidth]{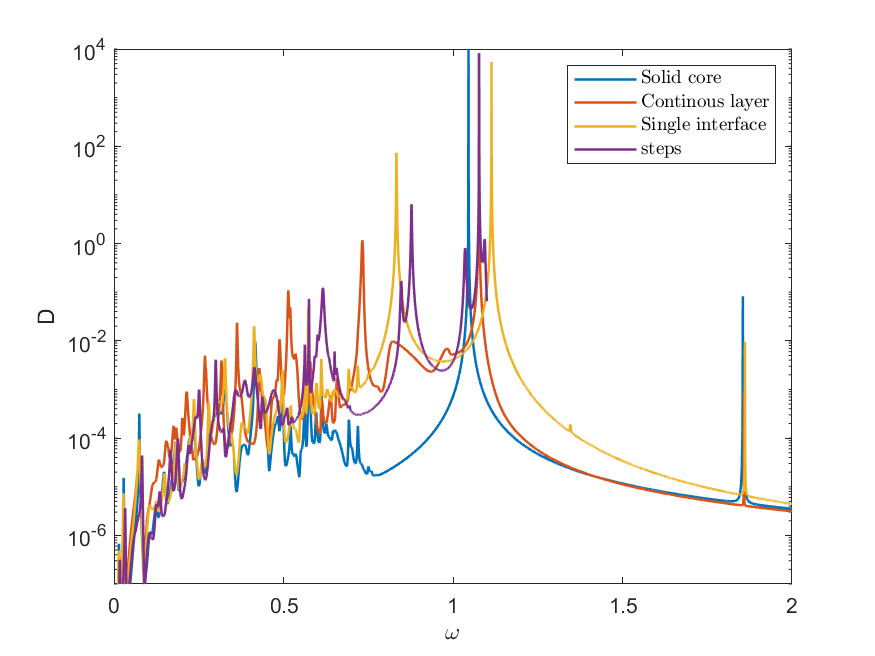}
	\caption{Total dissipation for different dilute core modes, where in all cases $\Omega=0.4$, $\nu=\kappa=10^{-6}$.}\label{fig:diss_outercore_comp}
\end{figure}

Motivated by the relevance of giant planet interior models that consist of an extended ``dilute" core, and that the properties of this core are highly uncertain, in this section we explore further the consequences of different buoyancy frequency profiles describing stable stratification. We consider four cases that have different buoyancy profiles that each represent a stratified outer core extending to half of the planetary radius surrounded by a convective envelope. We compare:

\begin{itemize} \setlength \itemsep{0pt}
\item \textbf{Case 1}: a large solid core with a convective envelope, $\alpha=0.5$, $\bar{N}=0$.
\item \textbf{Case 2}: a uniformly stably stratified layer extending from an inner core boundary to an outer core boundary, $\alpha=0.1$, $\beta=0.5$, $\bar{N}=1$.
\item \textbf{Case 3}: a single stable interface at the outer core boundary $\beta r_0$, $\alpha=0.1$, $\beta=0.5$, $\bar{N}=1$.
\item \textbf{Case 4}: a staircase extending from an inner core boundary to an outer core boundary, $\alpha=0.1$, $\beta=0.5$, $\bar{N}=1$, steps $=3$.
\end{itemize}
All other parameters are kept constant with $\Omega=0.4$ and $\nu=\kappa=10^{-6}$. We plot the total dissipation in each case in Figure \ref{fig:diss_outercore_comp}. 

We find that when considering the low frequency inertial range (i.e.~observed to be $|\omega|\lesssim \Omega=0.4$), the frequency-dependent dissipation is remarkably similar in all four cases. The inertial wave behaviour in the convective envelope then appears to dominate the behaviour and is little affected by the form of the stratified layer (or solid core) beneath it. We see that the stable layer, single interface, and staircase, each act like a solid boundary for the propagation of inertial waves in the convective envelope, and they enhance the dissipation for low frequencies in a very similar manner.

	\begin{figure*}\centering
   \underline{\textbf{Case 1}}\\
	\subfigure[$\omega=0.21$]{
        \includegraphics[width=0.22\textwidth, trim=86mm 9.5mm 25mm 55mm, clip]{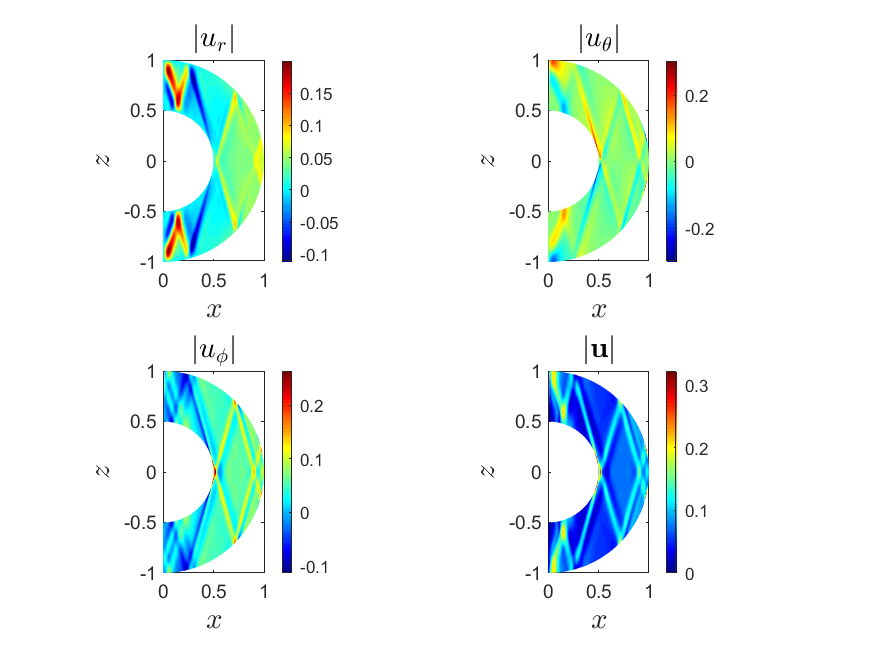}}
	\subfigure[$\omega=0.73$]{
        \includegraphics[width=0.22\textwidth, trim=86mm 9.5mm 25mm 55mm, clip]{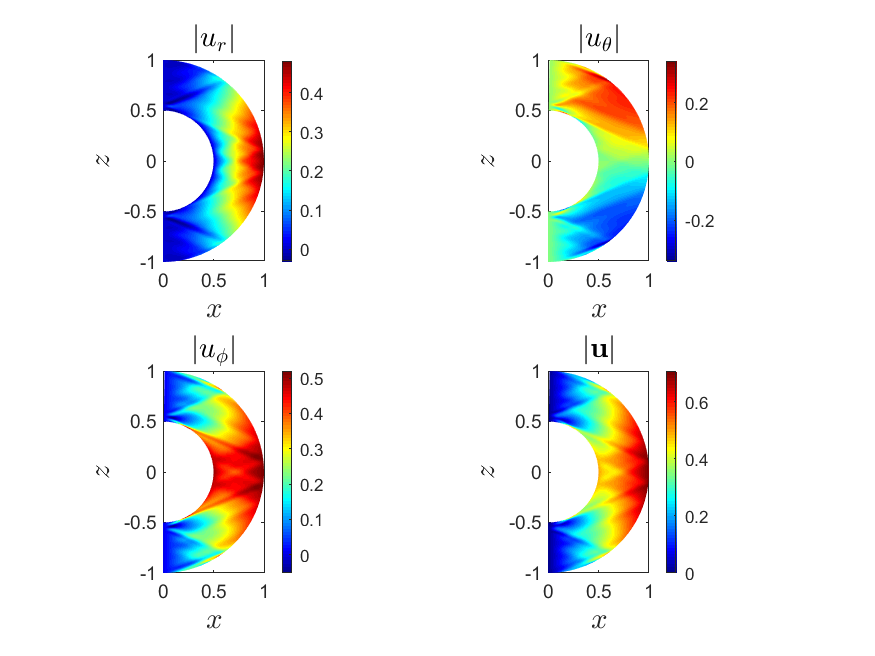}}
        \subfigure[$\omega=0.90$]{
        \includegraphics[width=0.22\textwidth, trim=86mm 9.5mm 25mm 55mm, clip]{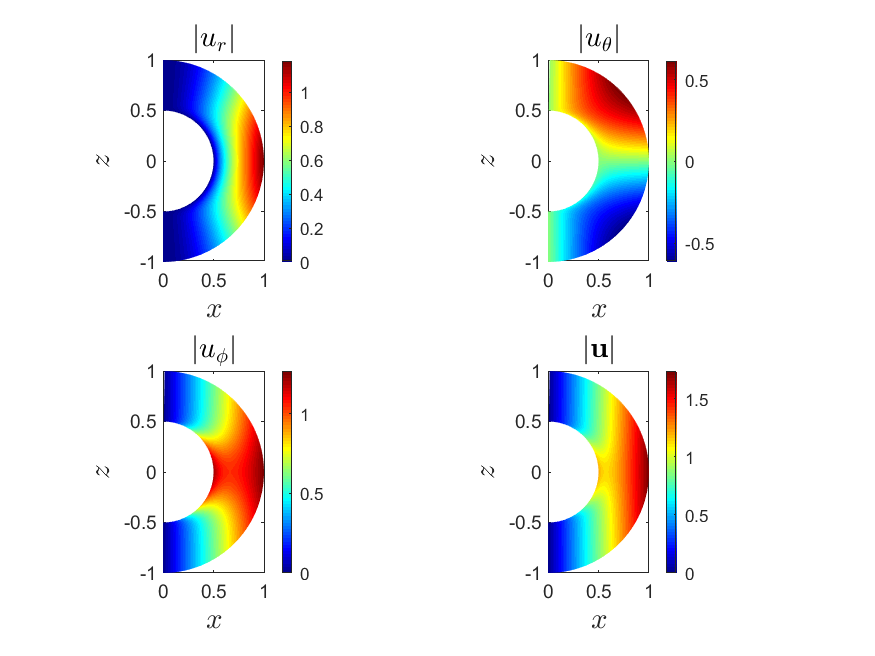}} \\
\underline{\textbf{Case 2}}\\
		\subfigure[$\omega=0.21$]{
        \includegraphics[width=0.22\textwidth, trim=86mm 9.5mm 25mm 61.8mm, clip]{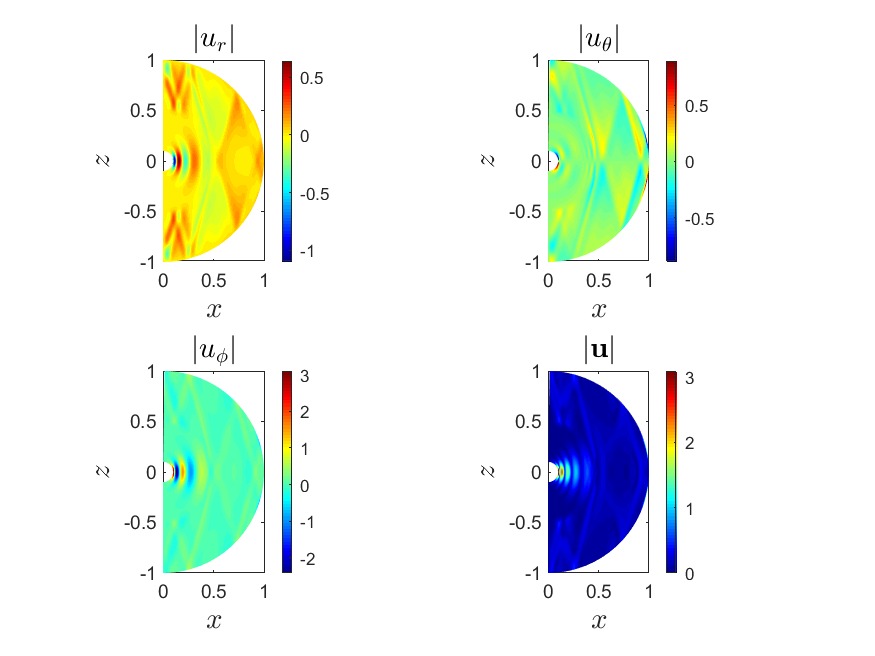}}
	\subfigure[$\omega=0.73$]{
        \includegraphics[width=0.22\textwidth, trim=86mm 9.5mm 25mm 61.8mm, clip]{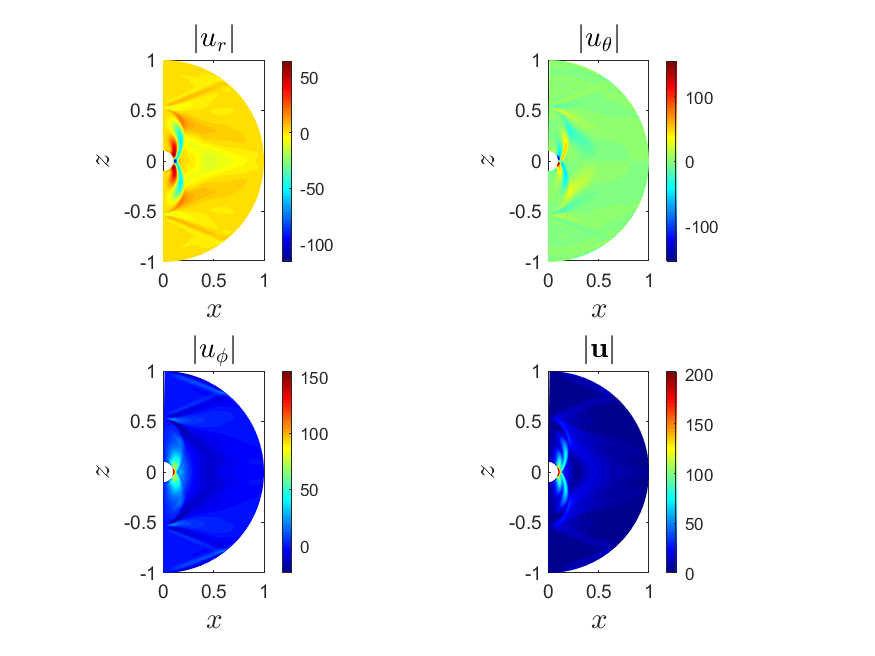}}
    \subfigure[$\omega=0.90$][$\omega=0.73$]{
        \includegraphics[width=0.22\textwidth, trim=86mm 9.5mm 25mm 61.8mm, clip]{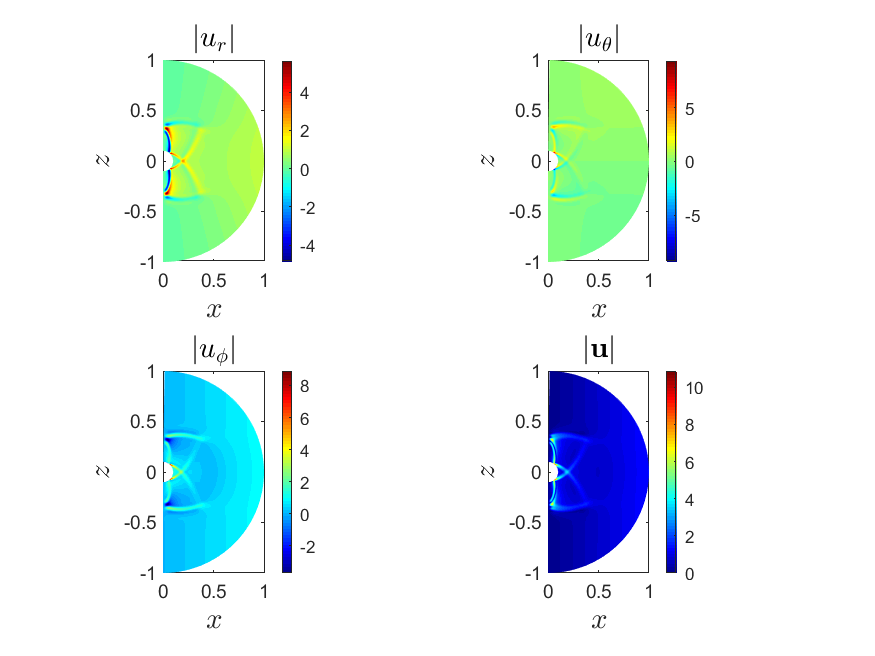}} \\
	\underline{\textbf{Case 3}}\\
	\subfigure[$\omega=0.21$]{
        \includegraphics[width=0.22\textwidth, trim=86mm 9.5mm 25mm 61.8mm, clip]{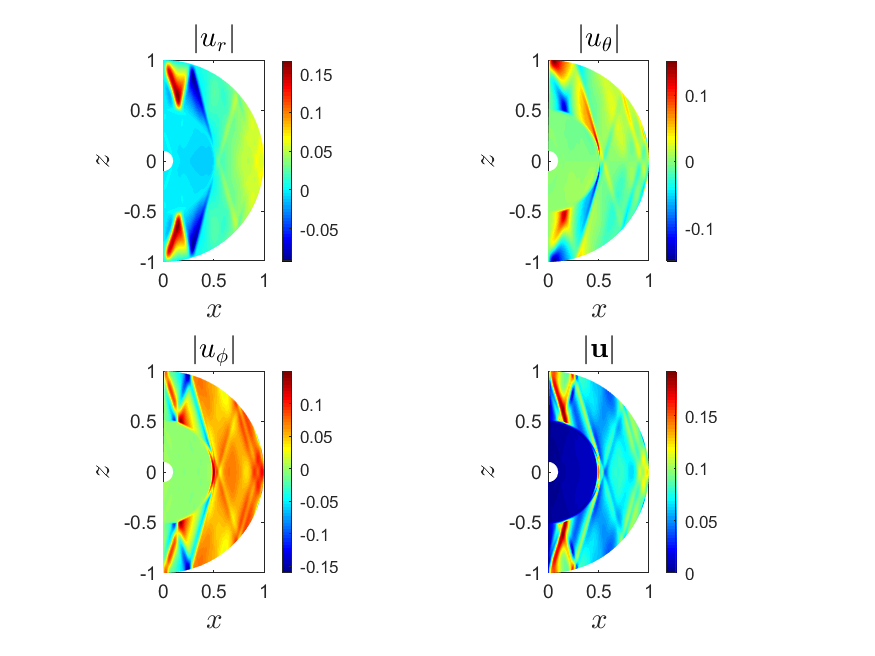}}
	\subfigure[$\omega=0.73$]{
       \includegraphics[width=0.22\textwidth, trim=86mm 9.5mm 25mm 61.8mm, clip]{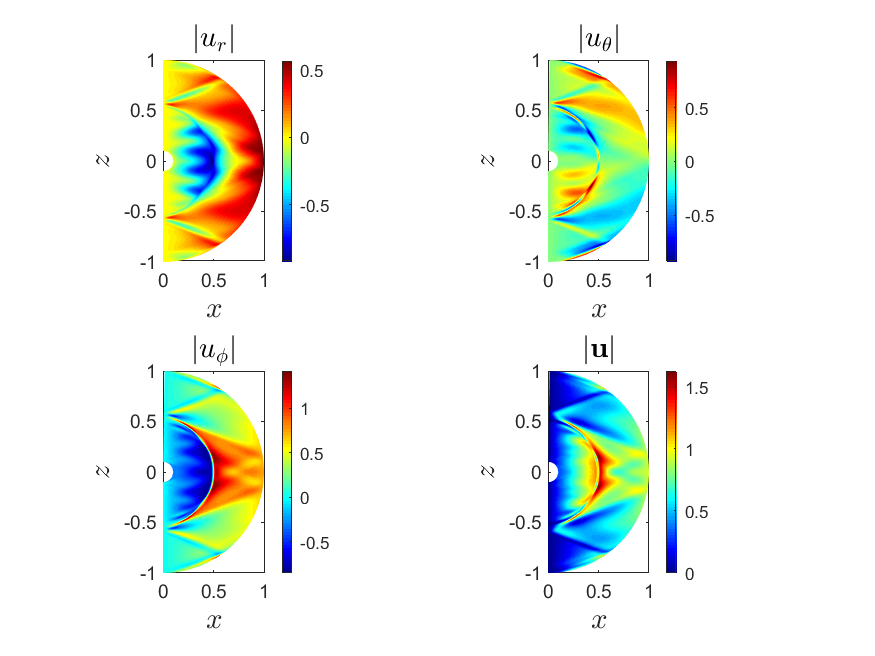}}
        \subfigure[$\omega=0.90$]{
        \includegraphics[width=0.22\textwidth, trim=86mm 9.5mm 25mm 61.8mm, clip]{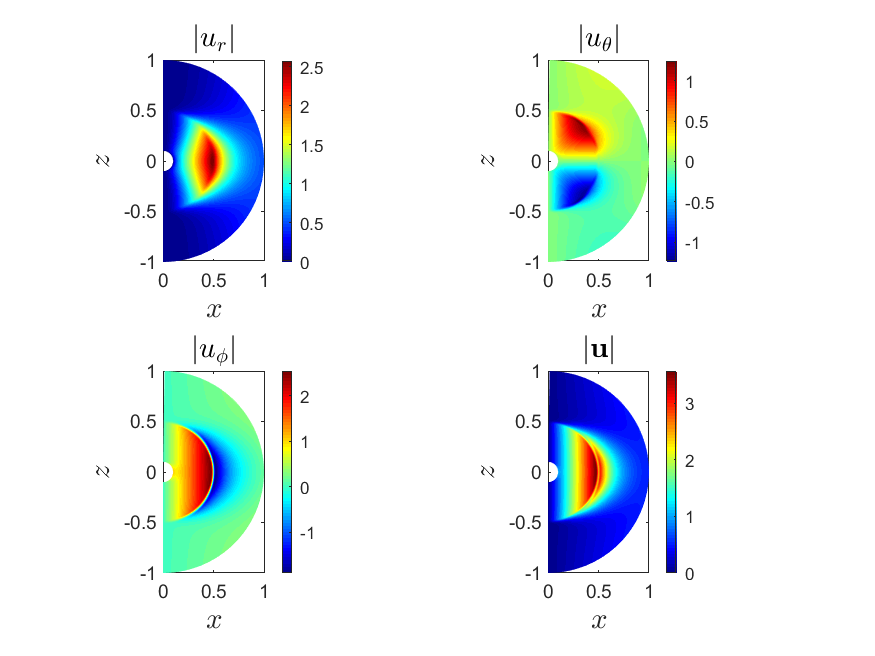}} \\
	\underline{\textbf{Case 4}}\\
\subfigure[$\omega=0.21$]{
        \includegraphics[width=0.22\textwidth, trim=86mm 0mm 25mm 61.8mm, clip]{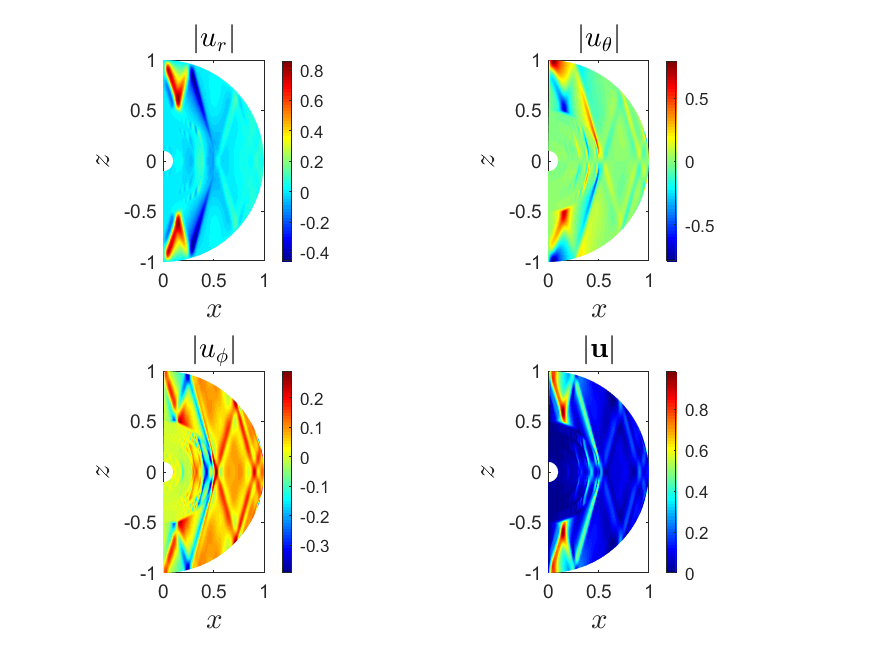}}
	\subfigure[$\omega=0.73$]{
       \includegraphics[width=0.22\textwidth, trim=86mm 0mm 25mm 61.8mm, clip]{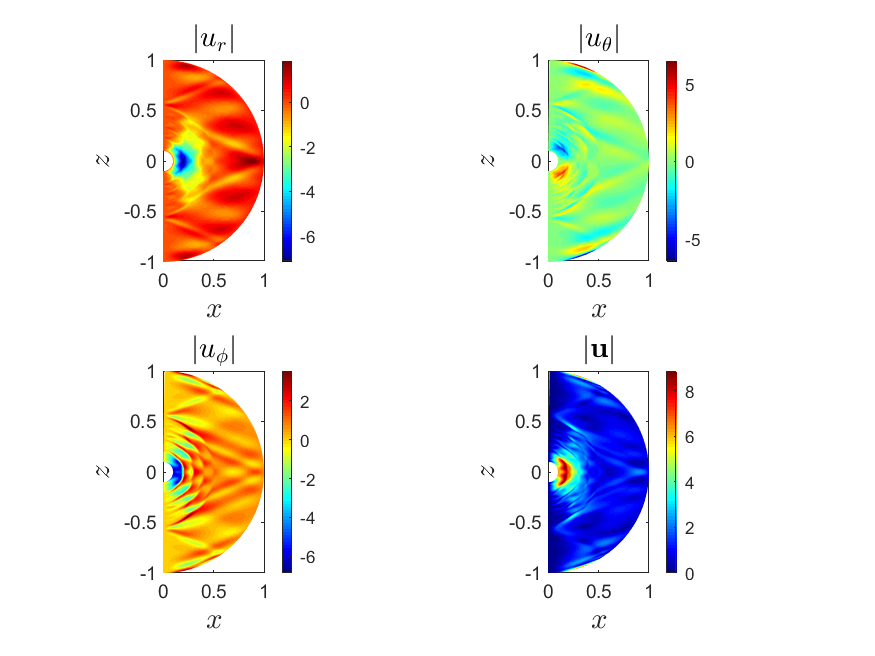}}
   \subfigure[$\omega=0.90$]{
        \includegraphics[width=0.22\textwidth, trim=86mm 0mm 25mm 61.8mm, clip]{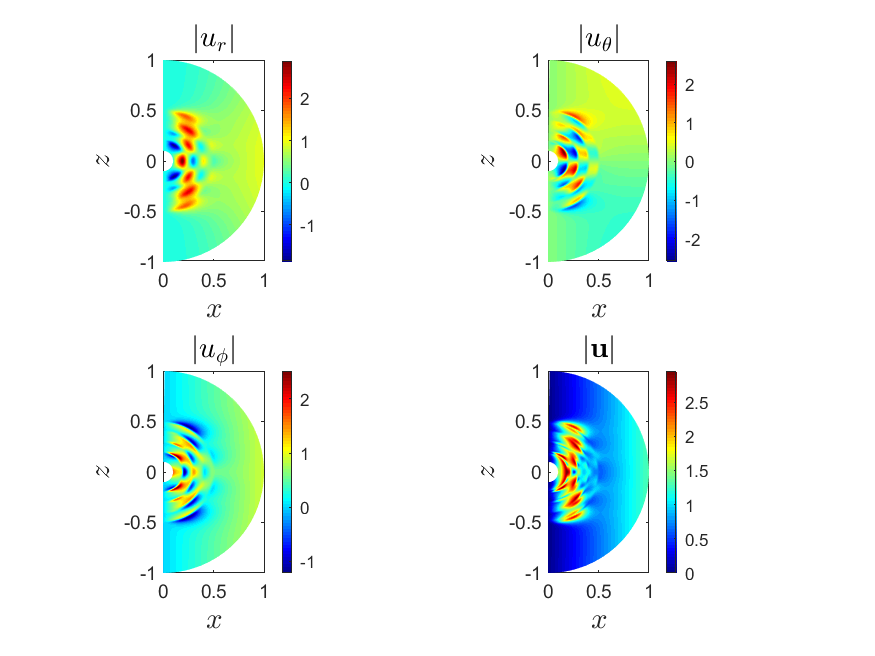}}
	\caption{Examples of the spatial structure for different dilute cores and forcing frequencies, with $\nu=\kappa=10^{-6}$ and $\Omega=0.4$.}\label{fig:_outercore_u}
\end{figure*}

Within the mid-frequency range between $0.8 \lesssim \omega \lesssim 1$, the behaviour varies significantly, as it is in this frequency range that the gravito-inertial modes within the stratified region are dominant when $\bar{N}=1$. These modes are sensitive to the form of the stratification adopted, and we observe peaks corresponding to gravito-inertial modes in the case of a uniformly stratified layer, as well as the interfacial modes that are characteristic of a staircase structure in cases with interfaces. 

In Figure~\ref{fig:_outercore_u}, we compare the spatial structure of the response in all four cases at three different forcing frequencies, using colour-scales that differ in each panel to most clearly illustrate the variation in space. Considering the first column for which the forcing frequency is low ($\omega=0.21$), within the inertial wave range, we observe the solution in the convective envelope to be similar in each case. Their structures are very similar, and their amplitudes are mostly similar but do differ to some extent. In all examples the stable stratification is acting effectively as a solid boundary for the propagation of inertial waves, showing that an extended (sufficiently stably) stratified core acts like a large solid core, enhancing the dissipation over cases with a small core. Inertial waves are presumably excited in a similar way at the critical latitudes on the interface at the outer boundary of the core.

In the second column, at a higher forcing frequency of $\omega=0.73$, the spatial structure observed in the convective region again remains consistent. However, we now see the different modes that form within the stratified region, which vary significantly, contributing to the differences between the resultant dissipation. At both forcing frequencies, for the staircase and interface cases we faintly observe additional inertial modes in the deeper convective layers, as well as gravito-inertial modes in the uniformly stratified case. These additional modes can explain the increase in dissipation observed for Cases 2--4 over that of a solid core (Case 1). 

Finally, in the last column we show the solution with the highest forcing frequency $\omega=0.9$, at which we are outside the inertial wave range ($|\omega|<2\Omega=0.8$) but within the gravito-inertial wave ranges. We clearly observe different responses in each case that  depend on the properties of the stratified region, with wavelike behaviour only observable in the cases with a stratified layer and a staircase structure. Purely non-wavelike behaviour is observed in the unstratified case, as expected since inertial waves do not propagate with this frequency.

To summarise this section and \S~\ref{4.3}, we find that a sufficiently strongly stratified core behaves very similarly to a rigid core for the inertial wave response, and for the corresponding dissipation at low frequencies ($|\omega|\ll \bar{N}$), and it is largely insensitive to the properties of the stratified layer beneath. For larger frequencies $\Omega \lesssim |\omega|\sim \bar{N}\sim \omega_d$, we find more substantial differences between these cases due to the decreasing importance of rotation relative to internal and surface buoyancy forces, and the transmission of wave energy between the convective and stably stratified layers. For the forcing due to Jupiter's and Saturn's moons, we typically expect $|\omega|\sim \Omega$, but it is not certain how large $\bar{N}$ should be. Weaker values of $\bar{N}\sim \Omega$, such as those considered by \cite{Lin2023,Dewberry2023} for Jupiter would permit substantial connection between convective and stable layers for these tidal frequencies, such that the response would be expected to differ somewhat from the case of a solid core. On the other hand the larger values inferred for Saturn \citep{Mankovich2021} for which $\bar{N}\gg \Omega$, would typically predict that the inertial wave response in the envelope may be better represented by a solid core at the outer interface of the stratified ``dilute core". Similarly, extrapolating our result to slowly rotating solar-type stars, for which $\bar{N}\gg \Omega$, the radiative zone would be expected to behave quite similarly to a rigid core for the inertial waves in the envelope, thereby motivating use of rigid boundary conditions in studies of their properties \citep[e.g.][]{AB2022}.

\section{Application to Saturn's tidal dissipation}
\label{Saturn}

\begin{figure*}
	\centering
	\subfigure{
	\includegraphics[width=\textwidth, trim=0mm 13mm 0mm 13mm, clip]{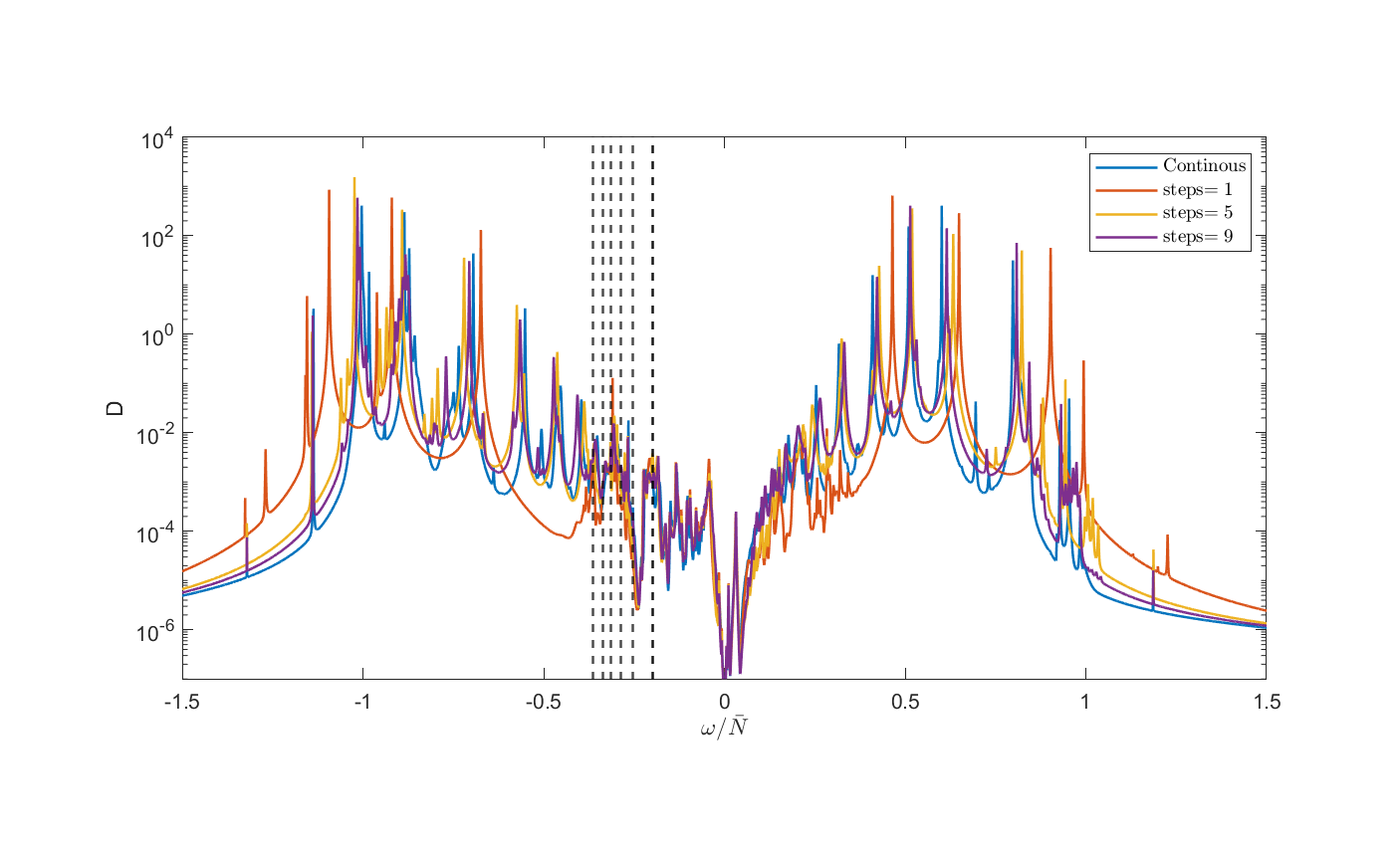}\label{fig:Saturn_diss}}
	\subfigure{
	\includegraphics[width=\textwidth, trim=0mm 13mm 0mm 13mm, clip]{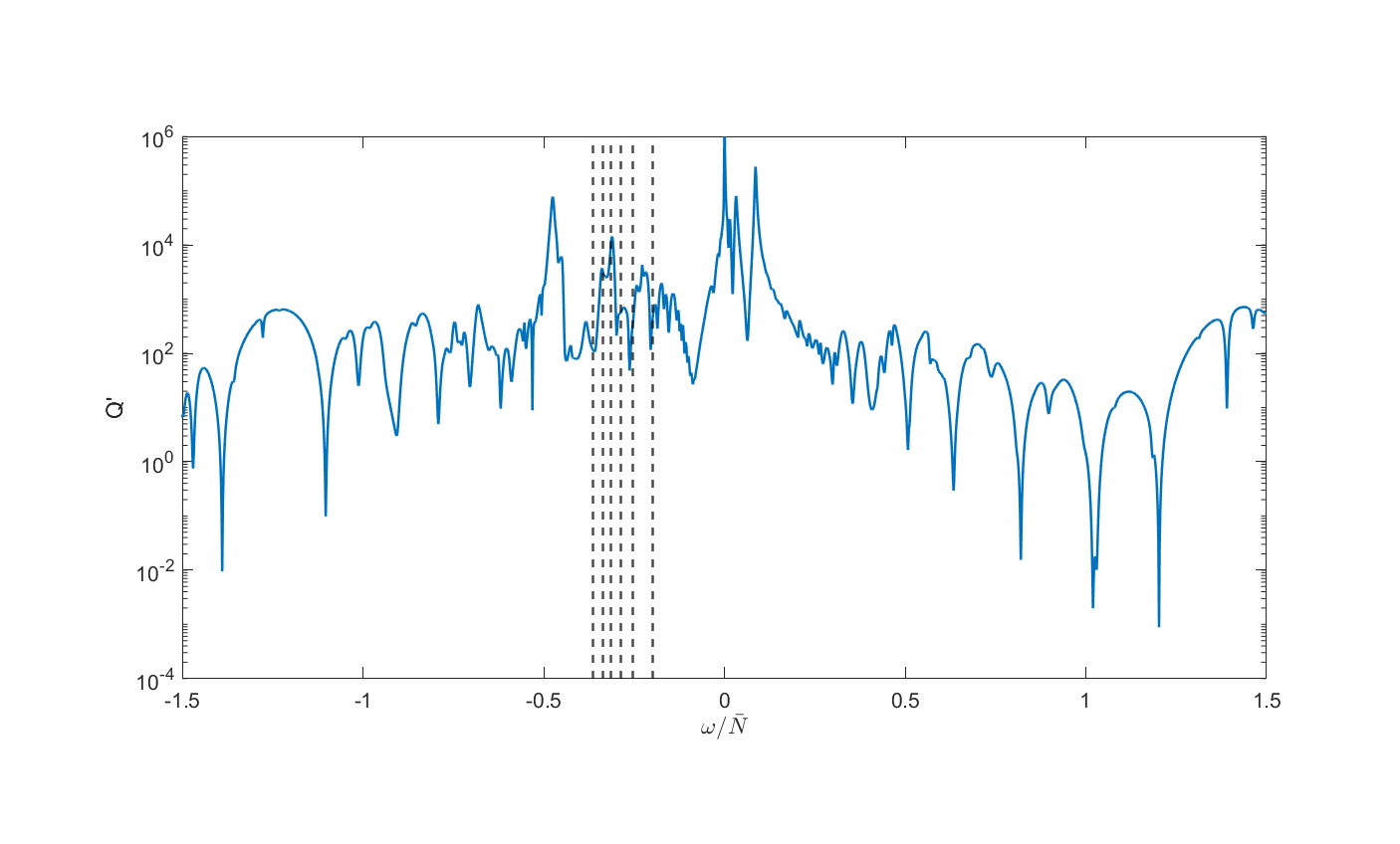}\label{fig:Saturn_Q}}
\caption{Example of total dissipation and modified tidal quality factor for Saturn-like parameters, with the tidal forcing frequency of six of Saturn's moons over-plotted: Mimas, Enceladus, Tethys, Dione, Rhea and Titan.}\label{fig:Saturn}
\end{figure*}

Finally, we explore cases with parameter values that are as consistent as possible with the latest observational constraints for Saturn. Our model is the simplest global one able to capture the dynamics of stratified layers, since we adopt the Boussinesq approximation, and hence neglect realistic variations in density throughout the planet. It is still instructive however to compute the tidal response in this model and to compare with observational constraints. To do this, we adopt values similar to those suggested by \cite{Mankovich2021} and consider an example where $\alpha=0.1$ (tiny solid core for which the precise value of $\alpha$ is unlikely to be important), $\beta=0.6$ (extended stably stratified ``dilute core"), $\bar{N}=2$ and $\Omega=0.4$. We consider both a uniformly stratified case as well as a staircase structure with one, five and nine steps. Figure~\ref{fig:Saturn_diss} shows the dissipation rate using these four profiles and Figure~\ref{fig:Saturn_Q} shows the modified tidal quality factor $Q'$ for the case of a uniformly stratified layer, where $Q'=15|\omega|/(16\pi D_{total})$ -- see section 4 of \citetalias{Pontin2023} for further details. The black vertical lines show the tidal frequency of six of Saturn’s major moons: Mimas, Enceladus, Tethys, Dione, Rhea and Titan, as points of reference for the relevant frequency regimes. The tidal forcing frequencies are also reported in Table~\ref{tab:Saturn_moons}.

\begin{table}\flushleft
\begin{tabular}{|l|l|l|l|l|}
\hline
Satellite & Period& $\Omega_o$& $\omega$   & $\omega$ \\ \hline
 Units & (days)  & ($\omega_d$) & $(\omega_d)$ & $(\omega_d / \bar{N})$ \\ \hline
Mimas     & 0.942  & 0.176          & -0.395                                         & -0.198      \\ %\hline
Enceladus & 1.37  & 0.121         & -0.505                                        & -0.253      \\ %\hline
Tethys    & 1.89  & 0.0879         & -0.572                                         & -0.286      \\ %\hline
Dione     & 2.74  & 0.0606         & -0.626                                         & -0.313      \\ %\hline
Rhea      & 4.52  & 0.0367        & -0.674                                         & -0.337      \\ %\hline
Titan     & 15.9 & 0.0104         & -0.727                                         & -0.363      \\ \hline
\end{tabular}\caption{Tidal forcing frequencies $\omega=2(\Omega_o-\Omega_s)$, for six of Saturn's major moons. Data taken from \cite{JPL}. In this case $\bar{N}=2\omega_d$.}\label{tab:Saturn_moons} 
\end{table}

We see that all four cases show qualitatively and quantitatively similar dissipation profiles, and the typical level of dissipation (e.g.~quantified by the frequency average) is almost unchanged as the step number is varied. In these examples all three wave frequency ranges -- gravito-inertial, inertial, and surface gravity modes -- overlap, making it difficult to separate the behaviour of each. However, given the sensitivity to the tidal frequency due to the moons exhibited, stably stratified layers could have important implications for Saturn's tidal dissipation rates.

The tidal quality factor $Q'$ obtained in Figure~\ref{fig:Saturn_Q} ranges from approximately $10^2$ to $10^4$ at the frequencies relevant for Saturn's moons. This is comparable to observational constraints from the migration rates of Saturn's moons \citep[e.g.][]{Lainey2017}, which provide $Q'\approx (0.94\pm 0.44)\times 10^4$. Our idealised calculations therefore highlight the importance of considering stably-stratified layers on the excitation and dissipation of inertial and internal waves in planets. This figure demonstrates that efficient tidal dissipation rates -- sufficient to explain the observed migration rates of Saturn’s moons -- are predicted at the frequencies of the orbiting moons due to the excitation of inertial and gravito-inertial waves in our models with stable layers. Note that we do not require resonance locking to operate \citep[e.g.][]{Fuller2016}, though the possibility of resonance locking and whether or not it is a viable mechanism in giant planets should be explored further. The presence of a stably stratified dilute core, and its effects in enhancing inertial wave excitation in the overlying convective envelope, as well as -- to a lesser extent -- the additional gravito-inertial mode excitation in the fluid core itself, may be the key mechanisms of tidal dissipation in Saturn, and which could explain observations. We envisage similar results may apply to Jupiter also \citep[see also][]{Lin2023,Dewberry2023}, though see the caveats in \S~\ref{dilute}.

\section{Conclusions}
\label{conclusions}

We have presented new theoretical models of giant planets containing stable layers similar to those constrained observationally for Saturn, and hypothesised for Jupiter, to explore dissipation of tidal flows inside these planets. We have studied the role of stably-stratified and semi-convective layers on tidal dissipation in rotating giant planets, extending our prior work without rotation \citep[in paper 1,][]{Pontin2023} to account for Coriolis forces. Rotation permits the propagation of inertial waves that can significantly enhance tidal dissipation in neutrally-stratified convective regions. These can be tidally forced for frequencies $|\omega|\leq 2|\Omega|$, which is typically the most relevant range for solar and extrasolar giant planets. Rotation also modifies the properties of internal, surface and interfacial gravity modes (the former are then commonly referred to as gravito-inertial waves). With our idealised (Boussinesq) model of a rotating and tidally-forced planet, we analysed the dissipative fluid response in a spherical shell using both linear theoretical analysis and numerical calculations (with high-resolution spectral methods). Our parameter study analysed the properties of both the dissipative forced response and free oscillation modes as we varied the properties of any stably-stratified layers in the planet (including their sizes, strengths and compositions -- layered or smooth), the sizes of any solid core, the viscosity and thermal diffusivity and the rotation rate (relative to the dynamical frequency), in addition to scanning the full range of relevant tidal frequencies.

We found the presence of an extended stably-stratified fluid core in a giant planet significantly enhanced tidal wave excitation of both inertial waves in the convective envelope and gravito-inertial waves in the core. We have demonstrated that efficient tidal dissipation rates -- sufficient to explain the observed migration rates of Saturn’s moons -- are predicted at the frequencies of the orbiting moons due to the excitation of inertial waves in convective envelopes in our models with interior stable layers, and to a lesser extent gravito-inertial waves in the fluid core itself. Stable layers could also be important for tidal evolution of hot and warm Jupiters, and hot Neptunes, providing efficient tidal circularisation rates \citep[see discussion in][]{Pontin2023}.

We analysed both the frequency-dependent and frequency-averaged response to establish some overarching trends in this problem, building upon our non-rotating study in \cite{Pontin2023}. We established that increasing the rotation rate typically enhances the inertial wave response, in turn increasing the total dissipation. Increasing the size of the core, whether it is a solid core or a stably stratified layer (layered or smooth), significantly increases the dissipation rate. We find that gravito-inertial waves excited in a stably stratified layer can enhance the dissipation compared to that of a solid core with the same radius, depending on the tidal frequency that is relevant. As in non-rotating cases, we established that provided a sufficient number of steps are present in a staircase-like density structure, the region will behave like a uniformly stably stratified layer when considering any frequency-integrated quantities. However, important differences in the frequencies of the free modes, and hence the enhancement of dissipation at the frequencies of these resonances, are found, which can lead to significant differences in the response to different stratified models at a given tidal frequency.

We found that a key parameter in the excitation of inertial waves is the size of the outer convective envelope (i.e.~the radius to which the dilute core extends). The dissipative properties of the envelope were shown to be approximately independent of the properties of the stratified layer beneath it, whether it is a stably stratified layer (layered or smooth), single interface or solid core for Saturn-like parameter values. The buoyancy frequency profile beneath this layer can alter the frequencies of the free modes however. This conclusion is valid for low frequencies relative to the buoyancy frequency of the stable layer, as expected for Saturn. Different results might be found for more weakly stratified layers.

Future work should study more sophisticated planetary models that also account for magnetism and differential rotation \citep{Baruteau2013,Guenel2016,Wei2018,LO2018,Astoul2021,AB2022}, as well as the interaction of inertial waves with turbulent convection \citep[as opposed to large-scale tidal flows e.g.,][]{Duguid2020}. It would also be of interest to separate the thermal and compositional contributions to the buoyancy and study whether double-diffusive effects could be important in this problem.

\noindent
%\begin{acknowledgements}
CMP was supported by STFC PhD studentship 2024753. AJB was supported by STFC grants ST/R00059X/1, ST/S000275/1 and ST/W000873/1. RH was supported by STFC grants ST/S000275/1 and ST/W000873/1. We thank St\'{e}phane Mathis and Quentin Andr\'{e} for discussions at an early stage in this project, and Aur\'{e}lie Astoul, Chris Jones and Gordon Ogilvie for helpful feedback. We would like to thank the two reviewers for their constructive and positive referee reports.
%\end{acknowledgements}

\appendix

\section{Analytical derivation of f-mode frequencies with rotation}\label{analyticalfmode}

Here we outline a derivation of the f-mode frequencies of a uniformly rotating homogeneous spherical fluid body i.e.~neglecting centrifugal deformations, consistently with the model adopted throughout our paper. We follow \cite{Barker2016} and use the elegant Lagrangian perturbation theory of \cite{Lebovitz1989,Lebovitz1989a}. For this section only we define $\boldsymbol{\xi}\in \mathbb{R}^3$ to be the Lagrangian displacement vector, satisfying (in the rotating frame)
\begin{align}
\label{app1}
&\partial_t^2\boldsymbol{\xi}+2\boldsymbol{\Omega}\times \partial_t\boldsymbol{\xi}-\nabla(\boldsymbol{\xi}\cdot\nabla p)+\nabla \Delta p=0, \\
& \nabla\cdot\boldsymbol{\xi}=0,
\end{align}
where $\Delta p$ is the Lagrangian pressure perturbation (which vanishes on the free surface) and we neglect perturbations to the gravitational potential. For our basic state, we assume a fixed spherically-symmetric gravitational potential such that $\nabla p = -\omega_d^2 \boldsymbol{r}$. We seek solutions in the form of solenoidal vector fields whose components are polynomials in the Cartesian coordinates up to a specified harmonic degree $\ell=2$ (for the pressure perturbation), such that
\begin{equation} \label{basis}
\boldsymbol{\xi}(\boldsymbol{x},t)=\sum_{i=1}^{i_\mathrm{max}}\alpha_i(t) \boldsymbol{\xi}_i(\boldsymbol{x}),
\end{equation}
where $\alpha_i(t)\in\mathbb{C}$ is an amplitude and $i_\mathrm{max}\in\mathbb{Z}^+$ is the number of basis elements considered. To analyse surface gravity modes, we consider irrotational motions that perturb the boundaries of the body (i.e.~basis elements belonging to the subspace $\mathrm{U}_2$ in \citealt{Barker2016}). To exactly represent all surface gravity modes with $\ell=1$, we require $i_\mathrm{max}=3$ linearly-independent vectors,
\begin{align}
\label{l1basis}
& \left(1,0,0\right)^T, \quad \left(0,1,0\right)^T, \quad \left(0,0,1\right)^T.
\end{align}
To exactly represent all such modes up to $\ell=2$, we require $i_\mathrm{max}=8$, of which the first 3 are those in (\ref{l1basis}), with 5 additional linearly independent vector fields 
\begin{align}
& \left(y,x,0\right)^T, \quad \left(0,z,y\right)^T, \quad \left(z,0,x\right)^T, \quad \left(x,-y,0\right)^T, \quad \left(x,y,-2z\right)^T.
\end{align}
We can project (\ref{app1}) onto the basis given in (\ref{basis}) using the inner product (involving integration over volume $V$)
\begin{equation}
\langle\boldsymbol{\xi}_i,\boldsymbol{\xi}_j\rangle = \int_V\boldsymbol{\xi}^*_i\cdot\boldsymbol{\xi}_j\, \mathrm{d}V,
\end{equation}
to obtain
\begin{equation}
\langle \boldsymbol{\xi}_i,\boldsymbol{\xi}_j\rangle \ddot{\alpha}_j + \langle \boldsymbol{\xi}_i,2\boldsymbol{\Omega}\times \boldsymbol{\xi}_j\rangle\dot{\alpha}_j+\langle \boldsymbol{\xi}_i,-\nabla(\boldsymbol{\xi}_j\cdot\nabla p)\rangle\alpha_j=0,
\end{equation}
where a sum over $j$ is implied. Seeking solutions $\alpha_j\propto \mathrm{e}^{-\mathrm{i}\omega t}$, this problem is converted to a quadratic eigenvalue problem of the form
\begin{equation}
(-\omega^2 \mathrm{M} - \mathrm{i}\mathrm{A}\omega +\mathrm{B})\boldsymbol{\alpha}=\boldsymbol{0},
\end{equation}
for appropriate matrices M, A and B and column vector $\boldsymbol{\alpha}$ with elements $\alpha_j$. To obtain all $\ell=1$ surface gravity modes, we restrict $i_\mathrm{max}=3$ to obtain the 6 eigenvalues
\begin{equation}
\omega = \underbrace{\pm\Omega\pm \sqrt{\Omega^2+\omega_d^2}}_{m=1}, \quad \mathrm{or}\quad \underbrace{\pm \omega_d}_{m=0},
\end{equation}
where we have identified the corresponding azimuthal wavenumber magnitude $m$. These are unimportant but strictly unphysical modes that involve the body oscillating about a fixed position in space, and arise because we have fixed the gravitational potential. On the other hand for $i_\mathrm{max}=8$ we obtain the additional eigenvalues
\begin{equation}
\label{l2modes}
\omega =\underbrace{\pm\Omega\pm \sqrt{\Omega^2+2\omega_d^2}}_{m=2}, \quad\mathrm{or}\quad \underbrace{\pm\frac{\Omega}{2}\pm \frac{1}{2}\sqrt{\Omega^2+8\omega_d^2}}_{m=1}, \quad\mathrm{or}\quad \underbrace{\pm \sqrt{2}\omega_d}_{m=0}.
\end{equation}
These are the frequencies of all surface gravity modes with $\ell=2$. When $\Omega^2\ll \omega_d^2$, i.e. for slow rotation, we find the latter can be written
\begin{equation}
\omega =\pm\sqrt{2}\omega_d\pm \frac{m}{2}\Omega,
\end{equation}
which agree with the results of \cite{Lebovitz1961} for a (self-gravitating) Maclaurin spheroid in the same limit. In this paper we have adopted a fixed gravitational potential for simplicity, and because the effects of surface perturbations on the gravitational potential are unrealistically enhanced in a Maclaurin spheroid over a more realistic planet model that is denser at its centre than near its surface \citep[see Appendix A of][]{Barker2016}.

\bibliographystyle{aasjournal}
\bibliography{paperIII}

\end{document}